\documentclass[10pt,journal,compsoc]{IEEEtran}
\ifCLASSOPTIONcompsoc
  \usepackage[nocompress]{cite}
\else
  \usepackage{cite}
\fi
\ifCLASSINFOpdf
\else
\fi

\usepackage{pgfplots} 
\usetikzlibrary{patterns}
\usepackage{booktabs} 
\usepackage{url}
\usepackage{xspace}
\usepackage[font=normal]{caption}
\usepackage{subcaption}
\usepackage{float}
\usepackage{textcomp}
\usepackage{multirow,makecell}
\usepackage{pifont}
\usepackage{tikz}
\usepackage{bbm}
\usepackage[inline]{enumitem}
\usepackage{adjustbox}
\usepackage{listings}
\usepackage{tikz}
\usetikzlibrary{tikzmark}
\usetikzmarklibrary{listings}
\usepackage{amssymb}
\usepackage{amsmath}
\usepackage{graphicx}
\usepackage{makecell}
\usepackage{xspace}
\usepackage{multicol}
\usepackage{colortbl}

\newcommand{\etal}{{\em et al.}\xspace}

\newcommand{\BfPara}[1]{{\vspace{1.5mm}\noindent\bf#1.}\xspace}

\newcommand{\EMP}[1]{{\vspace{0.5mm}\noindent\em#1.}\xspace}

\usepackage{xcolor}

\definecolor{darkgreen}{rgb}{0.0, 0.3, 0.13}
\definecolor{darkred}{rgb}{0.2, 0.0, 0.13}

\usepackage{tcolorbox}
\tcbset{textmarker/.style={%
        parbox=false,boxrule=0mm,boxsep=0mm,arc=0mm,
        outer arc=0mm,left=3mm,right=3mm,top=3pt,bottom=3pt,
        toptitle=1mm,bottomtitle=1mm}
        }
        
\newtcolorbox{blueBox}{textmarker,
    colback=blue!10!white}

\lstdefinestyle{myStyle}{
  belowcaptionskip=1\baselineskip,
  breaklines=true,
  language=C++,
  showstringspaces=false,
  basicstyle=\footnotesize\ttfamily,
  keywordstyle=\bfseries\color{green!40!black},
  commentstyle=\itshape\color{purple!40!black},
  identifierstyle=\color{blue},
  stringstyle=\color{orange},
  numbers=left,
  firstnumber=1,
}

\usepackage{hyperref}

\hypersetup{
	plainpages=false,
	colorlinks,
	urlcolor=blue,
	linkcolor=blue,
	citecolor=blue,
	bookmarksnumbered
}

\usepackage[framemethod=TikZ]{mdframed}
\definecolor{boxcolor}{RGB}{240, 248, 255} 

\newmdenv[
  backgroundcolor=boxcolor,
  linewidth=0.25pt,
  innerleftmargin=3pt,
  innerrightmargin=3pt,
  innertopmargin=1pt,
  innerbottommargin=1pt
]{takeawaybox}

\usepackage{tikz}
\usepackage{tabularx}
\usepackage{colortbl}

\usepackage{amsmath}
\newtheorem{takeaway}{Takeaway}

\usepackage{orcidlink}

\begin{document}

\title{Empirical Evaluation of Concept Drift in ML-Based Android Malware Detection}

\author{Ahmed Sabbah~\orcidlink{0000-0001-5034-8038}, Radi Jarrar~\orcidlink{0000-0003-2692-8096}, Samer Zein~\orcidlink{0000-0003-3720-4384},  David~Mohaisen~\orcidlink{0000-0003-3227-2505}\IEEEmembership{, Senior Member,~IEEE}

\IEEEcompsocitemizethanks{\IEEEcompsocthanksitem Ahmed Sabbah, Radi Jarar, and Samer Zein are with the Department of Computer Science, Birzeit University, Palestine.
E-mail: asabah@birzeit.edu, rjarrar@birzeit.edu, szain@birzeit.edu. \IEEEcompsocthanksitem David Mohaisen is with the Department of Computer Science, University of Central Florida, Orlando, FL 32816 USA. E-mail: mohaisen@ucf.edu (Corresponding author).}}


\IEEEtitleabstractindextext{%
\begin{abstract}
Despite outstanding results, machine learning-based Android malware detection models struggle with concept drift, where rapidly evolving malware characteristics degrade model effectiveness. This study examines the impact of concept drift on Android malware detection, evaluating two datasets and nine machine learning and deep learning algorithms, as well as Large Language Models (LLMs). Various feature types—static, dynamic, hybrid, semantic, and image-based—were considered. The results showed that concept drift is widespread and significantly affects model performance. Factors influencing the drift include feature types, data environments, and detection methods. Balancing algorithms helped with class imbalance but did not fully address concept drift, which primarily stems from the dynamic nature of the malware landscape. No strong link was found between the type of algorithm used and concept drift, the impact was relatively minor compared to other variables since hyperparameters were not fine-tuned, and the default algorithm configurations were used. While LLMs using few-shot learning demonstrated promising detection performance, they did not fully mitigate concept drift, highlighting the need for further investigation.
\end{abstract}

}

\maketitle

\IEEEdisplaynontitleabstractindextext
\IEEEpeerreviewmaketitle

\section{Introduction}\label{sec:introducion}
In today's fast-paced and information-driven world, mobile apps running on smart devices are central to our modern life. According to Kaspersky, in the first quarter of 2025, over 12 million mobile attacks involving malware, adware, or potentially unwanted applications were blocked, and more than 180,000 malicious and potentially unwanted installation packages were identified~\cite{Kivva2025}. Additionally, according to a report presented at Mobile World Congress 2025, Trojan banker attacks on smartphones increased by 196\% compared to the previous year. Over 33.3 million global attacks targeting smartphone users were detected, as cybercriminals increasingly rely on mass malware distribution to steal banking credentials~\cite{Kaspersky2025}.
However, mobile malware is on the rise, highlighting the need to develop effective approaches for efficiently analyzing, understanding, and detecting such threats. Security analysis of mobile applications focuses on understanding their behavior and intent to assess whether an application is malicious with approaches that use network traffic~\cite{WangYCYZC18}, inner software interactions~\cite{CaiMRY19}, and permissions utilization~\cite{AroraPC20}. Each of these approaches can be implemented using static, dynamic, and hybrid analysis to generate the intended mobile app artifacts. These approaches are used to gain information and insights that can be utilized for the classification task~\cite{Alzubaidi21}.  

The static, dynamic, and hybrid techniques provide a wealth of analysis modalities and data artifacts that can be used to understand the intent of the software. Manually sniffing through the generated artifacts is cumbersome and does not scale to the size of the problem at hand. To this end, there has been a growing trend in utilizing machine learning (ML) algorithms to automatically understand the intent of mobile apps using static, dynamic, and hybrid analysis artifacts as features. Those approaches have been shown to have excellent accuracy in extrapolating malware labels, sometimes reaching perfect accuracy. For example, studies that used static features~\cite{TchakounteNKU21} achieved an accuracy of $0.978$, while another study~\cite{DhalariaG20} reached an accuracy score of $0.959$. The dynamic approach achieved high accuracy scores of $0.97$ in~\cite{AmerE22} and $0.99$ in~\cite{ZhangMZRNJY22}. 

ML is heavily utilized to detect malware in the context of Android apps, which brings about an arms race between malware authors and defenders. With new malware samples and families emerging, ML algorithms are no longer effective. This was shown by Chen \etal, who found that after training an Android malware classifier on data for one year, the F1 score dropped from $0.99$ to $0.76$ within just six months of deployment in new test samples~\cite{ChenDW23}. In the real world, this idea is manifested in a conflicting reality with the promised accuracy: despite the reported accuracy of near-perfect detection, malware evolution is still a serious threat. The main reason is that new malware strains frequently adapt to maximize profit gains, variations emerge as new vulnerabilities are uncovered, and adversaries quickly change tactics when encountering defenses. Consequently, the new test distribution diverges from the initial training distribution, a phenomenon termed concept drift~\cite{Moreno-TorresRACH12}. As a result, the classifier performance slowly declines as the model fails to classify the new sample accurately. 

Concept drift has been recognized in the literature on malware analysis and detection, and several preliminary strategies have been developed to address the issue that affects feature space or data space. There are two primary strategies to address concept drift per the limited literature. The initial strategy involves creating systems that are inherently more resistant to drift by developing stronger feature spaces. Due to their latent feature spaces, neural networks have been proposed to offer better generalization to new variants, thus showing greater resilience to concept drift~\cite{PendleburyPJKC19}. However, creating robust feature spaces remains an unresolved research issue, and it is unclear whether a malware representation immune to concept drift can be developed~\cite{BarberoPPC22}. Therefore, it is essential to understand the factors that have a role beyond concept drift in ML and deep learning detection models, a gap that this work recognizes and understands by measurement.  

{\BfPara{Contributions} Our contributions are as follows:}
\begin{enumerate}[leftmargin=*]

\item {\bf Understanding the effect of feature type}. We examine the impact of different feature types---static, dynamic, and hybrid—as well as the data collection environment (real device vs. emulator) on resilience to concept drift, using the Kronodroid~\cite{Guerra-ManzanaresBN21} and Troid~\cite{KinoonAAM24} datasets.

\item {\bf Investigating the effect of algorithm types on detection models}. To investigate the impact of different algorithms on detection performance, we employed a variety of classifiers, including traditional ML models such as Random Forest (RF) and Gradient Boosting (GB), as well as deep learning models like Convolutional Neural Networks (CNN) and Recurrent Neural Networks (RNN). We analyzed which algorithms demonstrate greater resilience to concept drift over time. Additionally, we evaluated two LLMs using a few-shot approach.

\item {\bf Understanding the impact of detection approach}. We investigate the effects of various detection methods on concept drift, including image-based techniques, semantic-based approaches using Term Frequency-Inverse Document Frequency (TF-IDF) with API calls, and numeric-based methods (e.g., permissions and system calls).

\item {\bf Studying the role of temporal data imbalance} We simulate concept drift using a cross-year strategy and show how temporal imbalance affects model performance. We evaluate balancing strategies and quantify their ability to reduce drift effects over time.
\end{enumerate}

\BfPara{Organization} The remainder of this paper is structured as follows. Section~\ref{sec:Motivation} outlines the motivation and research questions. The background on concept drift is covered in Section~\ref{DriftSection}. Related work is discussed in Section~\ref{sec:related}. Section~\ref{sec:DataRepresentationLearning} details the data collection and analysis methodology. Experimental results and a discussion are in Section~\ref{sec:Experiments} followed by the limitations in Section~\ref{sec:limitations} and conclusion in Section~\ref{sec:conclusion}.

\section{Motivation Research Questions}\label{sec:Motivation}

Android malware detection models often struggle with concept drift, wherein the characteristics of malware evolve over time, diminishing the effectiveness of existing models. As malware developers continuously introduce new obfuscation techniques and functionalities, both static and dynamic detection approaches experience a decline in accuracy. This presents a critical challenge in maintaining the security of Android devices, as current detection methods fail to adapt to the evolving threat landscape. This research investigates the impact of concept drift on various ML models for Android malware detection, with the goal of developing strategies to understand and mitigate its root causes. Specifically, our work seeks to answer the following research question:

\BfPara{RQ}~{How prevalent is concept drift across different ML–based Android malware detection approaches, including both binary (detection) and multi-class (family-labeling) classification settings?}

We divide this question into three subquestions: {\bf RQ-1.1.} To what extent do different feature types and data collection environments influence the resilience of detection models to concept drift?  {\bf RQ-1.2.} To what extent do different classification algorithms and detection approaches influence model performance and resilience to concept drift?  {\bf RQ-1.3.} How does temporal data imbalance contribute to concept drift, and to what extent can data balancing techniques mitigate its effects on detection accuracy over time?

Although the model demonstrated high accuracy on a dataset without accounting for temporal factors, variations in accuracy across classifiers stem from their inherent characteristics and the features employed. These differences can be mitigated through parameter tuning. The primary focus is on the impact of temporal factors, particularly when evaluating the model on new samples. For example, a model trained on data from 2008-2015 may be tested on data from 2022. To investigate concept drift, we consider:

\begin{enumerate}[leftmargin=*] \item \textbf{Ignoring temporal factor.} Algorithms were tested on all features with timestamps. Seven algorithms evaluated static, dynamic, and hybrid features using machine and deep learning methods and assessed them without drift.
\item \textbf{Cross years strategy.} 
Models were trained on data from one year and tested on others (e.g., trained on data from 2008, tested on data from 2009-2020). Static, dynamic, and hybrid features were evaluated, both with and without balancing (i.e., Synthetic Minority Oversampling Technique, or simply SMOTE).

\item \textbf{Incremental strategy.} 
Models were trained incrementally by adding years to the training set and testing on new data. For example, the model trained on data from 2008-2009 was tested on data from 2010-2020, then retrained with 2010 added, and so forth.

\item \textbf{Grouping strategy.} 
Due to the lack of evenly distributed samples across years, for malware family classification, we grouped multiple years into subsets for training and testing. In this strategy, we created three subsets of years: 2008-2012, 2013-2016, and 2017-2020.

\end{enumerate}

\section{Concept Drift}
\label{DriftSection}
Concept drift was first introduced in 1986 by Schlemmer \etal~\cite{SchlimmerG86} to refer to the unexpected change in the statistical properties or defining features of the target variable over time in non-stationary data distributions. This change presents a significant challenge for ML models that assume stationary input data distributions, where training and testing data are expected to be very similar~\cite{XiangZCW23}. In real-world scenarios, such as malware detection, the evolving data can lead to concept drift, which impacts the accuracy of the model over time. Concept drift can arise in multiple cases, including changes in feature distributions. 

\BfPara{Root Causes} Xiang \etal~\cite{XiangZCW23} illustrated three causes of concept drift based on joint probability distribution.
\begin{itemize}[leftmargin=*]
    \item {\em Virtual concept drift:} This type of concept draft happens in cases where the probability of $x$ changes while the probability of $y$ given $x$ remains unchanged. In this case, the decision boundary remains unaffected, and only the feature space changes. In the malware context, this cause occurs when the malware evolves, and the adversaries change the code of the app (static features) or behavior (dynamic features). However, the malware still belongs to the same type and family. The case is captured by $P_{t0}(x) \neq P_{t1}(x) \quad \text{and} \quad P_{t0}(y|x) = P_{t1}(y|x)$.

    \item {\em Real concept drift:} When the probability of
$y$ given $x$ changes while the probability of $x$ remains constant. This case can be expressed by $P_{t0}(y|x) \neq P_{t1}(y|x) \quad \text{and} \quad P_{t0}(x) = P_{t1}(x)$. This scenario directly affects the ML model, changing both the feature space and the decision boundary, for example, the emergence of a new malware family.

    \item {\em Hybrid concept drift:} This scenario includes both virtual and real concept drift and can exist in the data stream simultaneously. This case can be expressed as $P_{t0}(x) \neq P_{t1}(x)$ and $P_{t0}(y|x) \neq P_{t1}(y|x)$.
    
\end{itemize}

\BfPara{Concept Drift Types} Concept drift can take different shapes over time: abrupt, incremental, gradual, and recurring drift~\cite{XiangZCW23}. Each type represents a different shape of change in the fundamental concept of the data stream. Abrupt drift denotes sudden shifts from one concept to another in a short time frame. However, incremental drift is similar but slow and there are continuous shifts between concepts. Gradual drift presents periodic shifts between concepts. The last type is recurring drift, which includes the periodic reappearance of previous concepts over time~\cite{XiangZCW23}.

\section{Related Work}\label{sec:related}

\subsection{Malware Analysis and Detection}
Malware detection can be carried out using three main approaches: static, dynamic, or hybrid, and there has been a plethora of work on each direction, which we review in the following, then highlight the issue of concept drift in malware analysis and detection.

\BfPara{Static Approach} The static analysis approach decompiles and disassembles code without executing the application, extracting features from APK files for malware classification. Alzubaidi~\cite{Alzubaidi21} identified three primary feature extraction methods: signature-based, permission-based, and Dalvik bytecode. Karbab \etal proposed a resource-based method, categorizing it as semantic-based rather than Dalvik bytecode~\cite{KarbabDDM21}. Vishnoi \etal associated misuse detection with knowledge-based methods, whereas anomaly detection aligns with the behavior-based dynamic approach~\cite{VishnoiMNP21}. The primary objective of static analysis remains feature extraction for malware detection models.

\EMP{Signature-based method.} Signature-based detection creates unique signatures for known malware families by extracting features such as permissions and content strings~\cite{LiFWCZYWG22}. Ngamwitroj \etal achieved $0.865$ accuracy in malware detection with a method using permissions and transmission data from the manifest file~\cite{NgamwitrojL18}. Tchakounte \etal introduced LimonDroid, which combines fuzzy hashing with YARA rules, achieving $0.978$ accuracy on 341 applications~\cite{TchakounteNKU21}.

\EMP{Permission-based method} Ilham \etal extracted permissions from the manifest file, and used them to achieve $0.98$ accuracy using RF and SMO algorithms in  detection~\cite{IlhamAA18}. Katos \etal's method, based on the composition ratio of permission pairs, achieved $0.97$ accuracy on the Drebin dataset~\cite{KatoSS21}. Other studies combined permission features with APK features, achieving varying accuracies~\cite{UroojSMAR22, ShatnawiYY22}.

\EMP{Resource-based method} This method relies on meta-data in the manifest file. Urooj \etal showed that dangerous permissions indicate malware behavior~\cite{UroojSMAR22}. Millar \etal presented a CNN-based network, achieving detection rates of $0.91$ on Drebin and $0.81$ on AMD datasets~\cite{MillarMRM21}. Dhalaria \etal achieved  $0.959$ accuracy by combining features from \textit{classes.dex} and \textit{AndroidManifest} files~\cite{DhalariaG20}.

\EMP{Semantic-based method} This approach uses various data sources for semantic information extraction. Bai \etal proposed a scheme converting network traffic into text for feature representation~\cite{BaiLLQH21}. Zhang \etal used the method-level correlation of API calls for Android malware detection~\cite{ZhangLZP19}. Related to that is the image-based approach, where static features are converted into grayscale images for further processing~\cite{XingJEJW22}. Unver \etal transformed features from Manifest.xml and DEX files into grayscale images for malware detection~\cite{UnverB20}. Although efficient, the image-based approach remains highly vulnerable to code manipulation and obfuscation techniques that thwart detection~\cite{UnverB20}. Hasib~\etal proposed MCNN-LSTM, a hybrid model combining CNNs for spatial feature extraction and LSTMs for sequence learning, achieving 0.9971 accuracy and a 0.98 F1 score on multi-class text classification with imbalanced data~\cite{HasibAKMSMYJAR23}. However, their domain differs from ours, and performance on imbalanced data in one context does not guarantee similar results elsewhere. Our study of imbalanced data is a step toward understanding its impact on concept drift, not classification. While not applied to malware detection, the architecture is transferable to static analysis tasks with sequential patterns in tokenized code, API calls, or permissions.

\EMP{Transformer and LLM-based semantic methods}  Recent advances in Android malware detection increasingly use transformer-based models and LLMs to extract contextual semantic features, improving generalization to new malware. While not explicitly targeting concept drift, these models offer contextual awareness and adaptability useful for evolving threats. MalBERT applies BERT to static features like permissions, intents, and API calls, modeling apps as token sequences to capture contextual relationships. It achieved 0.976 accuracy in binary and 0.91 in multi-class classification~\cite{RahaliA2021}. García-Soto~\etal used CodeT5-generated embeddings from decompiled Java code and trained an LSTM classifier, achieving an average accuracy of 0.81 over ten runs despite sequence length constraints~\cite{Garcia-SotoMHC22}. Extending the semantic paradigm, Li~\etal proposed a multimodal malware detection approach that fuses features from both source code and binaries. Java code is segmented by GUI structure and processed using a pre-trained language model, while binary code is converted to grayscale images and analyzed with a fine-tuned vision model. The method achieved 0.977 precision and 0.984 recall on two benchmark datasets~\cite{LiLLZZL25}. Taking an alternative approach, Tang~\etal developed an unsupervised anomaly detection approach using low-level hardware performance counters. Their method does not rely on labeled data, making it inherently adaptable to evolving threats. Although not framed in terms of drift, this unsupervised design aligns well with the core goals of drift-resistant malware detection, achieving up to 0.995 detection accuracy for shellcode injections~\cite{TangSS14}. Additionally, recent surveys by Wang \etal~\cite{WangNLZ2025}, Al-Karaki \etal~\cite{AlKarakiKO2024}, and Lin and Mohaisen~\cite{LinM25} provide broad overviews of LLMs in software and malware analysis. While these works do not focus on concept drift, they highlight key factors such as model robustness, zero-shot generalization, and adaptability---challenges directly relevant to designing effective drift-aware malware detectors.

\begin{table}
\begin{center}
\caption{\normalfont \normalfont A comparison of a set of the literature works on static, dynamic, and hybrid analysis. The static techniques are broken down into signature (S), permission (P), resources (R), semantic (Se), and image (I) based techniques.}\vspace{-3mm}
\scalebox{0.80}{
\begin{tabular}{p{10em} p{2em} p{6em} p{5em}  p{5em}}
\hline
\textbf{Reference} &\textbf{Year} & \textbf{Approach} & \textbf{Method} & \textbf{Accuracy}   \\ \hline
Ngamwitroj \etal \cite{NgamwitrojL18} & 2018 & Static (S) & Statistical & 0.865 \\
Tchakounte \etal \cite{TchakounteNKU21} & 2021& Static (S) & Rule & 0.978 \\ 
Ilham \etal\cite{IlhamAA18} & 2018  &  Static (P) & ML & 0.98  \\ 
Sahin \etal \cite{SahinKAK23} & 2021 & Static (P) & ML & 0.960 \\
Millar \etal \cite{MillarMRM21} & 2021 & Static (P/R) & ML & 0.959 \\
Shatnawi \etal \cite{ShatnawiYY22} & 2022 & Static (P/R) & ML & 0.940 \\
Bai \etal \cite{BaiLLQH21} & 2021 & Static (Se) & DL & 0.926 \\
Xing \etal \cite{XingJEJW22} & 2022 & Static (I) & ML,DL & 0.96  \\
Unver \etal \cite{UnverB20} & 2020 & Static (I) & ML & 0.987  \\
Garcia-Soto \etal~\cite{Garcia-SotoMHC22} & 2022 & Static (Se) & DL, LLM & 0.810 \\
Li \etal~\cite{LiLLZZL25} & 2025 & Static (Se+I) & Multimodal & Pr: 0.977\\

\hline
Bhatia \etal \cite{BhatiaK17} & 2017& Dynamic & Statistical & 0.88 \\ 
Hu \etal \cite{HuJC20} & 2020 & Dynamic & ML & 0.90 \\
Zhang \etal \cite{ZhangMZRNJY22} & 2022 & Dynamic & Fuzzy & 0.993 \\
Mahindru \etal \cite{MahindruS17} & 2017 & Dynamic & ML & 0.997 \\
Casolare \etal \cite{CasolareDIMMS21} & 2021 & Dynamic & ML & 0.89  \\
Mahdavifar \etal \cite{MahdavifarKFAG20} & 2020 & Dynamic & ML, DL & 0.978  \\
Wit \etal \cite{WitBH22} & 2022 & Dynamic & ML & 0.72  \\ 
Tang \etal~\cite{TangSS14} & 2014 & Dynamic & Unsupervised & 0.995 \\

\hline
Rahali \etal~\cite{RahaliA2021} & 2021 & Hybrid & BERT & 0.976 \\
Wang \etal~\cite{WangZH22} & 2022 & Hybrid & ML & F1: 0.975  \\
Tidke \etal~\cite{TidkeKT18} & 2018& Hybrid & ML & NA\\ 
Zhang \etal~\cite{ZhangXMZLT21} & 2021 & Hybrid & ML & 0.973   \\
Amer \etal \cite{AmerE22} & 2022 & Hybrid & ML & 0.970  \\
\hline
\end{tabular}}
\label{tab1:Static}\label{tab:dynamicTable}\label{tab:Hybrd1}
\end{center}\vspace{-5mm}
\end{table}

\BfPara{Dynamic Approach} Dynamic analysis focuses on the runtime app's behavior. Features can be extracted at both hardware (e.g., memory, CPU, sensors) and software (e.g., network traffic, API calls)~\cite{Alzubaidi21}. Sihag \etal used kernel-level Android logs to generate app signatures, identifying malware based on dangerous permissions~\cite{SihagSV020}. Bhatia \etal applied statistical analysis to classify apps using system calls~\cite{BhatiaK17}. Feng \etal developed EnDroid, extracting runtime behavior to detect malware using the Chi-square test~\cite{FengMSXM18}. 

\BfPara{Hybrid Approach} 
Hybrid approaches combine dynamic and static features for Android malware detection. In~\cite{WangZH22}, a hybrid method was proposed using static analysis to compare permission patterns and dynamic analysis through the memory heap to extract object relationships. This approach outperformed others on a dataset of $21,708$ apps. Jang \etal introduced Andro-Dumpsys, which combines malware and malware creator data for detection. It uses volatile memory acquisition and similarity matching with known malware and creators, enhancing detection accuracy~\cite{JangKWMK16}.

\EMP{Summary} A summary of various static, dynamic, and hybrid approaches from the literature is presented in~\autoref{tab1:Static}. These methods incorporate a range of techniques, from traditional ML algorithms to deep learning, fuzzing, and rule-based classification. In terms of performance, these approaches are competitive, achieving top accuracy levels as high as 0.997 in certain cases.

\subsection{Android Malware Concept Drift}

In supervised ML, a classifier predicts a target variable using a labeled dataset, where concept drift refers to changes in the relationship between input and target variables over time~\cite{GamaZBPB14}. Research on Android malware detection highlights the effectiveness of ML in identifying mobile malware while showing the importance of addressing concept drift~\cite{HuMZLYL17}. Previous studies have explored concept drift in both feature and data spaces. Chen~\etal conducted experiments to evaluate the impact of feature space drift compared to data space drift on the deterioration of malware detection models over time. Their findings were applied to two malware detectors---one for Android and another for PE (portable executable)---across different feature types and configurations~\cite{ChenZKYCPPCW23}.

Guerra-Manzanares \etal studied the influence of concept drift on Android malware detection by analyzing dynamic features (system calls) and highlighted the importance of timestamps in modeling concept drift~\cite{Guerra-Manzanares23}. Later, the same author examined the temporal data of malware and benign apps, developing a concept drift handling approach using a classifier pool, emphasizing the role of timestamping in detection accuracy without optimizing performance across feature sets~\cite{Guerra-ManzanaresB22a}. Their work focused on addressing concept drift by dynamically selecting the best classifier ensemble for each period, with results showing that timestamping choices significantly impact detection accuracy.

Chow \etal proposed a framework for analyzing datasets affected by concept drift, focusing on root causes, and revealed that performance drops are mainly due to the emergence of new malware families and the evolution of others~\cite{ChowKLCAP23}. Adversarial attacks are shown to mislead models by modifying malware features to appear benign~\cite{AbusnainaWAWCM23}, and Abusnaina \etal assessed the resilience of malware detectors to adversarial attacks over time, showing that such attacks can reduce accuracy by up to $0.70$, taking into account different drift directions influenced by time~\cite{AbusnainaAAAJNM22}.

Ceschin \etal investigated the effect of concept drift on Android malware classifiers using two datasets, DREBIN and AndroZoo, collected over nine years, employing Word2Vec and TF-IDF representations, along with adaptive random forest and stochastic gradient descent classifiers~\cite{CeschinBGPOG23}. Their results showed that malware evolution alters data distribution, requiring continuous classifier updates and feature extractors to maintain detection effectiveness. Their proposed method improved the F1 score by $22.05\%$ on the DREBIN dataset and $8.77\%$ on the AndroZoo dataset.

Qian~\etal~\cite{QianZHYC25} introduced LAMD, a context-driven framework for Android malware detection that addresses challenges posed by distribution drift, code noise, and structural complexity. LAMD isolates security-critical code using static analysis and backward slicing, then applies tier-wise code reasoning to guide an LLM from low-level instructions to high-level behavioral patterns. Evaluated on three time-sequenced test sets simulating increasing drift, LAMD maintained an F1 score around 0.9, while classical detectors like Drebin dropped sharply (from 0.813 to 0.616).

This study provides a comprehensive analysis of concept drift across various factors as shown in~\autoref{tab:SimilarStudies}, which highlights the differences between this and related studies.

\begin{table}[t]
\centering
\caption{\normalfont Literature comparison: features (F); static (S), dynamic (D), and hybrid (H), environment for data collection (E), balancing (B), machine and deep learning (MD), timestamps (T), and concept drift handling (A) are compared across different approaches.}\vspace{-2mm}
\begin{tabular}{lllllll}
\hline
\textbf{Work} & \textbf{F} & \textbf{E} & \textbf{B} & \textbf{MD} & \textbf{T} & \textbf{A} \\ \hline
Chen~\etal~\cite{ChenZKYCPPCW23} & S & $\times$ & $\times$ & $\times$ & $\checkmark$ & $\times$ \\ 
Guerra~\etal~\cite{Guerra-Manzanares23} &D &$\checkmark$ &$\times$ & $\times$ & $\checkmark$ & $\times$ \\ 

Guerra~\etal~\cite{Guerra-ManzanaresB22a} & S, D& $\times$& $\checkmark$ & $\times$ & $\checkmark$ & $\times$ \\ 

Chow~\etal~\cite{ChowKLCAP23} & -- & $\times$ & $\times$ & $\times$ & $\checkmark$ & $\times$ \\ 

Ceschin~\etal~\cite{CeschinBGPOG23} & S &  $\times$ &$\checkmark$ & $\times$ & $\checkmark$ & $\times$ \\ 
\hline
\textbf{This work} & S, D, H &$\checkmark$  & $\checkmark$ & $\checkmark$ & $\checkmark$ & $\checkmark$ \\ 
\hline
\end{tabular}
\label{tab:SimilarStudies}\vspace{-3mm}
\end{table}

\section{Data Representation \& Learning}\label{sec:DataRepresentationLearning}

\subsection{Dataset Overview}\label{sec:Dataset Overview}
\begin{figure*}[t]
    \centering
          \begin{tikzpicture}
        \begin{axis}[
            ybar,
            width=18cm,
            height=3cm,
            ylabel={\# Apps},
            symbolic x coords={2008, 2009, 2010, 2011, 2012, 2013, 2014, 2015, 2016, 2017, 2018, 2019, 2020},
            xtick=data,
            x tick label style={rotate=45, anchor=east},
            ymin=0,
            bar width=5pt,
            ymajorgrids=true,
            enlargelimits=false,
            tick label style={font=\scriptsize},
            legend style={at={(.75,1.2)}, anchor=north, legend columns=2,font=\tiny},
            nodes near coords,  
            every node near coord/.append style={font=\tiny, rotate=90, anchor=west}, 
            enlarge x limits=0.05
        ]
        \addplot[ybar, color=blue, pattern=north east lines, pattern color=blue] coordinates {
            (2008, 64) (2009, 622) (2010, 5074) (2011, 22873) 
            (2012, 342) (2013, 489) (2014, 632) (2015, 720) 
            (2016, 743) (2017, 650) (2018, 775) (2019, 200) (2020, 275)
        };
        \addlegendentry{Benign--Real Device}

        \addplot[ybar, color=blue, pattern=north west lines, pattern color=blue,draw=red] coordinates {
            (2008, 934) (2009, 20) (2010, 269) (2011, 3137) 
            (2012, 7564) (2013, 7487) (2014, 8005) (2015, 1424) 
            (2016, 2445) (2017, 4806) (2018, 4006) (2019, 1491) (2020, 1048)
        };
        \addlegendentry{Malware--Real Device}

        \addplot[ybar, color=blue, pattern=grid, pattern color=blue] coordinates {
            (2008, 66) (2009, 607) (2010, 4978) (2011, 22146) 
            (2012, 309) (2013, 444) (2014, 596) (2015, 689) 
            (2016, 715) (2017, 788) (2018, 733) (2019, 180) (2020, 1002)
        };
        \addlegendentry{Benign--Emulator}
        
        \addplot[ybar, color=blue!60,pattern=crosshatch, pattern color=blue,draw=red] coordinates {
            (2008, 686) (2009, 19) (2010, 260) (2011, 2277) 
            (2012, 6498) (2013, 5641) (2014, 6286) (2015, 1031) 
            (2016, 1941) (2017, 625) (2018, 2299) (2019, 1388) (2020, 256)
        };
        \addlegendentry{Malware--Emulator}
        \end{axis}
    \end{tikzpicture}
        \vspace{-3mm}

    \caption{KronoDroid dataset distribution of both malware and benign samples across different years.}
    \label{fig:Kdataset}\vspace{-5mm}
\end{figure*}
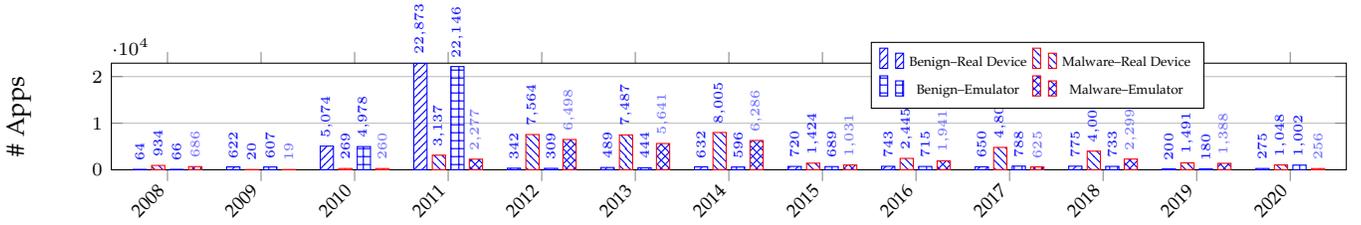

\BfPara{KronoDroid} 
The KronoDroid dataset \cite{Guerra-ManzanaresBN21} combines static and dynamic features from Android applications, sourced from VirusTotal, Drebin, VirusShare, AMD (malware), and APKMirror, F-Droid, MARVIN (benign). It spans 2008–2020, with 489 extracted features. The emulator dataset contains 28,745 malware samples (209 families) and 35,246 benign samples, while the real device dataset includes 41,382 malware samples (240 families) and 36,755 benign samples, encompassing all emulator data. Static features include 185 attributes (permissions, intents), while dynamic features consist of 288 system calls. KronoDroid also labels malware families for both real and emulator samples. ~\autoref{fig:Kdataset} shows the yearly distribution of malware and benign apps.

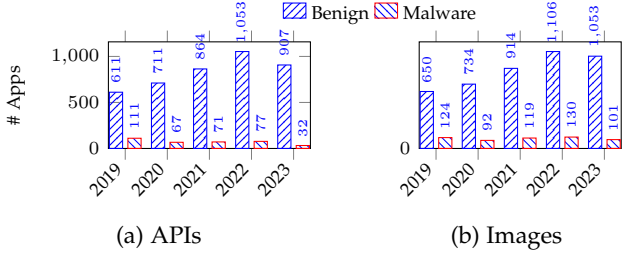
\begin{figure}[t]
    \centering

    \begin{minipage}[t]{0.48\columnwidth}
        \centering
        \begin{tikzpicture}
        \begin{axis}[
            ybar,
            bar width=5pt,
            width=\linewidth,
            height=3cm,
            ylabel={\# Apps},
            symbolic x coords={2019,2020,2021,2022,2023},
            xtick=data,
            x tick label style={rotate=45, anchor=east},
            ymin=0,
            enlarge x limits=0.1,
            nodes near coords,
            every node near coord/.append style={font=\tiny, rotate=90, anchor=west, color=blue},
            tick label style={font=\scriptsize},
            y label style={font=\scriptsize}
        ]
            \addplot+[ybar, pattern=north east lines, pattern color=blue] coordinates {
                (2019,611) (2020,711) (2021,864) (2022,1053) (2023,907)};
            \addplot+[ybar, pattern=north west lines, pattern color=blue] coordinates {
                (2019,111) (2020,67) (2021,71) (2022,77) (2023,32)};
        \end{axis}
            \node[anchor=west, font=\scriptsize, align=left] at (2.2,1.8cm) {
            \tikz \draw[pattern=north east lines, pattern color=blue, draw=blue] (0,0) rectangle (0.3,0.2); Benign 
            \tikz \draw[pattern=north west lines, pattern color=blue, draw=red] (0,0) rectangle (0.3,0.2); Malware
        };
        \end{tikzpicture}
        \subcaption{APIs}
    \end{minipage}
    \hfill
    \begin{minipage}[t]{0.48\columnwidth}
        \centering
        \begin{tikzpicture}
        \begin{axis}[
            ybar,
            bar width=5pt,
            width=\linewidth,
            height=3cm,
            symbolic x coords={2019,2020,2021,2022,2023},
            xtick=data,
            x tick label style={rotate=45, anchor=east},
            ymin=0,
            ytick={0},
            enlarge x limits=0.1,
            nodes near coords,
            every node near coord/.append style={font=\tiny, rotate=90, anchor=west, color=blue},
            tick label style={font=\scriptsize}
        ]
            \addplot+[ybar, pattern=north east lines, pattern color=blue] coordinates {
                (2019,650) (2020,734) (2021,914) (2022,1106) (2023,1053)};
            \addplot+[ybar, pattern=north west lines, pattern color=blue] coordinates {
                (2019,124) (2020,92) (2021,119) (2022,130) (2023,101)};
        \end{axis}

        \end{tikzpicture}
        \subcaption{Images}
    \end{minipage}\vspace{-2mm}
    \caption{Troid dataset distribution.}
    \label{fig:TroidDatasetSamples}\vspace{-5mm}
\end{figure}

\BfPara{Troid} Troid is a new Android malware dataset collected from the Google Play Store between 2019 and 2023~\cite{KinoonAAM24}. This dataset consists of $5,028$ malware samples, labeled using VirusTotal and tracking their availability and removal status on the Google Play Store. The dataset contains a set of features, including privacy policies, metadata, control flow graphs, permissions, API calls, strings, function names, hex dumps, and labels. The distribution of samples in the years is presented in~\autoref{fig:TroidDatasetSamples}. Two types of features were selected for this dataset. First, the API call sequences for each application were chosen. The number of benign apps was initially $4,459$, which was reduced to $4,146$ after removing apps that had no API calls. For malware samples, the count started at 569 and decreased to 358. The second type of feature was hex dumps, which were converted into grayscale and RGB images for the classification task. The final number of samples was $4,457$ benign and 566 malware samples.

\subsection{Malware Family} To analyze concept drift in a multiclass classification setting, we utilized the Knorodroid dataset, which includes 240 malware families from real devices and 209 from emulators. Two strategies were used: (1) ignoring the temporal factor when using RF, CNN, and RNN with static, dynamic, and hybrid features on both real and emulator data, and (2) a cross-year strategy, where data were divided into three time periods (2008-2012, 2013-2016, 2017-2020) to assess temporal shifts by training on a group and testing on others. 

\subsection{Balancing Algorithm}  
\begin{figure}[t]
    \centering
        \includegraphics[width=0.45\textwidth]{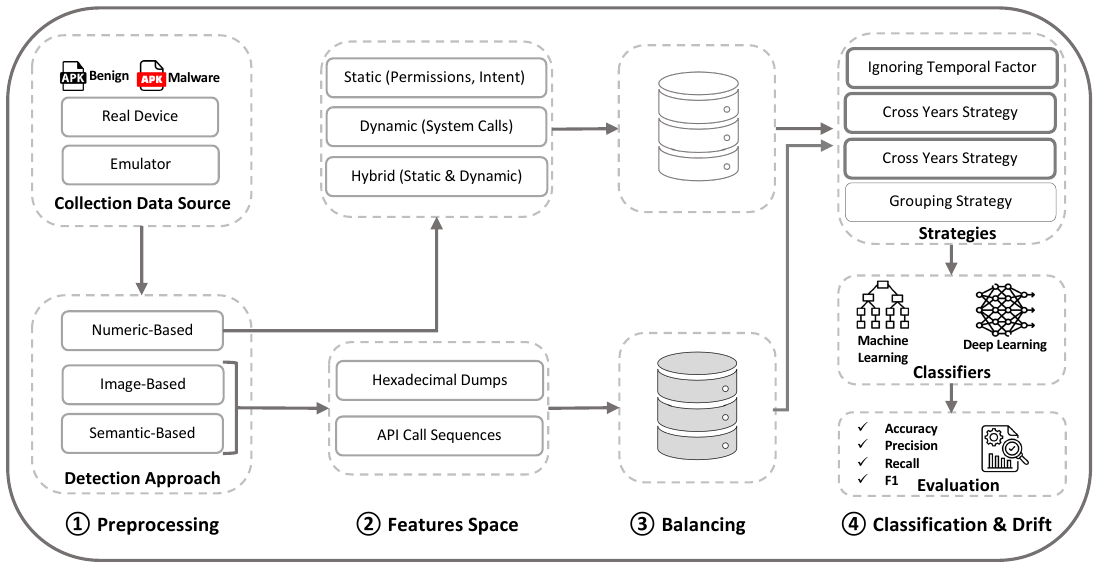}\vspace{-3mm}
        \caption{The overview of our approach operates in four stages: preprocessing, feature extraction, dataset balancing, and classification with concept drift detection.}
        \label{fig:Pipline}\vspace{-3mm}
\end{figure}

The datasets exhibit class imbalance, particularly when applying cross-year and incremental strategies, as illustrated in \autoref{fig:Kdataset} and \autoref{fig:TroidDatasetSamples}. Each year, the balancing algorithm partitions the data into features and labels, then determines the number of malware and benign samples. If the sample count for either class is below the maximum sample size (i.e., the highest number of malware samples in a given year), the algorithm employs SMOTE to augment the samples to the maximum. Conversely, if the count exceeds the maximum, RandomUnderSampler reduces the samples accordingly. This process is applied iteratively to both malware and benign samples. The balanced data for each year is then aggregated to form the final balanced dataset. The overall pipeline of this approach is depicted in \autoref{fig:Pipline}.

\subsection{Experiment Setup and Metrics}\label{sec:ExperimentalSetup}
\BfPara{Environment}
All experiments were conducted on Google Colab, using the default cloud configuration and local runtime with GPU support.\textit{ Scikit-learn} was used for ML tasks, including classification reports and confusion matrices, while \textit{Keras} (running on TensorFlow) was used for deep learning models. Additionally, we used two freely available instruct LLMs from the Together platform, accessed via public APIs: LLaMA-3.3-70B-Instruct-Turbo-Free (Meta) and Exaone-3-5-32B-Instruct (LG)~\cite{Together57:online}. For both models, the maximum token limit was set to 5, and the temperature was fixed at zero to minimize hallucinations.

\BfPara{Experiments Design} 
Each model was repeated with subsets of features selected from the Kronodroid and Troid datasets. Kronodroid features were used without any conversion. In contrast, two types of features were used from the Troid dataset: API call sequences, which were preprocessed for use with natural language processing techniques, and hexadecimal dumps, which were converted into images for an image-based detection approach.

\BfPara{Parameters} 
The default settings were used for RF, KNN, and GB, with a random state set to 42. The deep learning models utilized binary cross-entropy for loss, the Adam optimizer (learning rate = 0.001), early stopping, 15 epochs, a batch size of 64, and a 0.10 validation split. Model performance was evaluated using the accuracy and F1 score.

\BfPara{Architecture} The deep learning architectures are as follows: 

\EMP{CNN} 
A 1D convolutional Neural Network (CNN) architecture was used, including two convolutional layers (64 and 128 filters) with ReLU activation and max pooling. This was followed by a dense layer (128 units, ReLU), a dropout layer (rate = 0.2), and an output layer using sigmoid for binary classification and softmax for family classification.

 \EMP{RNN and Variants (LSTM, GRU)}
Recurrent Neural Network (RNN), Long Short-Term Memory (LSTM), and Gated Recurrent Unit (GRU) models were employed, each beginning with a 1D input layer (8 units) followed by a flattening layer. A dense layer with 128 units and ReLU activation was incorporated, along with a dropout layer (rate = 0.2). The output layer utilized sigmoid activation for binary classification and softmax activation for family classification.

\BfPara{CNN-Troid}
A 2D CNN architecture was employed for image classification, comprising three convolutional layers with 32, 64, and 128 filters, each followed by max pooling layers. A flattening layer transformed the 2D feature maps into a 1D vector, followed by a dense layer with 512 units and ReLU activation. To mitigate overfitting, a dropout layer with a rate of 0.2 was included. The final output layer utilized sigmoid activation.

\BfPara{LLM-Based Evaluation}
To assess the robustness of the LLM under concept drift, we adopt the same cross-year strategy. For each training year, a fixed set of 30 examples is selected as few-shot demonstrations and included in the prompt as labeled vectors. Each test sample is independently evaluated by appending it to the same prompt structure, resulting in one model query per test case.

LLM requests are executed in a stateless manner. Each prompt is self-contained, and the model has no memory of previous inputs or outputs. This ensures that predictions are not influenced by prior queries and that the evaluation reflects true generalization performance.

\EMP{Token Usage Estimation}
To estimate the computational efficiency of our LLM-based evaluation, we calculated the approximate number of tokens processed during few-shot classification across all temporal train-test configurations.

Each sample (either training or test) consists of a 30-dimensional feature vector derived from PCA, with each float formatted to 4 decimal places. On average, each float, along with its delimiter, consumes approximately 2 tokens, and the label line (\texttt{Label: Malware} or \texttt{Label: Benign}) adds approximately 2 tokens. Therefore, each sample contributes approximately 62 tokens.

For each train-test evaluation, we include 30 few-shot examples followed by 1 test example in the prompt, totaling $31 \times 62 = 1,922$ tokens per prompt. Each test year includes 10 test samples, resulting in approximately $10 \times 1,922 = 19,220$ tokens per train-test year pair.

Given 13 years of data (2008–2020), we train on one year and test on each of the remaining 12 years, resulting in $13 \times 12 = 156$ train-test combinations. Thus, the total token usage across the entire evaluation setup is approximately: $156 \times 19{,}220 \approx 2.99$ million tokens.

In each request, we used a prompt for both training and testing. For LLaMA, the training prompt was: ``This sample is an app feature vector of malware/benign.'' During testing, we passed an unseen sample using the prompt: "Now classify this sample: [sample vector]. Answer with exactly one word: Malware or Benign."

For Exaone, the training prompt was: "Answer with exactly one word: Malware or Benign. Only return the word — no punctuation, explanation, or extra text." The corresponding test vector was then appended for classification.

\BfPara{Evaluation Metrics} 
The evaluation metrics used in this study include precision, recall, accuracy, and F1 Score. \begin{enumerate*}
    \item \textbf{Precision:} Measures the accuracy of correctly classified malware apps among all predicted positive cases.
    \item \textbf{Recall:} Represents the proportion of correctly identified malware apps out of all actual malware cases.
    \item \textbf{Accuracy:} Computes the overall correctness of the model by measuring the ratio of correct predictions to total predictions.
    \item \textbf{F1 Score:} Provides a balanced measure of Precision and Recall by calculating their harmonic mean.
\end{enumerate*}

\section{Experiments Results and Discussion}\label{sec:Experiments}

\subsection{Ignoring the Temporal Factor}
In this set of experiments, we ignored the time and trained and tested on all datasets, which is a common practice in the literature. We use the results as a baseline to highlight the impact of correct implementation to capture concept drift. For the ML models, we split the dataset into 80\% for training and 20\% for testing. The same percentage was used for the deep learning models, with an additional 10\% of the training set used for validation. 

\BfPara{KronoDroid Dataset}
The results in~\autoref{tab:IgrnoreYearsRealDevice} compare the performance of RF, KNN, GB, CNN, RNN, LSTM, and GRU on data collected from a real device. The classifiers were evaluated for three feature extraction methods: static, dynamic, and hybrid, with accuracy and F1 as metrics. 

\begin{takeaway} Both deep and shallow learning models perform well, although feature complexity plays a role, with hybrid features yielding the most robust classification. Changing the algorithm minimally impacted the accuracy when using the same features and ignoring time.
\end{takeaway}

The results show the type of feature (static, dynamic, hybrid) plays a significant role in determining the performance of different classifiers. In the static feature category, all classifiers achieve high accuracy, with RF and RNN performing best,  RF achieving an accuracy and F1 score of 0.977 and RNN following at 0.970. In contrast, KNN shows the lowest performance with static features, with an accuracy and F1 score of 0.928. With dynamic features, performance decreases for most classifiers: RF and RNN deliver accuracies of 0.940 and 0.904, respectively, but KNN shows a larger drop, with an accuracy of 0.892. The hybrid feature approach, which combines static and dynamic data, produces the best results, particularly for deep learning models like LSTM and GRU, with an accuracy of 0.983. This suggests that deep learning models better capture complex patterns in hybrid data. 

\begin{table*}[t]
\begin{minipage}{.31\textwidth} %
   \centering
    \caption{\normalfont Performance of various models with real device features.} \vspace{-2mm}
    \scalebox{0.75}{
    \begin{tabular}{ccccccc}
        \hline
        \textbf{Model} & \multicolumn{2}{c}{\textbf{Static}} & \multicolumn{2}{c}{\textbf{Dynamic}} & \multicolumn{2}{c}{\textbf{Hybrid}} \\
        \hline
        & \textbf{A} & \textbf{F1} & \textbf{A} & \textbf{F1} & \textbf{A} & \textbf{F1} \\
        \hline
        RF   & 0.98 & 0.98 & 0.94 & 0.94 & 0.98 & 0.98 \\
        KNN  & 0.93 & 0.93 & 0.90 & 0.90 & 0.92 & 0.92 \\
        GB   & 0.95 & 0.95 & 0.89 & 0.89 & 0.97 & 0.97 \\
        CNN  & 0.96 & 0.96 & 0.90 & 0.90 & 0.97 & 0.97 \\
        RNN  & 0.97 & 0.97 & 0.90 & 0.90 & 0.98 & 0.98 \\
        LSTM & 0.97 & 0.97 & 0.93 & 0.92 & 0.98 & 0.98 \\
        GRU  & 0.97 & 0.97 & 0.93 & 0.93 & 0.98 & 0.98 \\
        \hline
    \end{tabular}}
    \label{tab:IgrnoreYearsRealDevice}
\end{minipage}\hspace{3mm}~
\begin{minipage}{.33\textwidth} %
    \centering
    \caption{\normalfont Performance of various models using emulator device features.} \vspace{-2mm}
    \scalebox{0.75}{
    \begin{tabular}{ccccccc}
        \hline
        \textbf{Model} & \multicolumn{2}{c}{\textbf{Static}} & \multicolumn{2}{c}{\textbf{Dynamic}} & \multicolumn{2}{c}{\textbf{Hybrid}} \\
        \hline
        & \textbf{A} & \textbf{F1} & \textbf{A} & \textbf{F1} & \textbf{A} & \textbf{F1} \\
        \hline
        RF   & 0.97 & 0.97 & 0.93 & 0.93 & 0.98 & 0.98 \\
        KNN  & 0.94 & 0.94 & 0.88 & 0.87 & 0.90 & 0.90 \\
        GB   & 0.95 & 0.95 & 0.85 & 0.84 & 0.96 & 0.96 \\
        CNN  & 0.95 & 0.95 & 0.83 & 0.82 & 0.97 & 0.96 \\
        RNN  & 0.97 & 0.97 & 0.90 & 0.90 & 0.99 & 0.98 \\
        LSTM & 0.97 & 0.97 & 0.89 & 0.89 & 0.97 & 0.97 \\
        GRU  & 0.97 & 0.97 & 0.93 & 0.93 & 0.98 & 0.98 \\
        \hline
    \end{tabular}
    }
    \label{tab:IgrnoreYearsEmulatorDevice}
\end{minipage}~
\begin{minipage}{.3\textwidth} %
    \centering
    \caption{\normalfont Performance of various algorithms/features types over Troid.} \vspace{-2mm}
    \scalebox{0.75}{
    \begin{tabular}{ccccccc}
        \hline
        \textbf{Model} & \multicolumn{2}{c}{\textbf{APIs Call}} & \multicolumn{2}{c}{\textbf{Grayscale}} & \multicolumn{2}{c}{\textbf{RGB}} \\
        \hline
        & \textbf{A} & \textbf{F1} & \textbf{A} & \textbf{F1} & \textbf{A} & \textbf{F1} \\
        \hline
        RF   & 0.93 & 0.64 & 0.88 & 0.48 & 0.88 & 0.48 \\
        KNN  & 0.92 & 0.54 & 0.85 & 0.49 & 0.89 & 0.47 \\
        GB   & 0.92 & 0.54 & 0.90 & 0.66 & 0.91 & 0.64 \\
        CNN  & 0.92 & 0.66 & 0.88 & 0.47 & 0.89 & 0.47 \\
        RNN  & 0.91 & 0.45 & 0.88 & 0.47 & 0.89 & 0.47 \\
        LSTM & 0.92 & 0.51 & 0.88 & 0.47 & 0.89 & 0.47 \\
        GRU  & 0.92 & 0.48 & 0.88 & 0.47 & 0.89 & 0.47 \\
        \hline
    \end{tabular}
    }
    \label{tab:IgnoreTimeTroid}
\end{minipage}
\end{table*}

\EMP{Emulator vs. Real Devices} Comparing the results of real device and emulator, we observe some key differences in performance across feature types as shown in \autoref{tab:IgrnoreYearsEmulatorDevice}. In particular, we found that models trained on real device data tend to perform slightly better than those trained on emulator data, especially in the hybrid feature type. For example, in real device data, RF achieved an accuracy of 0.986, while the performance dropped to 0.979 with the emulator. Secondly, despite that, the deep learning models, LSTM, GRU, and RNN, maintained high performance in real and in emulators. For instance, RNN achieved identical F1 scores of 0.982 for hybrid features in both environments. 

\begin{takeaway}
Per our results, deep learning models are more resilient to variations in real and emulator data collection methods, with minimal impact on model accuracy.
\end{takeaway}

\EMP{Malware Family Classification} 
In our experiments, we selected the top 10 families that dominate the dataset, resulting in a total of 30,522 malware samples out of 41,382 for real devices and 21,831 out of 31,046 for emulators. The results of the three classifiers are shown in~\autoref{tab:MultiPerformanceRealEmulator}. The results indicate that models perform better on real devices due to the more realistic execution environment, with RF consistently achieving the highest scores. However, dynamic analysis revealed a drop in performance across all models. Despite this trend, RF maintained its lead, while RNN outperformed CNN. The hybrid analysis significantly enhanced the performance of all models compared with the dynamic analysis. RF and RNN benefit the most from this approach.

\begin{table}[t]
    \centering
    \caption{\normalfont The performance for malware family classification using various models, data collection type, and feature type.} \vspace{-2mm}
    \scalebox{0.70}{
    \begin{tabular}{ccccccc|cccccc}
        \hline
        \multirow{3}{*}{\textbf{Model}} 
        & \multicolumn{6}{c|}{\textbf{Real Device}} 
        & \multicolumn{6}{c}{\textbf{Emulator}} \\
        \cline{2-13}
        & \multicolumn{2}{c}{\textbf{Static}} & \multicolumn{2}{c}{\textbf{Dynamic}} & \multicolumn{2}{c|}{\textbf{Hybrid}}
        & \multicolumn{2}{c}{\textbf{Static}} & \multicolumn{2}{c}{\textbf{Dynamic}} & \multicolumn{2}{c}{\textbf{Hybrid}} \\
        \cline{2-13}
        & \textbf{A} & \textbf{F1} & \textbf{A} & \textbf{F1} & \textbf{A} & \textbf{F1}
        & \textbf{A} & \textbf{F1} & \textbf{A} & \textbf{F1} & \textbf{A} & \textbf{F1} \\
        \hline
        RF   & 0.94 & 0.92 & 0.86 & 0.82 & 0.93 & 0.91 & 0.93 & 0.89 & 0.89 & 0.83 & 0.93 & 0.90 \\
        CNN  & 0.90 & 0.87 & 0.70 & 0.61 & 0.90 & 0.86 & 0.91 & 0.86 & 0.68 & 0.58 & 0.89 & 0.83 \\
        RNN  & 0.92 & 0.89 & 0.79 & 0.74 & 0.91 & 0.87 & 0.92 & 0.88 & 0.81 & 0.74 & 0.92 & 0.87 \\
        \hline
    \end{tabular}}
    \label{tab:MultiPerformanceRealEmulator}
\end{table}

\begin{takeaway}
Static features exhibit greater resilience in malware family classification, likely due to the consistent behavior of malware across environments, where feature convergence may influence performance.
\end{takeaway}

\BfPara{Troid Dataset}
The Troid dataset results, presented in \autoref{tab:IgnoreTimeTroid}, compare various feature types (API call, grayscale, and RGB images) under the same settings. Overall, RF performs best with API call features, achieving 0.93 accuracy and an F1 score of 0.64. However, its performance declines with grayscale and RGB images, reaching 0.88 accuracy and a significantly lower F1 score of 0.48.

GB performs better with RGB images, achieving 0.91 accuracy and an F1 score of 0.64, compared to its performance with API calls and grayscale images. Deep learning models, including CNN, RNN, LSTM, and GRU, maintain stable accuracy across feature types but generally exhibit lower F1 scores, particularly with grayscale and RGB images. This decline in F1 score is attributed to dataset imbalance, an issue addressed in the next strategy.

\begin{takeaway}
Classifier performance varies based on feature types and the chosen classification approach.
\end{takeaway}

\subsection{Cross Years Strategy}
\label{Corss_Years_Strategy}
In this set of experiments, we investigate the drift in accuracy of ML models over time. We trained the classifiers for each algorithm on all samples from specific years and tested the models on data from both future and past years in our dataset. To avoid the impact of data imbalance, we applied a balancing algorithm using undersampling or oversampling, depending on the number of samples in each year. In addition, to demonstrate concept drift and the effect of time on model performance, we adopted various types of features and approaches.

\BfPara{KronoDroid Dataset}
This dataset contains samples sorted over 13 years (2008--2020). The evaluation of each model is based on feature types (static, dynamic, hybrid) and the collection environment (real device vs. emulator).

\begin{figure*}[t]
    \centering
    \begin{subfigure}[t]{0.235\textwidth} 
        \centering
        \includegraphics[width=0.99\textwidth]{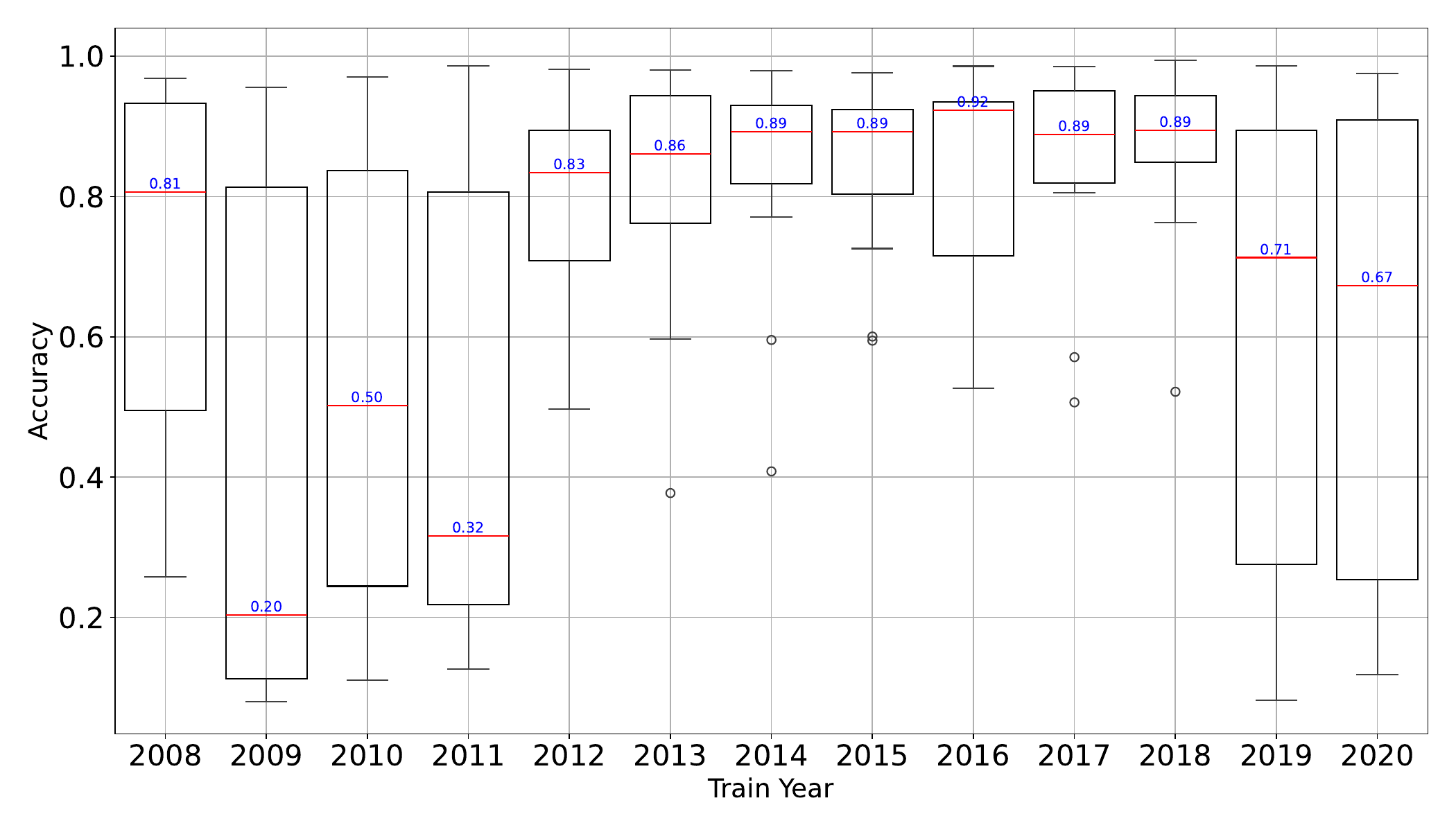}\vspace{-2mm}
        \caption{\normalfont Pre-Real-Acc.}
        \label{fig:Box_Plot_RF_Static_Pre_Real_Acc}
    \end{subfigure}
    ~
    \begin{subfigure}[t]{0.235\textwidth}  
        \centering
        \includegraphics[width=0.99\textwidth]{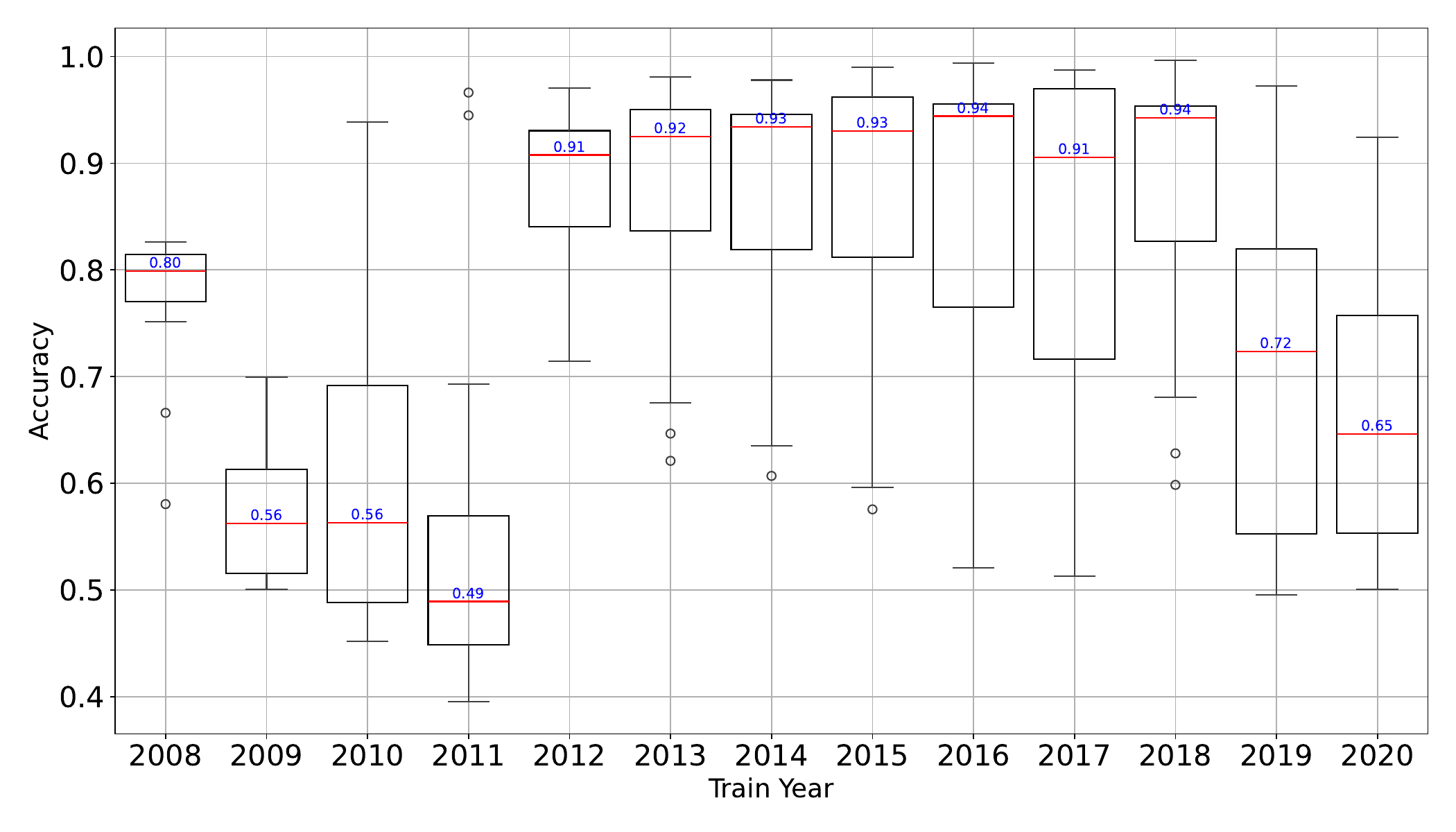}\vspace{-2mm}
        \caption{\normalfont Post-Real-Acc.}
        \label{fig:Box_Plot_RF_Static_Post_Real_Acc}
    \end{subfigure}
    ~
    \begin{subfigure}[t]{0.235\textwidth}
        \centering
        \includegraphics[width=0.99\textwidth]{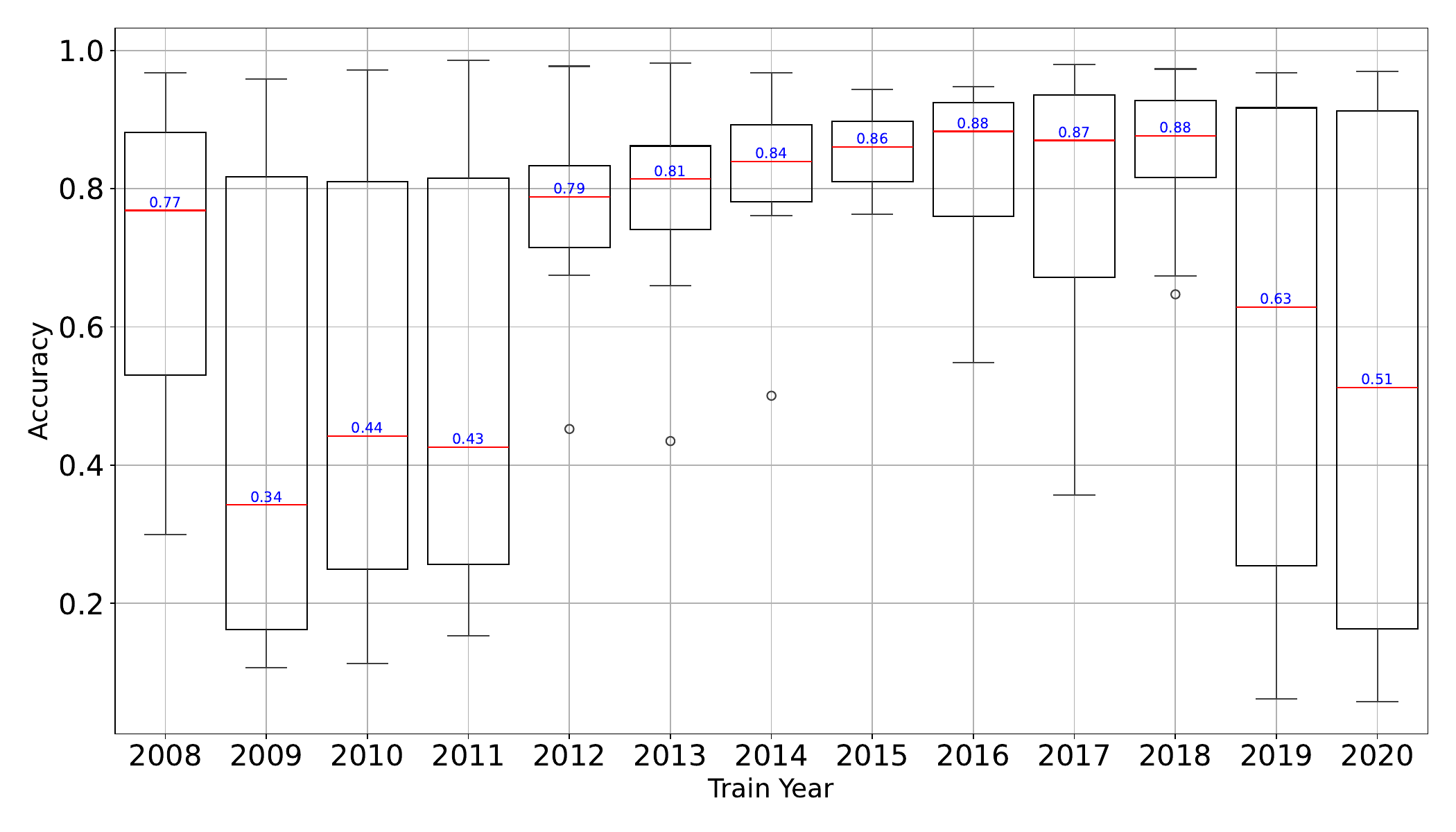}\vspace{-2mm}
        \caption{\normalfont Pre-Emu-Acc.}
        \label{fig:Box_Plot_RF_Static_Pre_Emu_Acc}
    \end{subfigure}
    ~
    \begin{subfigure}[t]{0.235\textwidth}
        \centering
        \includegraphics[width=0.99\textwidth]{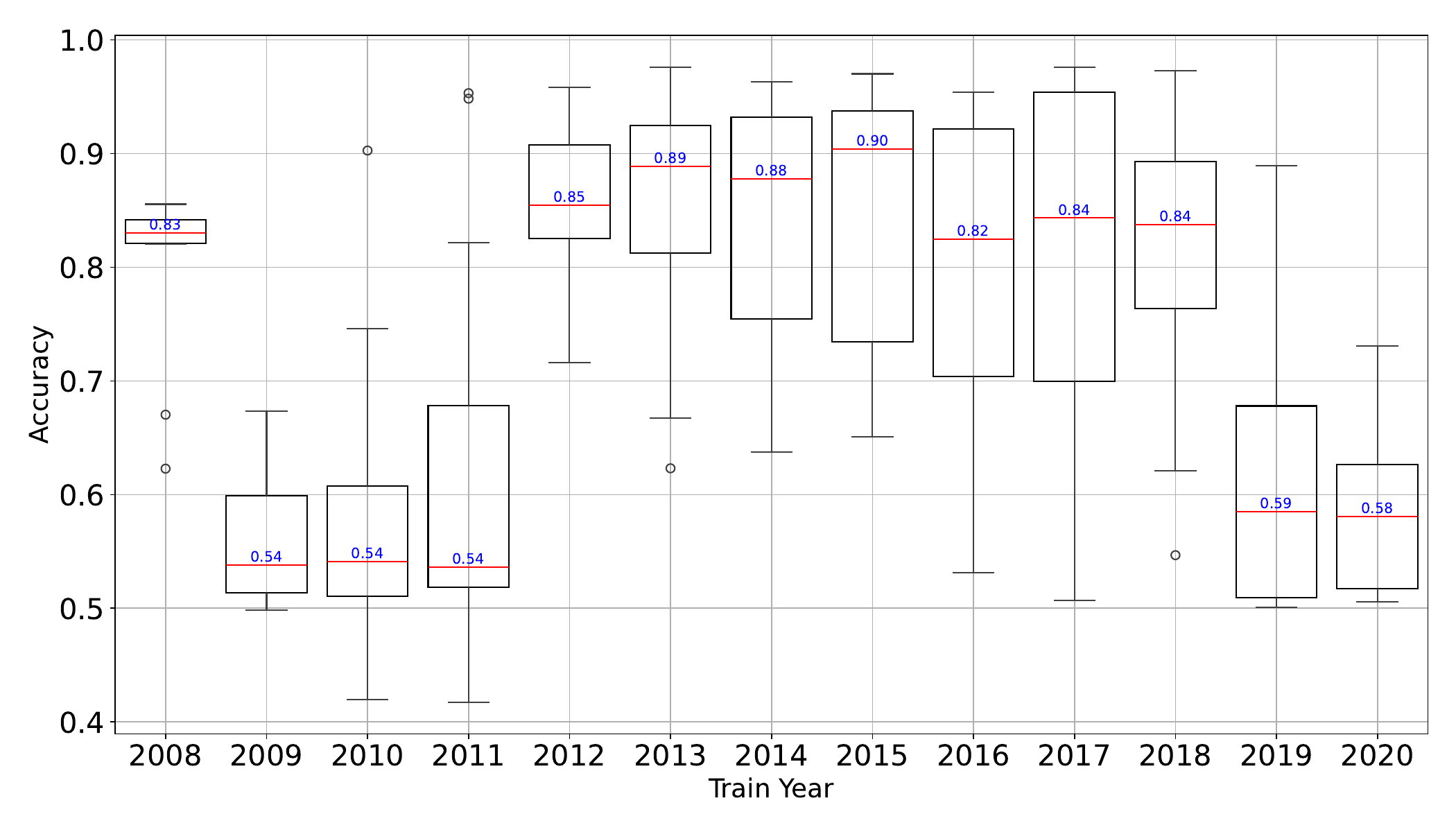}\vspace{-2mm}
        \caption{\normalfont Post-Emu-Acc.}
        \label{fig:Box_Plot_RF_Static_Post_Emu_Acc}
    \end{subfigure}

    \begin{subfigure}[t]{0.235\textwidth}
        \centering
        \includegraphics[width=0.99\textwidth]{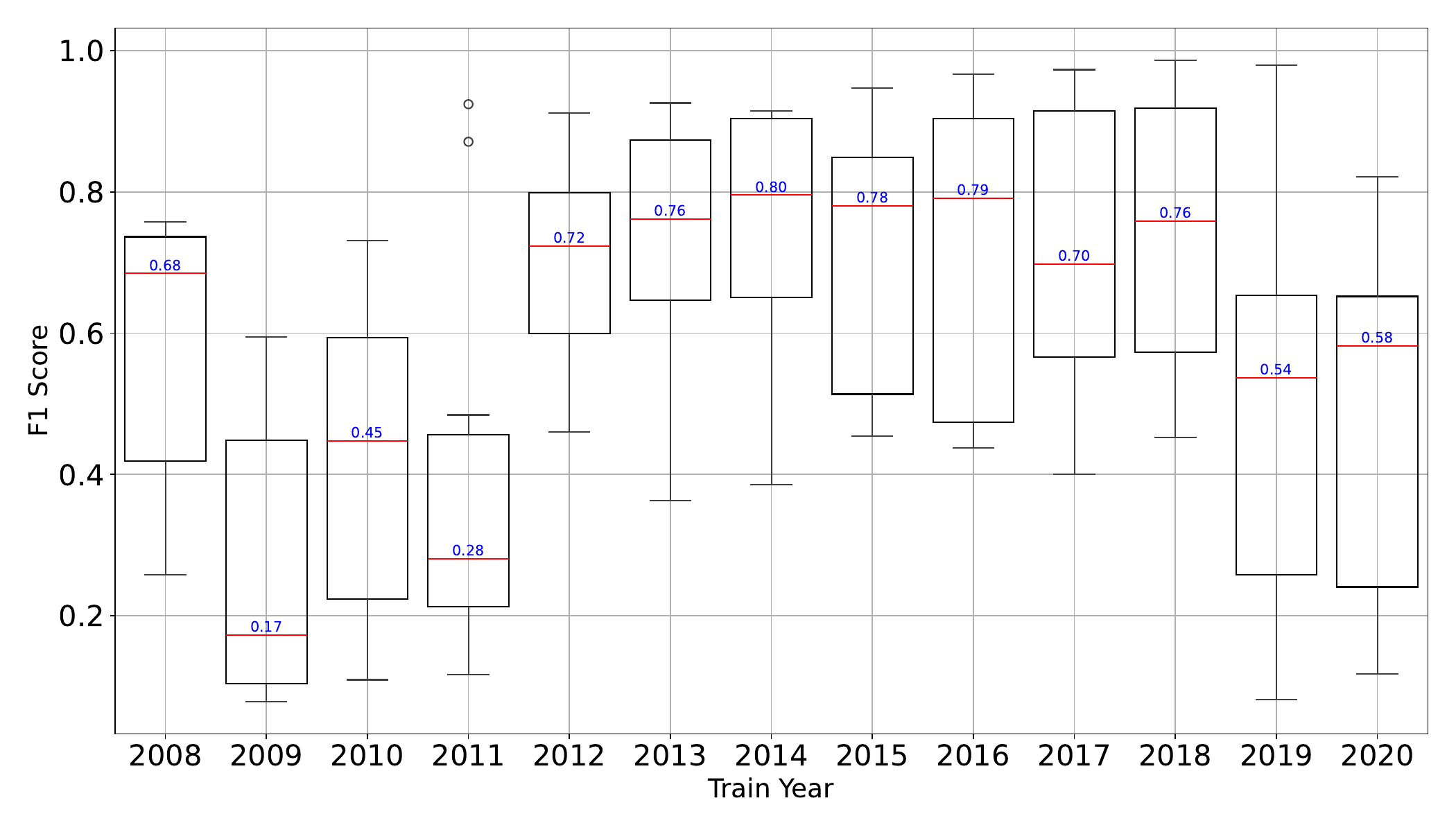}\vspace{-2mm}
        \caption{\normalfont Pre-Real-F1.}
        \label{fig:Box_Plot_RF_Static_Pre_Real_F1}
    \end{subfigure}
    ~
    \begin{subfigure}[t]{0.235\textwidth}
        \centering
        \includegraphics[width=0.99\textwidth]{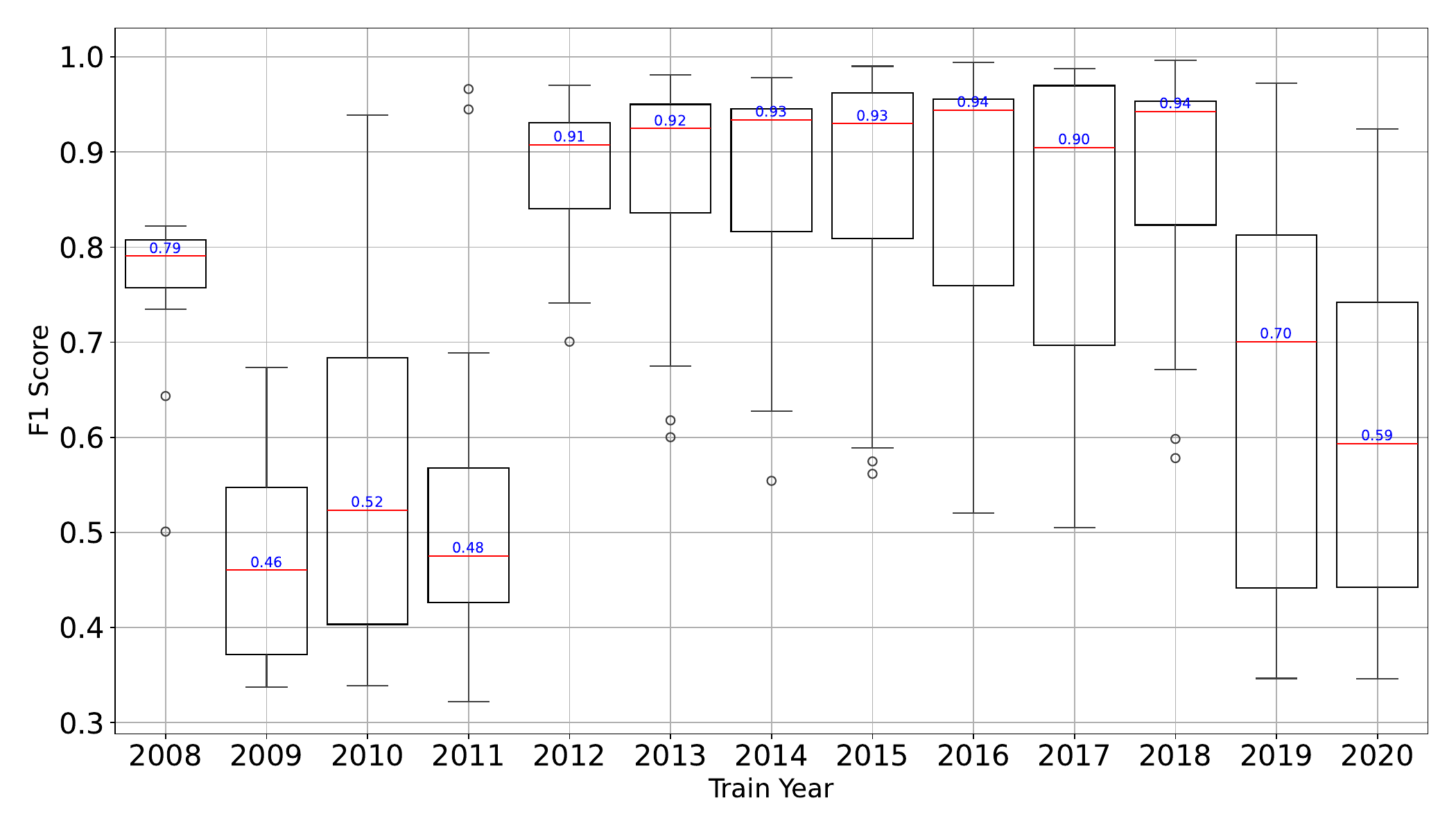}\vspace{-2mm}
        \caption{\normalfont Post-Real-F1.}
        \label{fig:Box_Plot_RF_Static_Post_Real_F1}
    \end{subfigure}
    ~
    \begin{subfigure}[t]{0.235\textwidth}
        \centering
        \includegraphics[width=0.99\textwidth]{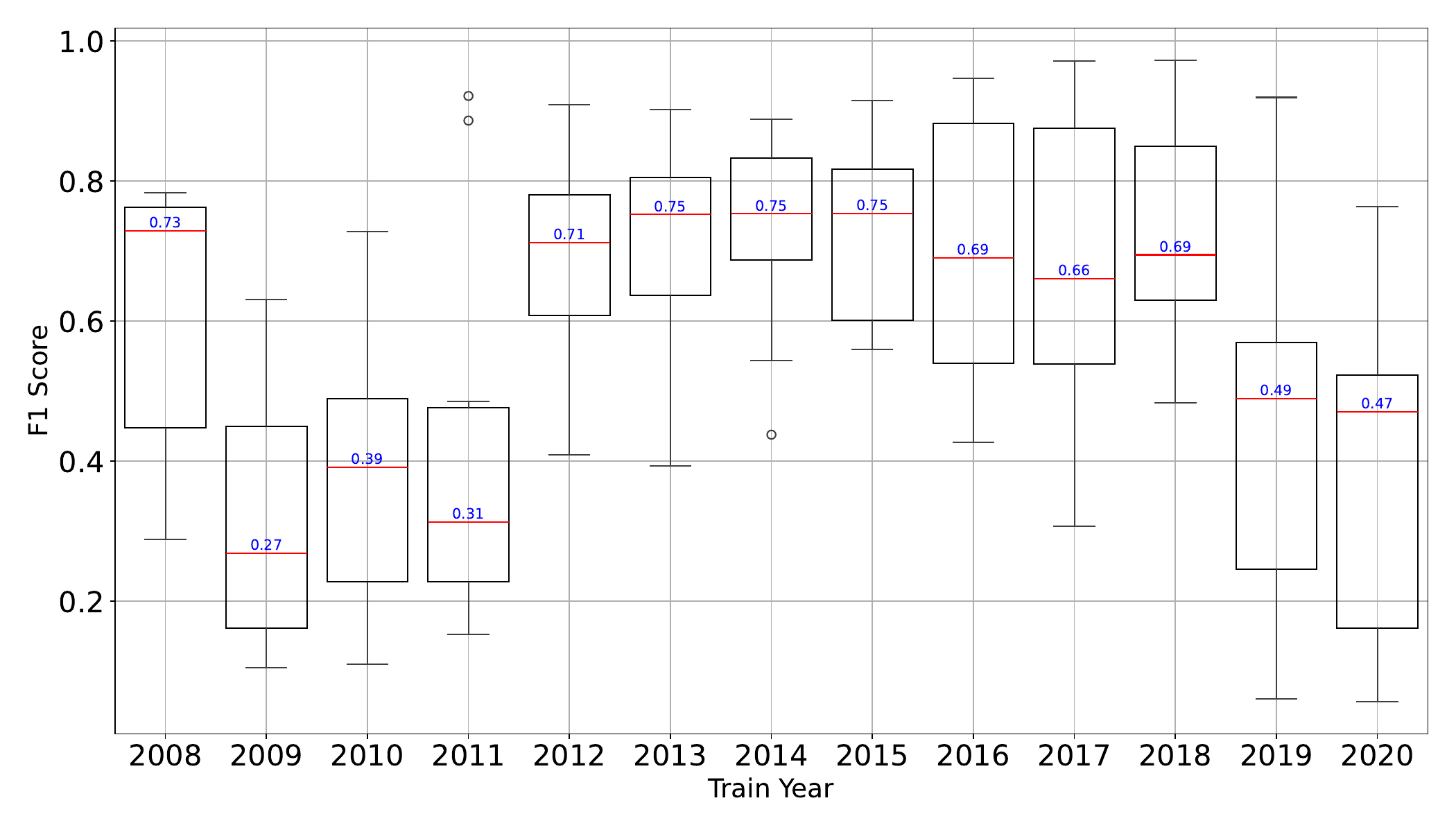}\vspace{-2mm}
        \caption{\normalfont Pre-Emu-F1.}
        \label{fig:Box_Plot_RF_Static_Pre_Emu_F1}
    \end{subfigure}
    ~
    \begin{subfigure}[t]{0.235\textwidth}
        \centering
        \includegraphics[width=0.99\textwidth]{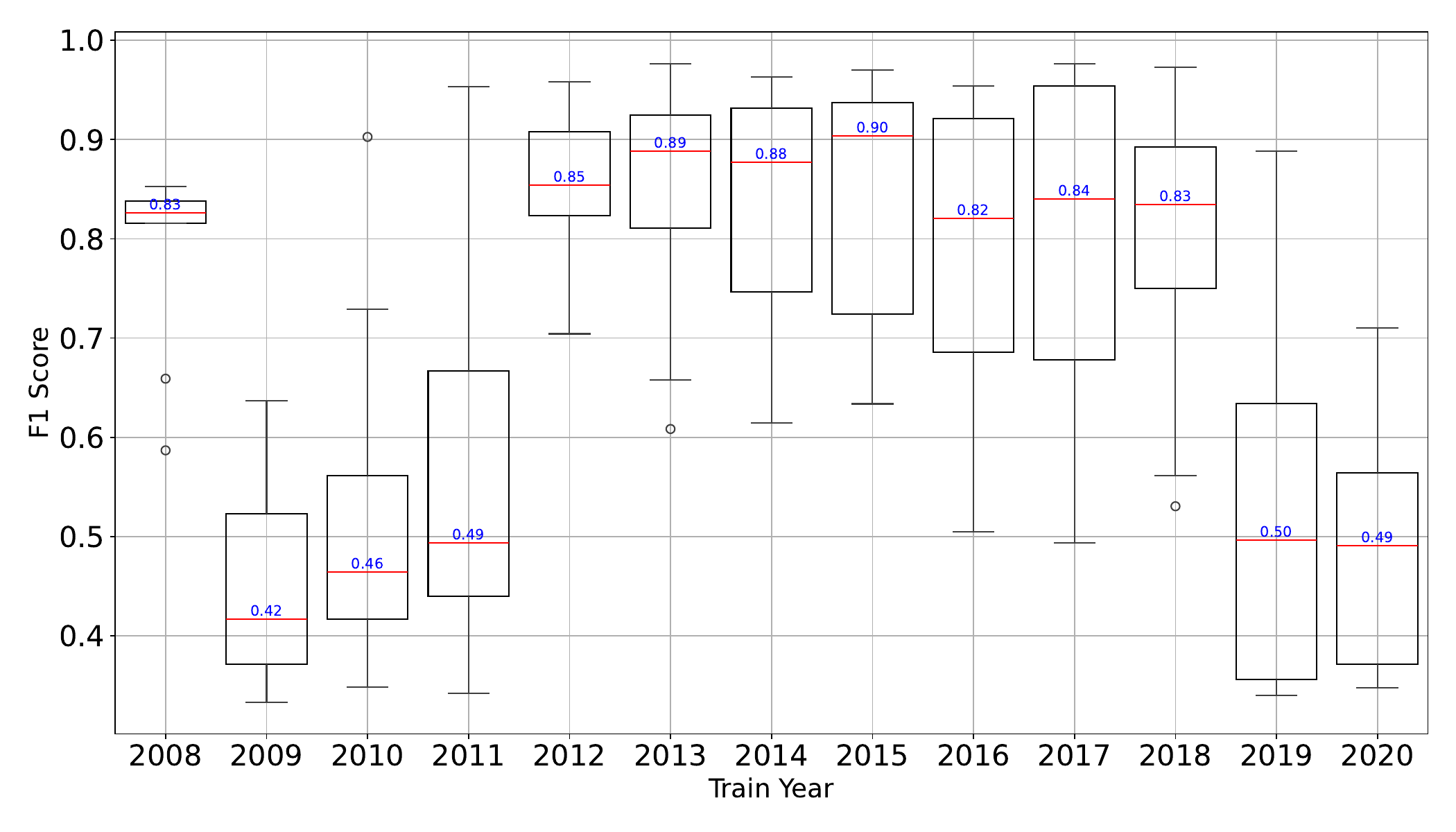}\vspace{-2mm}
        \caption{\normalfont Post-Emu-F1.}
        \label{fig:Box_Plot_RF_Static_Post_Emu_F1}
    \end{subfigure} \vspace{-2mm}

    \caption{\normalfont Cross-year RF pre- and post-balancing results with static features using real and emulator data.}
    \label{fig:BoxBlot_RF_Static_Real_Emu}
\end{figure*}

\EMP{Static Feature} \autoref{fig:BoxBlot_RF_Static_Real_Emu} illustrates the time effect on RF model performance, measured by accuracy and F1 score, before and after applying the balancing algorithm with real and emulator data. The results reveal temporal variations over 13 years, demonstrating how training in different years impacts performance. Pre-balancing, significant fluctuations appear, particularly when training on earlier years (2008–2013) and testing on later years. This suggests that despite achieving 0.977 accuracy with static features, the RF model struggles to generalize across distinct temporal contexts, indicating concept drift.

\EMP{Balancing} Post-balancing results are consistent, especially on the real device data, indicating that balancing the dataset helps mitigate the negative impact of temporal drift but fails to solve the problem entirely, as clearly shown in the results.

\EMP{Emulation} Emulator data exhibit greater performance instability, emphasizing the pronounced impact of time on model performance, particularly without balancing. Even after balancing, accuracy and F1 score fluctuate over time, indicating concept drift as the model's predictions degrade.

The instability and variations are captured in \autoref{fig:BoxBlot_RF_Static_Real_Emu}. The pre-balancing box plots show wider spreads in both accuracy and F1 scores for early years for emulator and real device testing. These wide spreads highlight high inconsistency and imply that models trained on older data cannot reliably predict on newer samples, emphasizing temporal instability. Moreover, while the spread of the results becomes narrower, showing that balancing improves the reliability of the classification, the persistence of variability, irrespective of the data extraction mechanism, indicates that time remains a factor that influences model performance.

\begin{takeaway}
Concept drift occurs in both real device- and emulator-based data collection. However, real device-based data collection shows more resistance to concept drift than emulators, especially after balancing.
\end{takeaway}

\begin{figure*}[t]
    \centering
    \begin{subfigure}[t]{0.235\textwidth} 
        \centering
        \includegraphics[width=0.99\textwidth]{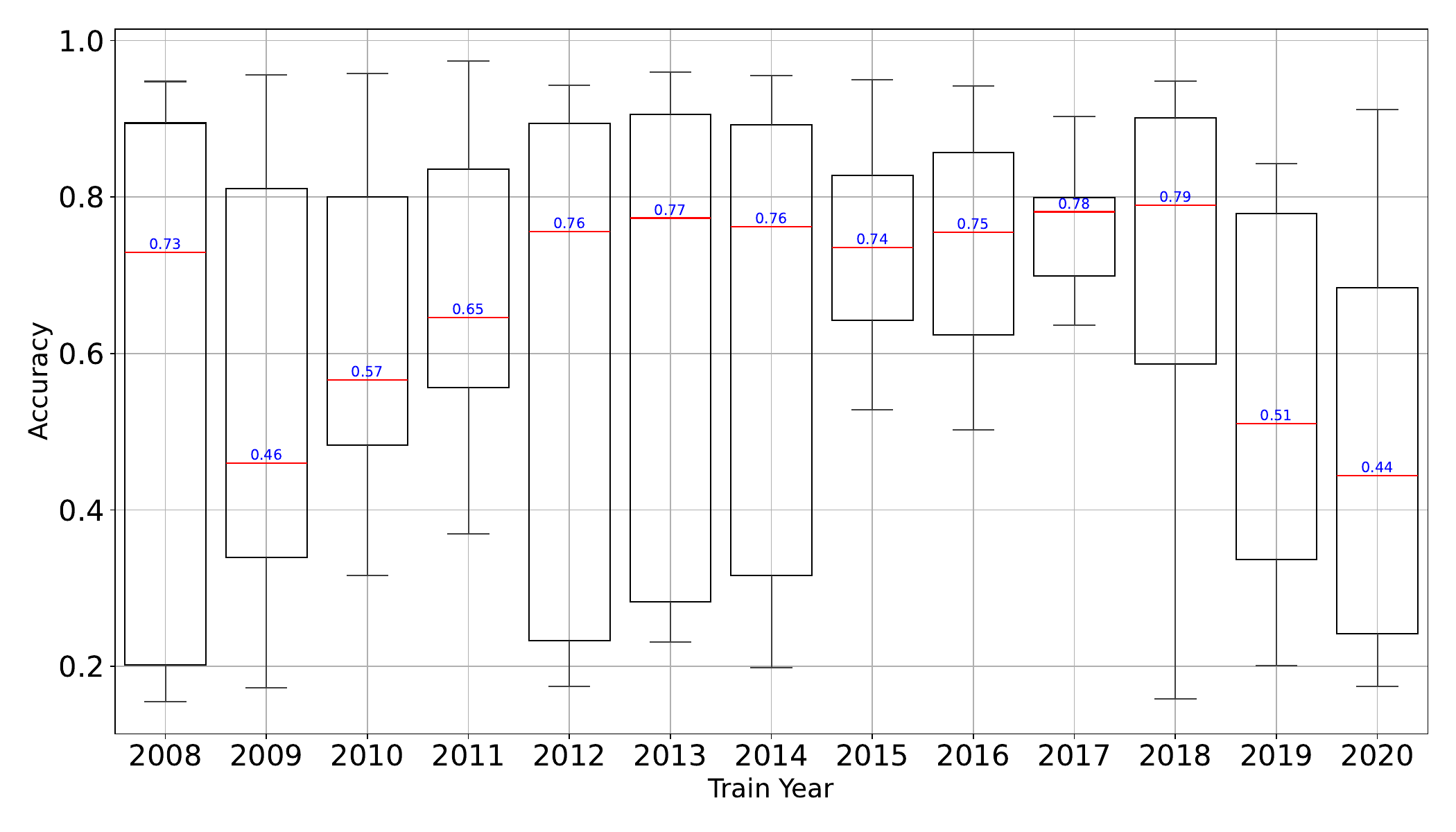}\vspace{-2mm}
        \caption{\normalfont Pre-Real-Acc.}
        \label{fig:Box_Plot_GRU_Dynamic_Pre_Real_Acc}
    \end{subfigure}
    ~
    \begin{subfigure}[t]{0.235\textwidth}  
        \centering
        \includegraphics[width=0.99\textwidth]{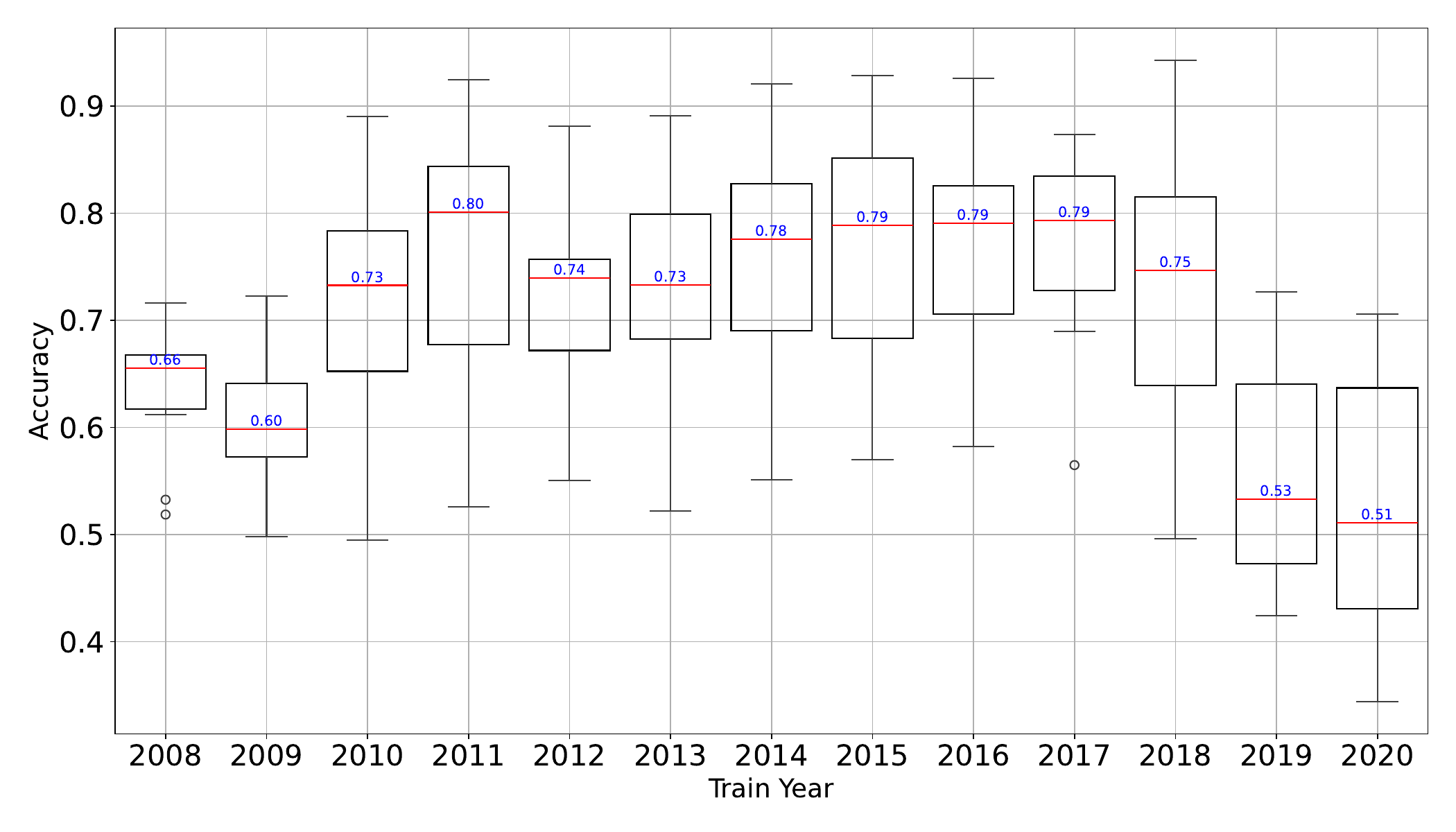}\vspace{-2mm}
        \caption{\normalfont Post-Real-Acc.}
        \label{fig:Box_Plot_GRU_Dynamic_Post_Real_Acc}
    \end{subfigure}
    ~
    \begin{subfigure}[t]{0.235\textwidth}
        \centering
        \includegraphics[width=0.99\textwidth]{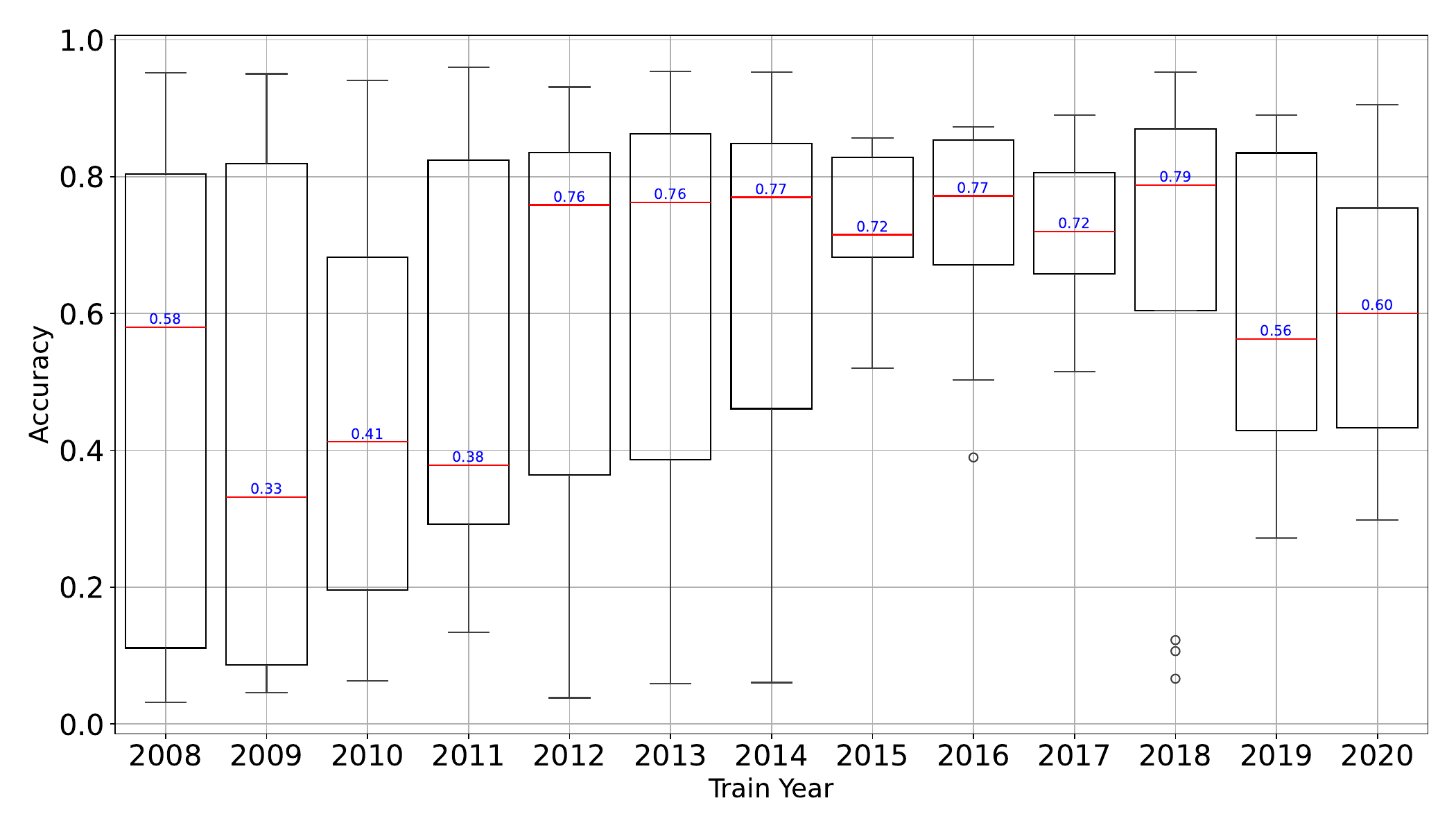}\vspace{-2mm}
        \caption{\normalfont Pre-Emu-Acc.}
        \label{fig:Box_Plot_GRU_Dynamic_Pre_Emu_Acc}
    \end{subfigure}
    ~
    \begin{subfigure}[t]{0.235\textwidth}
        \centering
        \includegraphics[width=0.99\textwidth]{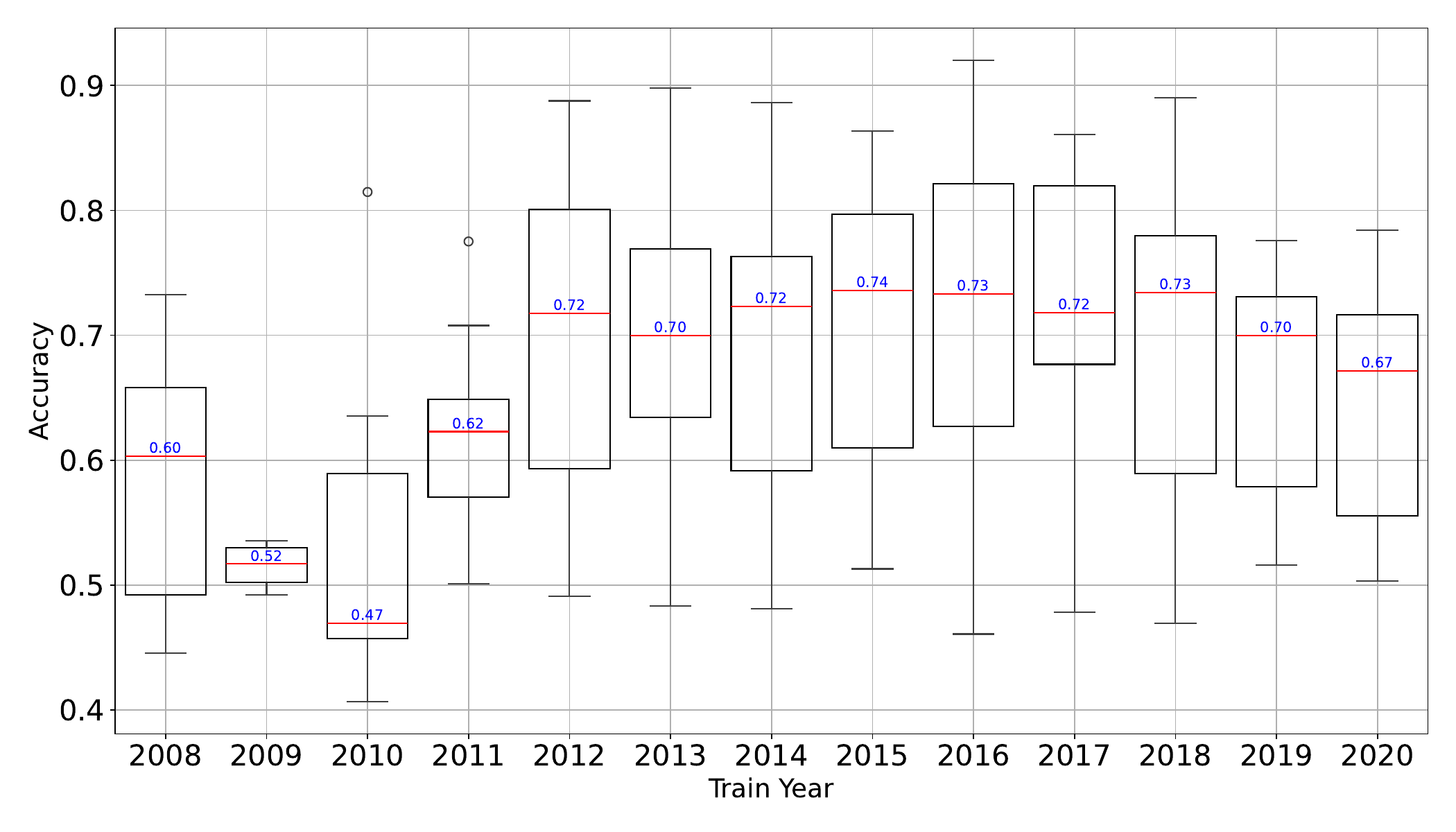}\vspace{-2mm}
        \caption{\normalfont Post-Emu-Acc.}
        \label{fig:Box_Plot_GRU_Dynamic_Post_Emu_Acc}
    \end{subfigure}
    
    \vspace{-2mm}
\caption{Cross-year pre- and post-balancing GRU accuracy with dynamic features from real/emulator data. (F1 in~\autoref{fig:Boxplot_Gru_Dynamic_Real_Emu_Cross_f1}).}\label{fig:Boxplot_Gru_Dynamic_Real_Emu_Cross}
\vspace{-3mm}\end{figure*}

\EMP{Dynamic Features}  
The GRU model achieves high performance on both real devices and emulators with dynamic features. \autoref{fig:Boxplot_Gru_Dynamic_Real_Emu_Cross} illustrates the performance evolution under the cross-year strategy, highlighting fluctuations in pre- and post-balancing accuracies. These variations indicate concept drift, particularly in the early years, where the accuracy drops significantly when trained on older data and tested on newer samples. Additionally, \autoref{fig:Boxplot_Gru_Dynamic_Real_Emu_Cross} further reflects this drift, showing inconsistent performance over time due to temporal shifts in the underlying data.

\begin{takeaway}
The models were susceptible to drift over time for both static and dynamic data collection mechanisms.
\end{takeaway}

\begin{figure*}[t]
    \centering
    \begin{subfigure}[t]{0.235\textwidth} 
        \centering
        \includegraphics[width=0.99\textwidth]{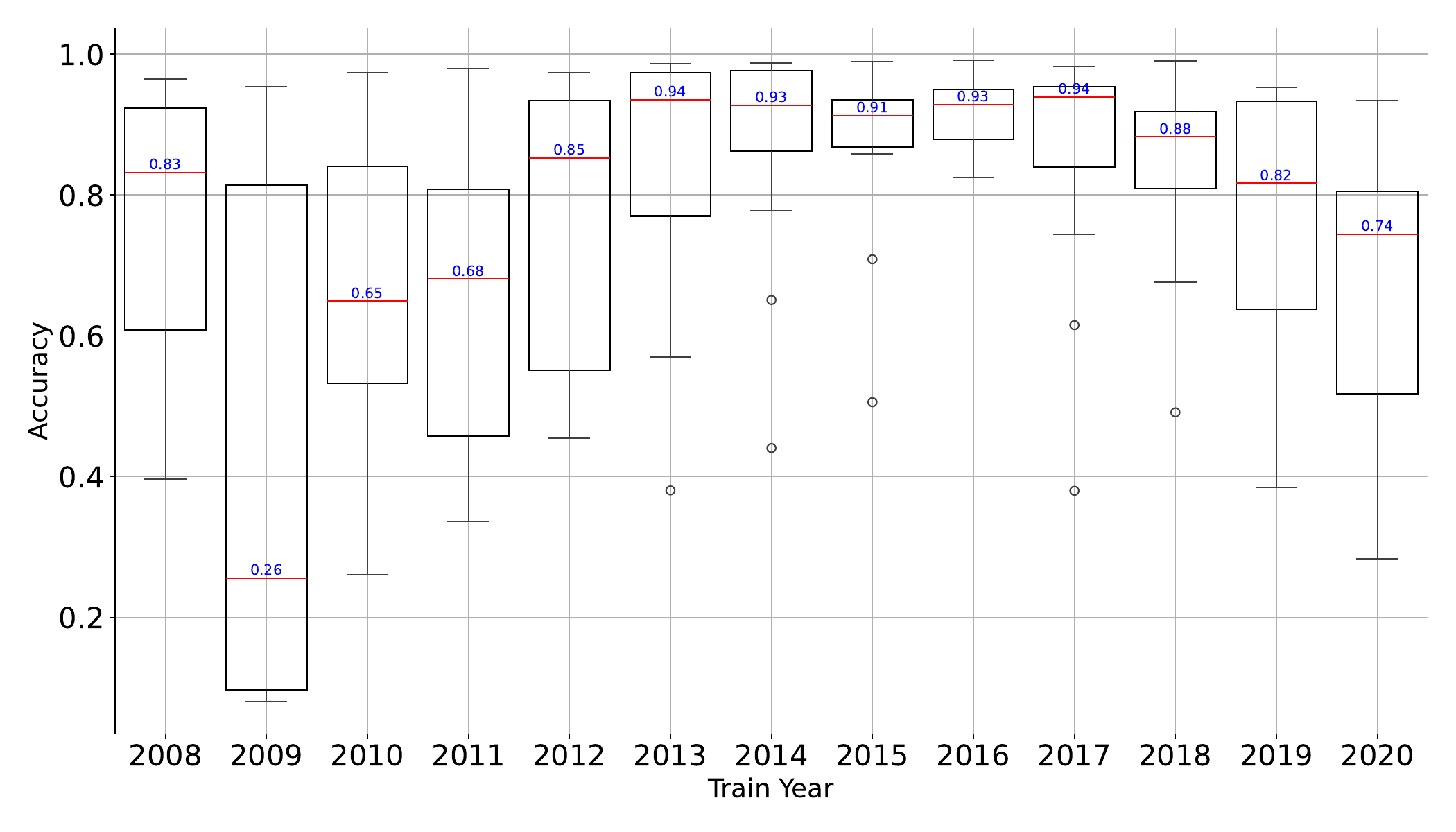}\vspace{-2mm}
        \caption{\normalfont Pre-Real-Acc.}
        \label{fig:Box_Plot_RNN_Hybrid_Pre_Real_Acc}
    \end{subfigure}
    ~
    \begin{subfigure}[t]{0.235\textwidth}  
        \centering
        \includegraphics[width=0.99\textwidth]{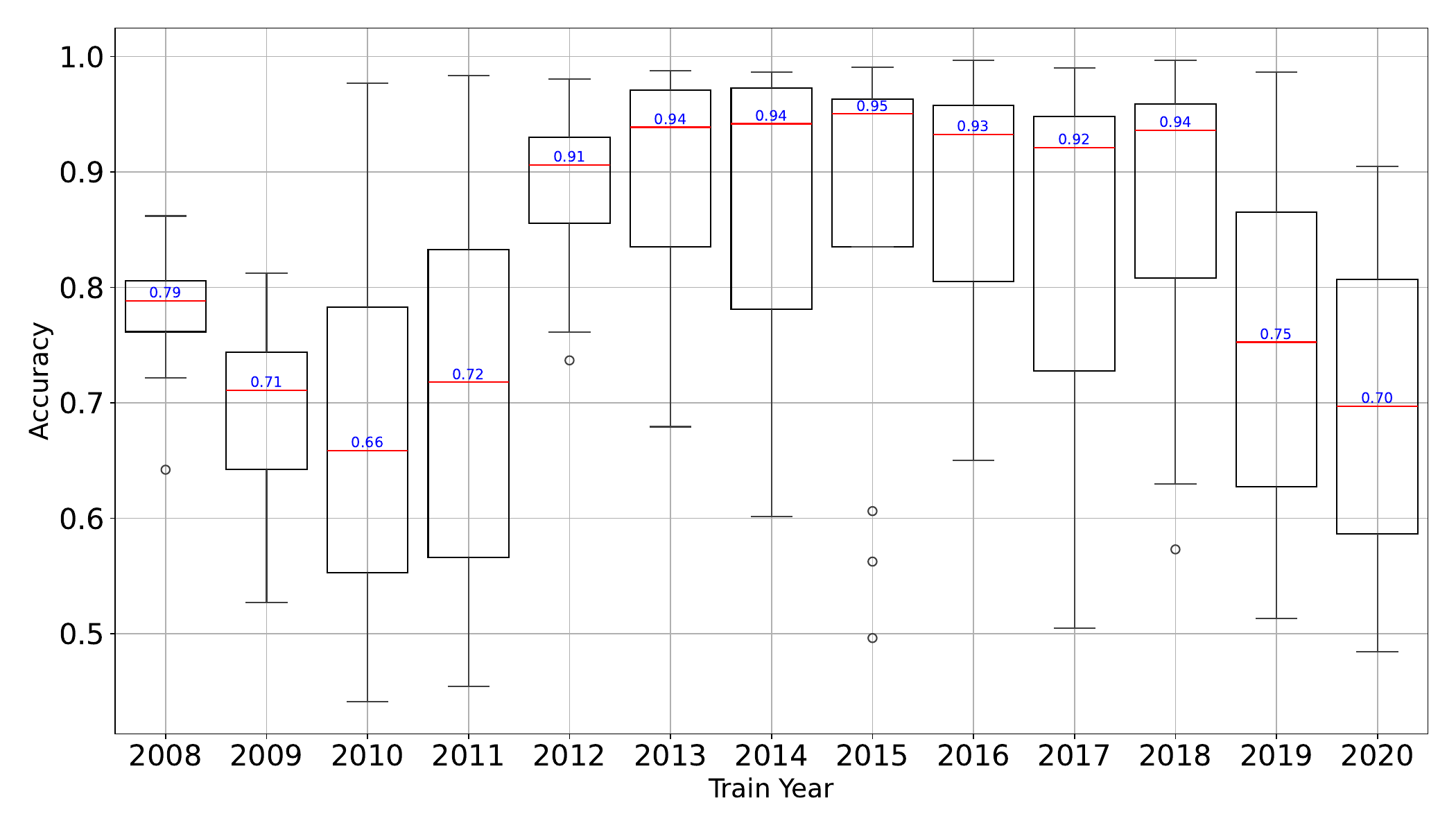}\vspace{-2mm}
        \caption{\normalfont Post-Real-Acc.}
        \label{fig:Box_Plot_RNN_Hybrid_Post_Real_Acc}
    \end{subfigure}
    ~
    \begin{subfigure}[t]{0.235\textwidth}
        \centering
        \includegraphics[width=0.99\textwidth]{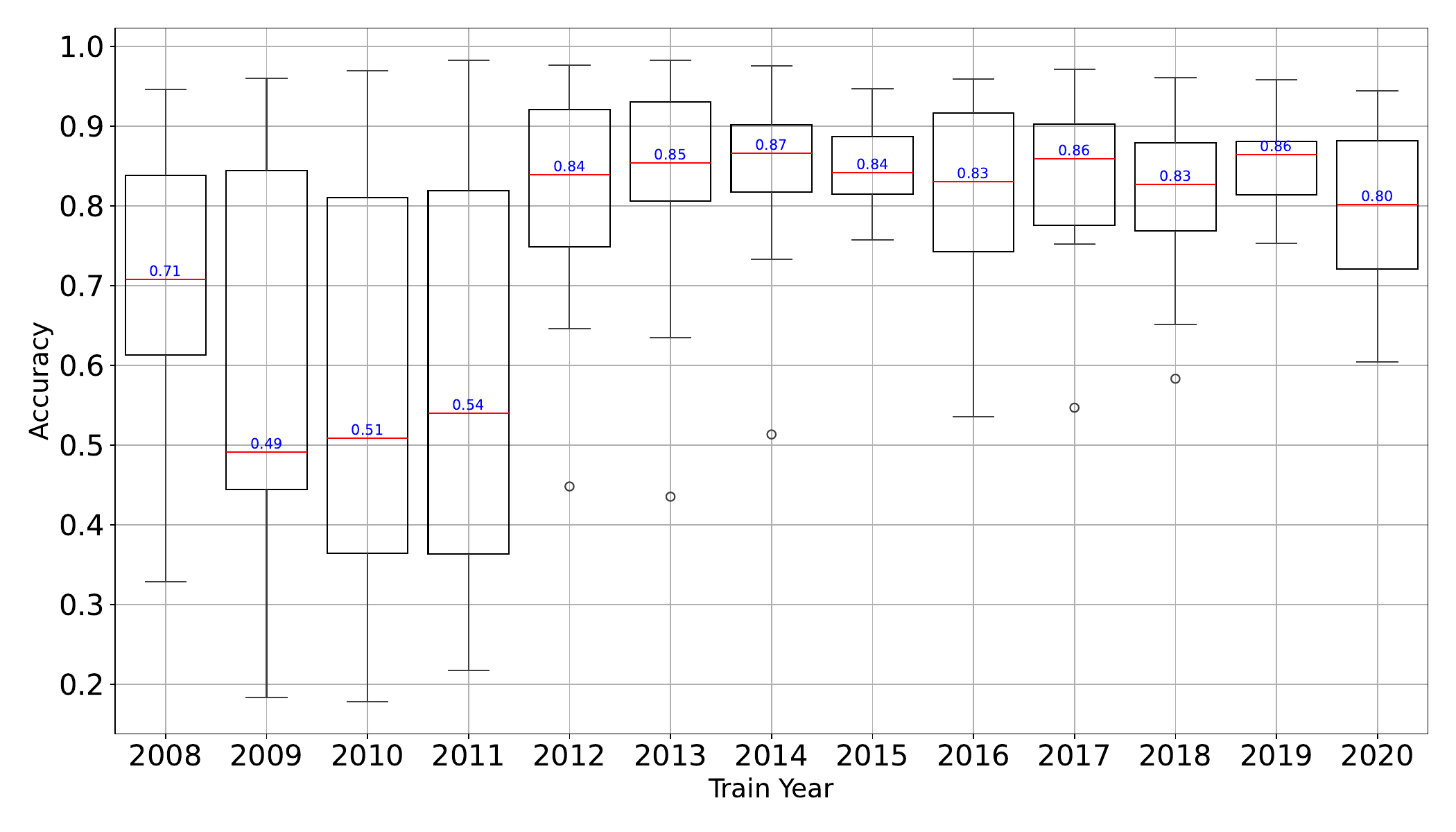}\vspace{-2mm}
        \caption{\normalfont Pre-Emu-Acc.}
        \label{fig:Box_Plot_RNN_Hybrid_Pre_Emu_Acc}
    \end{subfigure}
    ~
    \begin{subfigure}[t]{0.235\textwidth}
        \centering
        \includegraphics[width=0.99\textwidth]{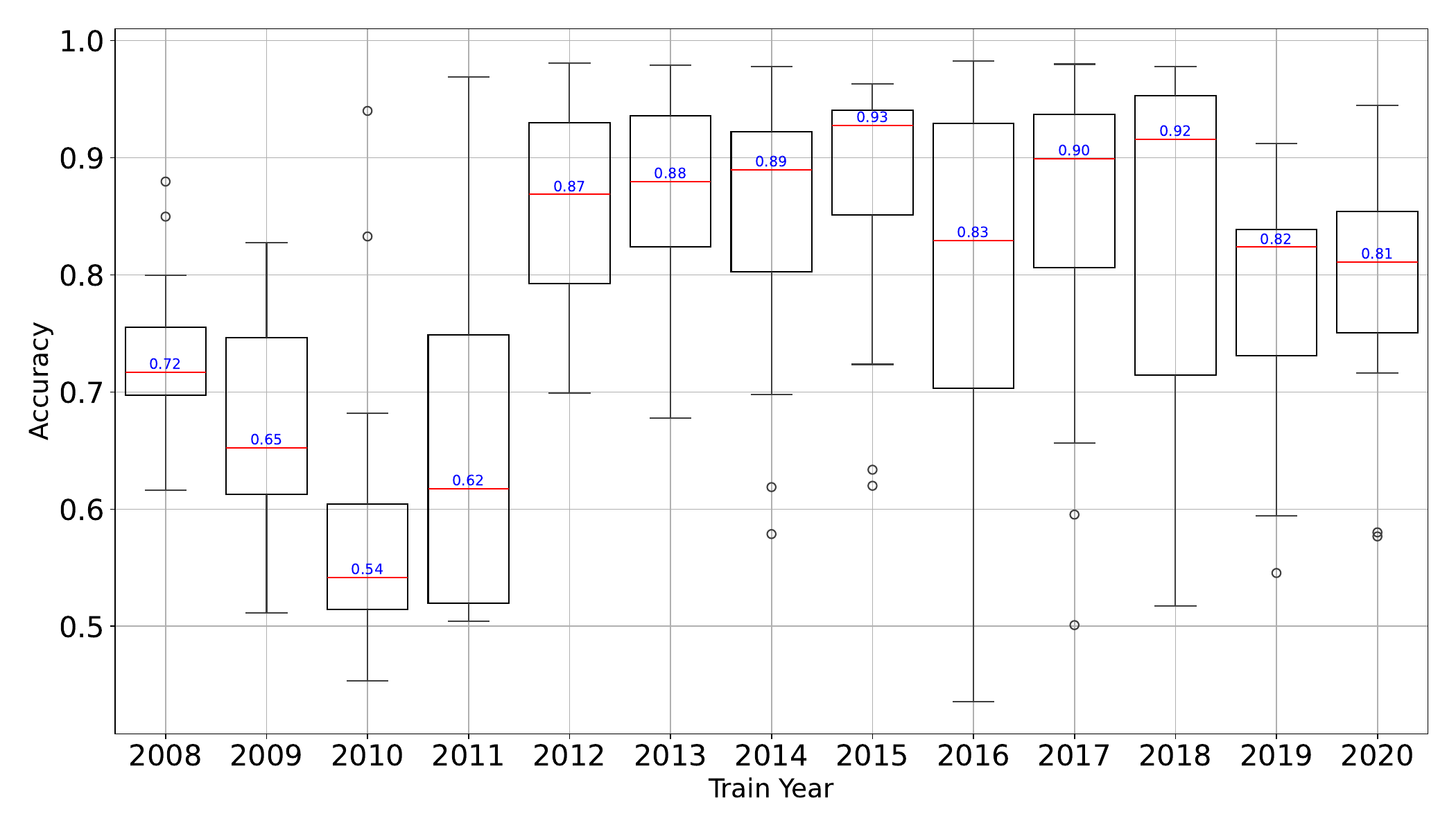}\vspace{-2mm}
        \caption{\normalfont Post-Emu-Acc.}
        \label{fig:Box_Plot_RNN_Hybrid_Post_Emu_Acc}
    \end{subfigure}
    \vspace{-2mm}
    \caption{\normalfont Cross-year pre- and post-balancing RNN accuracy with hybrid features from real/emulator data. (F1 in \autoref{fig:BoxBlot_RNN_Hybrid_Real_Emu_F1})}
    \label{fig:BoxBlot_RNN_Hybrid_Real_Emu}\vspace{-3mm}
\end{figure*}

\EMP{Hybrid Features}
The hybrid feature set nearly doubled in size to illustrate concept drift across various algorithms and its impact over time. When time was ignored, RNN achieved the highest accuracy of 0.982 on both real devices and emulators. \autoref{fig:BoxBlot_RNN_Hybrid_Real_Emu} highlights drift trends by plotting RNN model accuracy with hybrid features, revealing distinct pre- and post-balancing patterns. Before balancing, accuracy fluctuates significantly over the years, particularly early on (2008–2012), indicating performance instability. Upon balancing, accuracy stabilizes across both environments, as the balancing technique mitigates some adverse effects of the imbalance. However, variations persist, bracing that while balancing improves stability, it does not eliminate the drift's impact on performance over time.

For other algorithms employing this strategy, the results were similar, though detection accuracy varied across models. The Troid dataset was used, as the Kronodroid dataset does not support the relevant features.

\begin{takeaway}
Expanding the feature set did not mitigate concept drift. Additionally, while balancing improves performance, it does not eliminate the drift's impact on accuracy.
\end{takeaway}

\EMP{LLM-Based Results}
We evaluated two LLMs using few-shot learning approaches. Each model was assessed using three types of features: static (169 features), dynamic (288 features), and hybrid (457 features). Principal Component Analysis (PCA) was applied to each feature set for dimensionality reduction, with the number of components fixed at 30.

\begin{figure*}[t]
    \centering
    \begin{subfigure}[t]{0.23\textwidth} 
        \centering
        \includegraphics[width=0.99\textwidth]{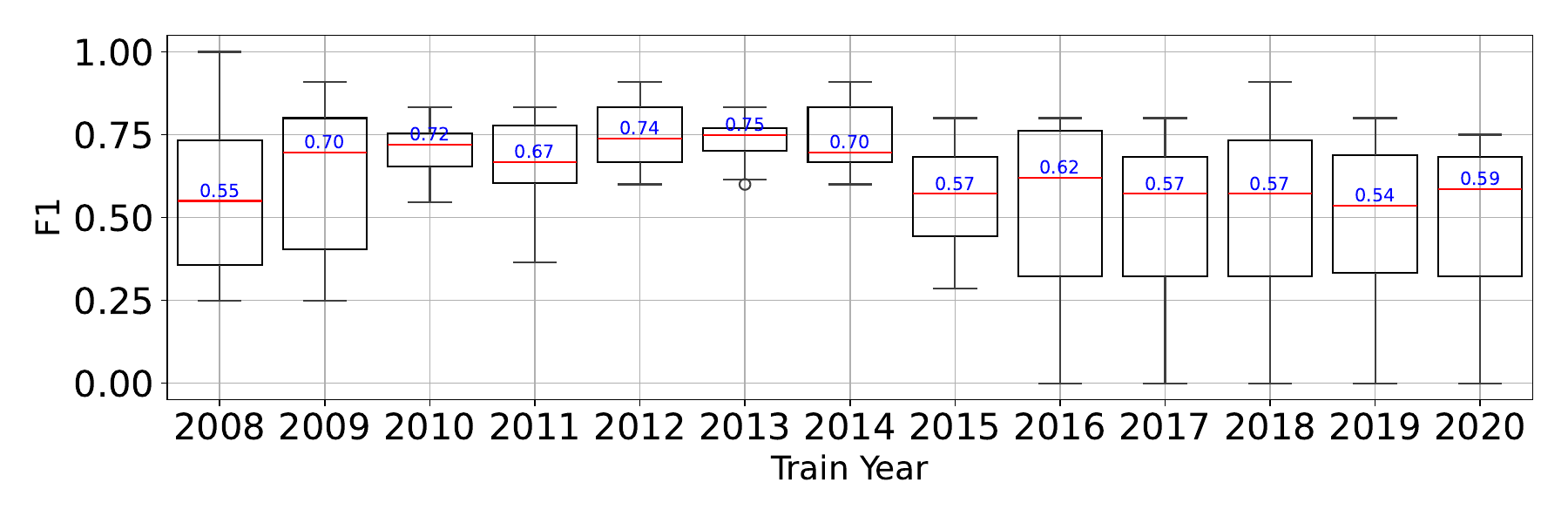}\vspace{-2mm}
        \caption{\normalfont L-Real-S-F1.}
        \label{fig:Box_Plot_LLM_Meta_F1_Real_Static}
    \end{subfigure}
    ~
    \begin{subfigure}[t]{0.23\textwidth}  
        \centering
        \includegraphics[width=0.99\textwidth]{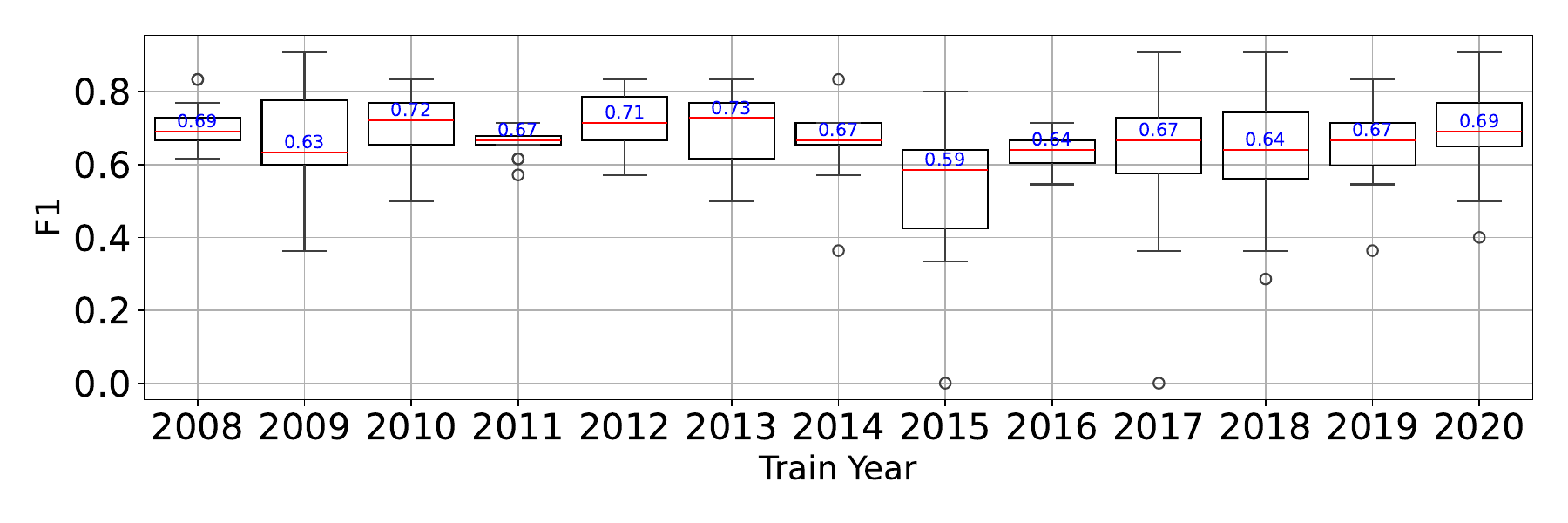}\vspace{-2mm}
        \caption{\normalfont E-Real-S-F1.}
        \label{fig:Box_Plot_LLM_LG_F1_Real_Static}
    \end{subfigure}
    ~
    \begin{subfigure}[t]{0.23\textwidth}
        \centering
        \includegraphics[width=0.99\textwidth]{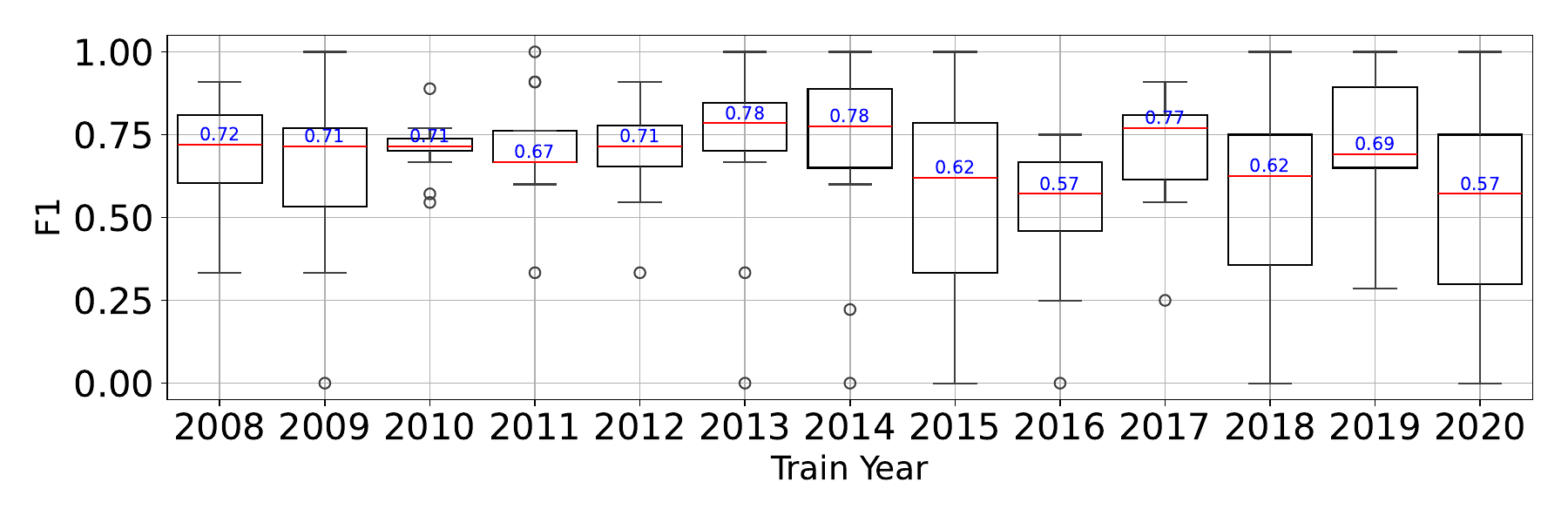}\vspace{-2mm}
        \caption{\normalfont L-Emu-S-F1.}
        \label{fig:fig:Box_Plot_LLM_Meta_F1_Emu_Static}
    \end{subfigure}
    ~
    \begin{subfigure}[t]{0.23\textwidth}
        \centering
        \includegraphics[width=0.99\textwidth]{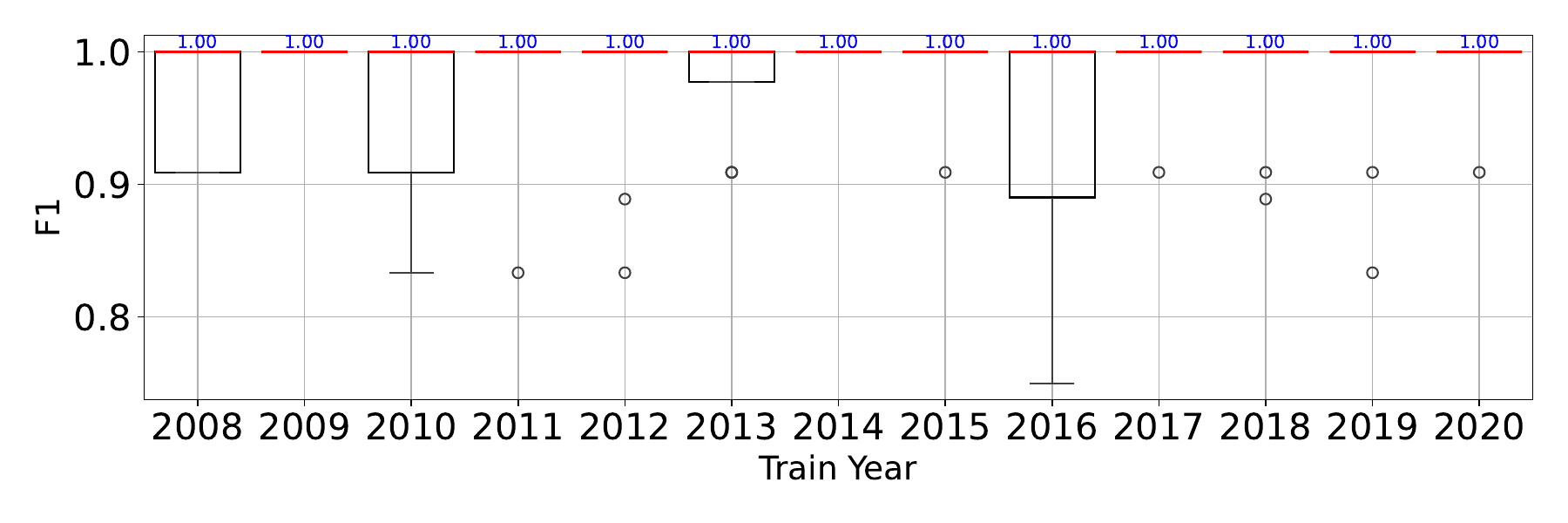}\vspace{-2mm}
        \caption{\normalfont E-Emu-S-F1.}
        \label{fig:Box_Plot_LLM_LG_F1_Emu_Static}
    \end{subfigure}

    ~
    
    \begin{subfigure}[t]{0.23\textwidth} 
        \centering
        \includegraphics[width=0.99\textwidth]{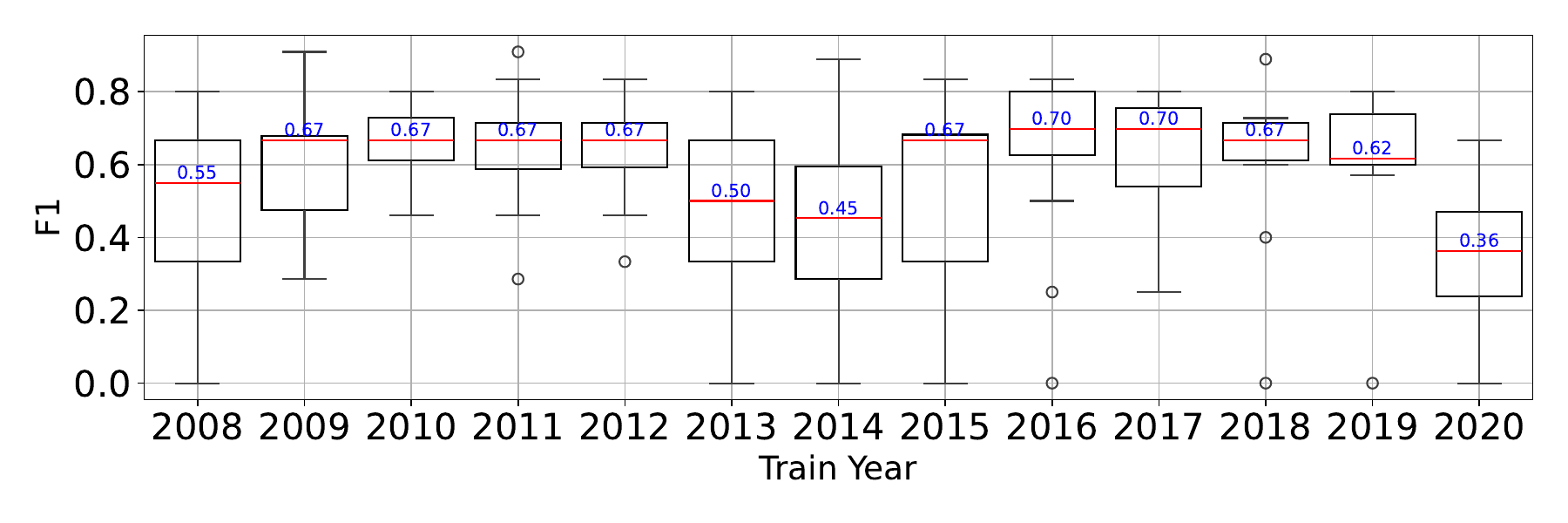}\vspace{-2mm}
        \caption{\normalfont L-Real-D-F1.}
        \label{fig:Box_Plot_LLM_Meta_F1_Real_Dynamic}
    \end{subfigure}
    ~
    \begin{subfigure}[t]{0.23\textwidth}  
        \centering
        \includegraphics[width=0.99\textwidth]{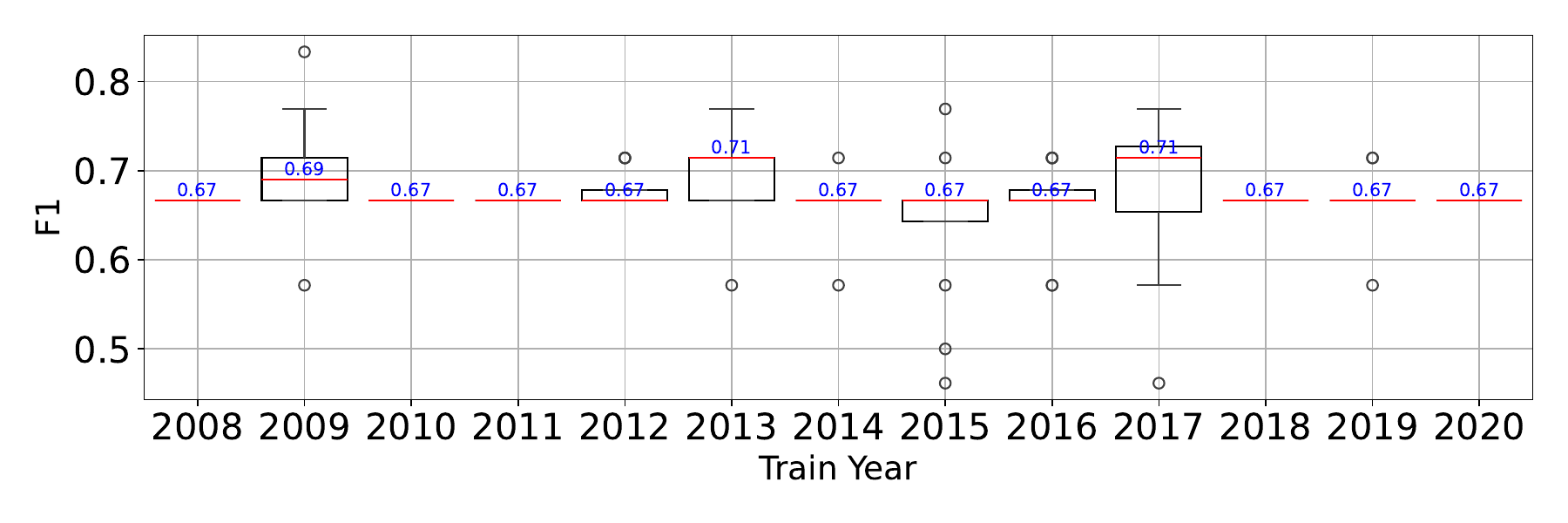}\vspace{-2mm}
        \caption{\normalfont E-Real-D-F1.}
        \label{fig:Box_Plot_LLM_LG_F1_Real_Dynamic}
    \end{subfigure}
    ~
    \begin{subfigure}[t]{0.23\textwidth}
        \centering
        \includegraphics[width=0.99\textwidth]{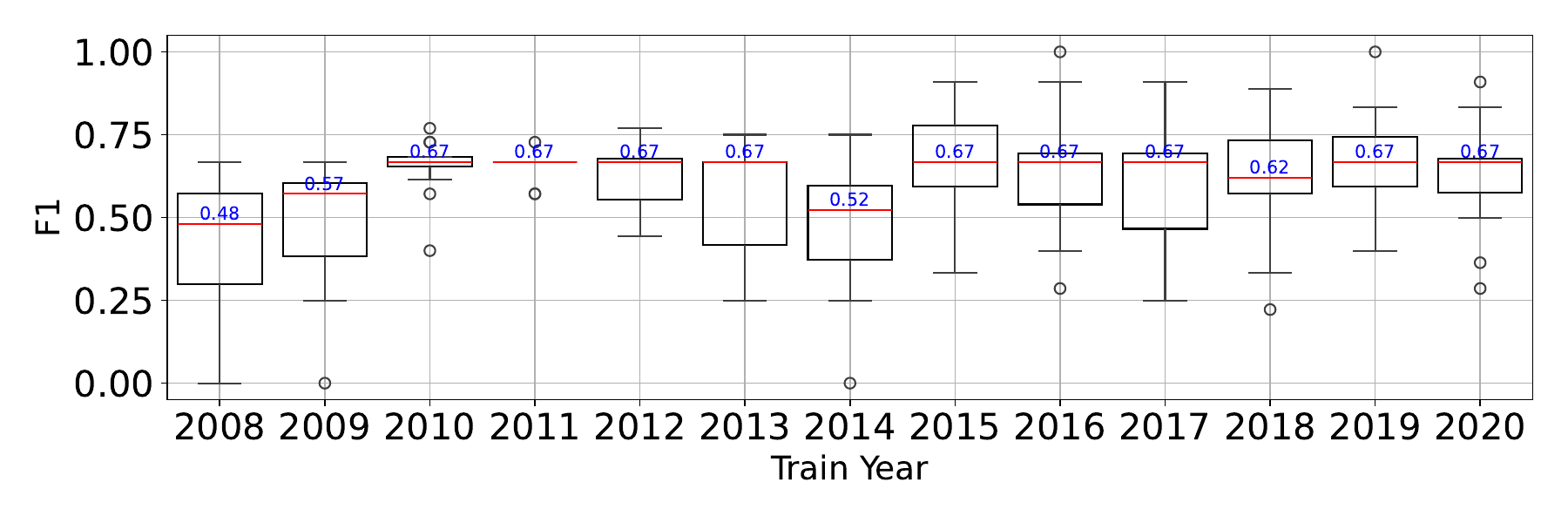}\vspace{-2mm}
        \caption{\normalfont L-Emu-D-F1.}
    \label{fig:fig:Box_Plot_LLM_Meta_F1_Emu_Dynamic}
    \end{subfigure}
    ~
    \begin{subfigure}[t]{0.23\textwidth}
        \centering
        \includegraphics[width=0.99\textwidth]{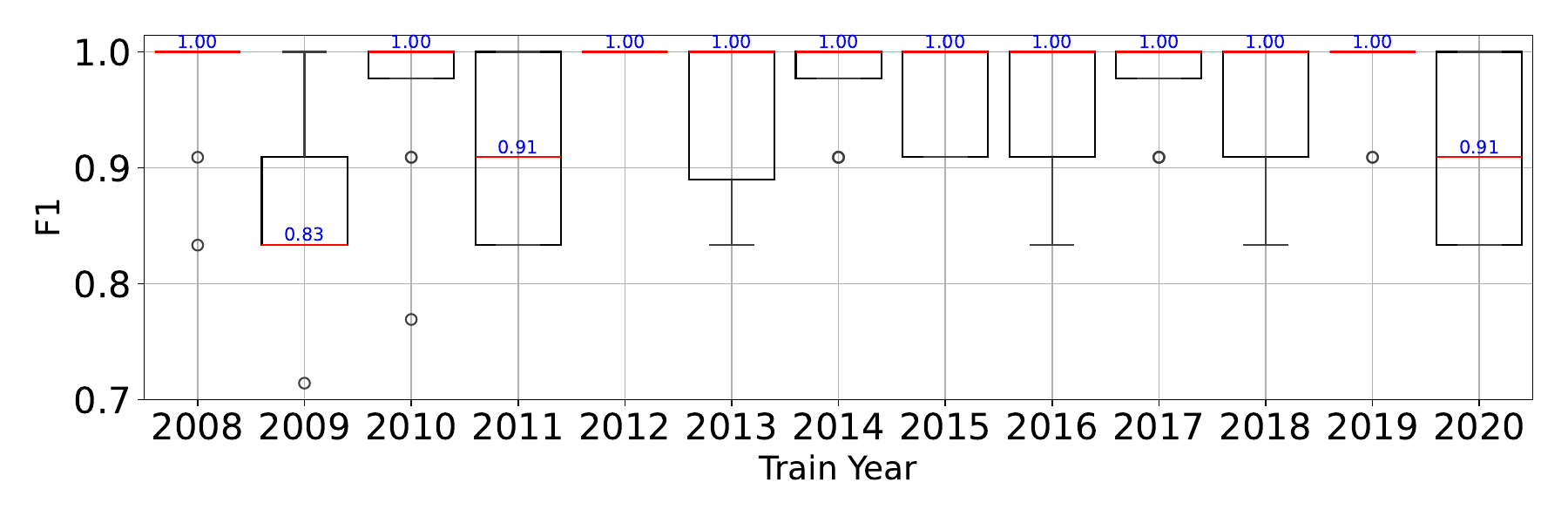}\vspace{-2mm}
        \caption{\normalfont E-Emu-D-F1.}
        \label{fig:Box_Plot_LLM_LG_F1_Emu_Dynamic}
    \end{subfigure}

    ~ 

    \begin{subfigure}[t]{0.23\textwidth} 
        \centering
        \includegraphics[width=0.99\textwidth]{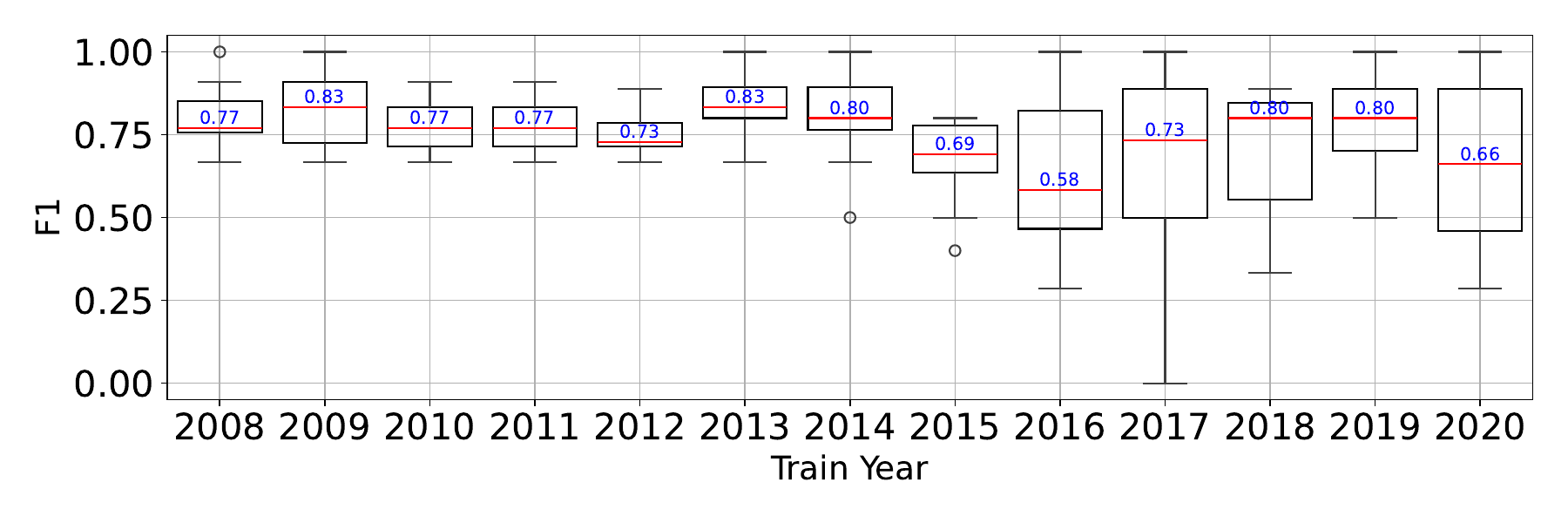}\vspace{-2mm}
        \caption{\normalfont L-Real-H-F1.}
        \label{fig:Box_Plot_LLM_Meta_F1_Real_Hybrid}
    \end{subfigure}
    ~
    \begin{subfigure}[t]{0.23\textwidth}  
        \centering
        \includegraphics[width=0.99\textwidth]{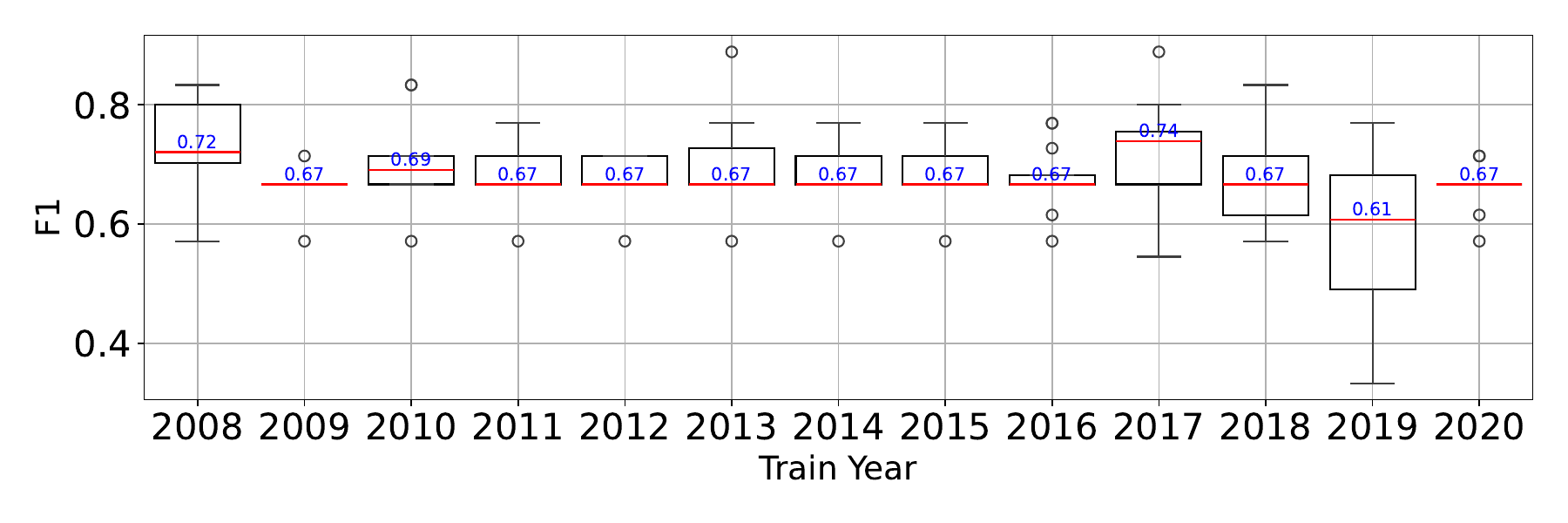}\vspace{-2mm}
        \caption{\normalfont E-Real-H-F1.}
        \label{fig:Box_Plot_LLM_LG_F1_Real_Hybrid}
    \end{subfigure}
    ~
    \begin{subfigure}[t]{0.23\textwidth}
        \centering
        \includegraphics[width=0.99\textwidth]{figsr/LLM/BP_F1_Meta_Emu_Hybrid.pdf}\vspace{-2mm}
        \caption{\normalfont L-Emu-H-F1.}
        \label{fig:fig:Box_Plot_LLM_Meta_F1_Emu_Hybrid}
    \end{subfigure}
    ~
    \begin{subfigure}[t]{0.23\textwidth}
        \centering
        \includegraphics[width=0.99\textwidth]{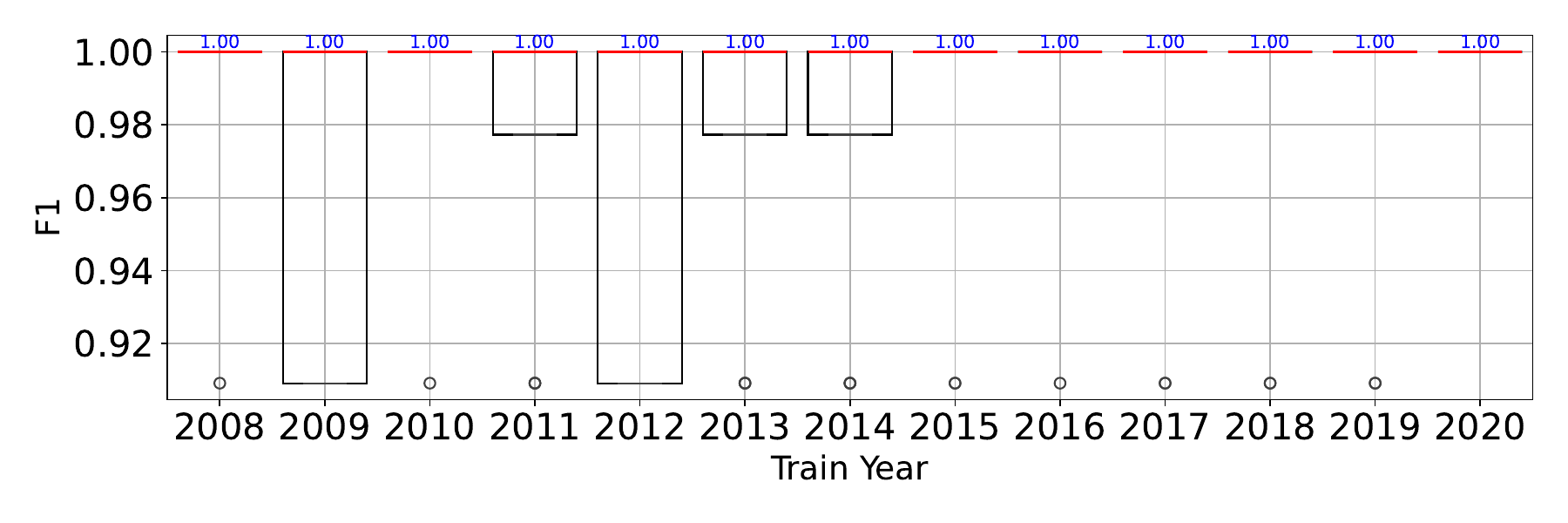}\vspace{-2mm}
        \caption{\normalfont E-Emu-H-F1.}
        \label{fig:Box_Plot_LLM_LG_F1_Emu_Hybrid}
    \end{subfigure}
    \vspace{-3mm}

    \caption{\normalfont LLaMA (L) and Exaone (E) results with Static (S), Dynamic (D), and Hybrid (H) features using real/emulator data.}
    \label{fig:BoxBlot_LLM_models}\vspace{-3mm}
\end{figure*}

Figure~\ref{fig:BoxBlot_LLM_models} presents the F1 score distributions for all feature types across real and emulator datasets. Since the primary objective is to investigate whether LLMs can help mitigate concept drift, we observe that drift effects remain visible across different data types and feature categories. Although the models were trained using few-shot samples, the results demonstrate their ability to capture meaningful patterns and perform reasonably well on the classification task.

\begin{figure*}[t]
    \centering
    \begin{subfigure}[t]{0.23\textwidth} 
        \centering
        \includegraphics[width=0.99\textwidth]{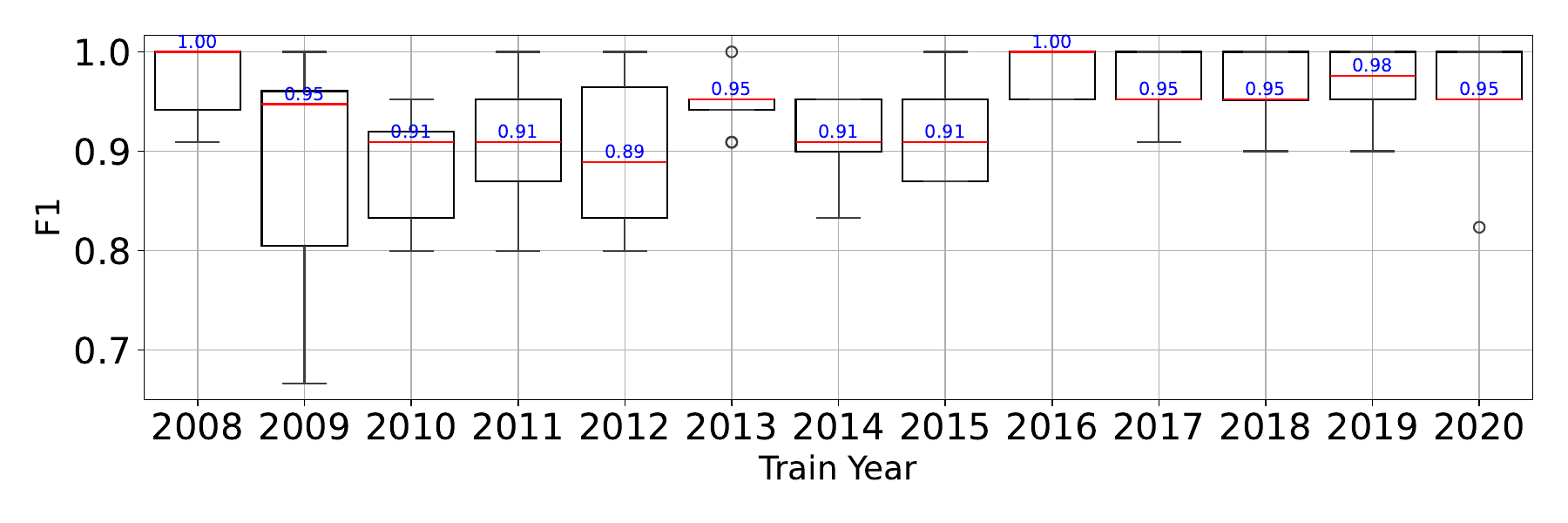}\vspace{-2mm}
        \caption{\normalfont Static}
        \label{fig:Box_Plot_LLM_5020LG_F1_Real_Static}
    \end{subfigure}
    ~
    \begin{subfigure}[t]{0.23\textwidth}  
        \centering
        \includegraphics[width=0.99\textwidth]{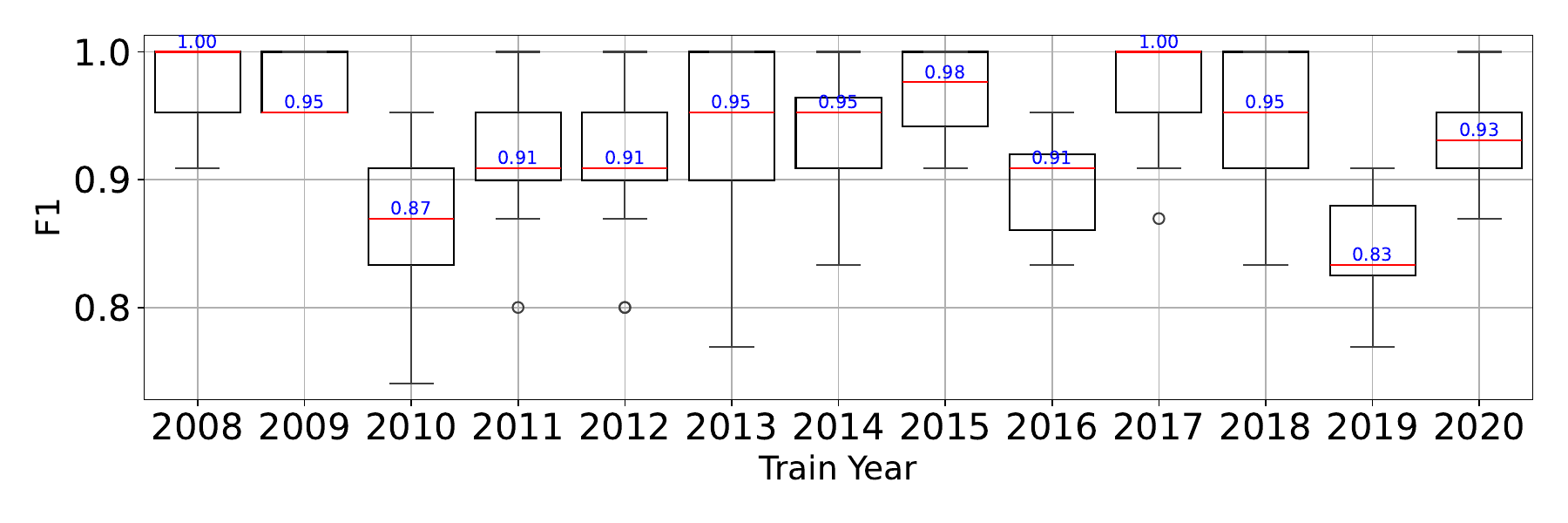}\vspace{-2mm}
        \caption{\normalfont Dynamic}
        \label{fig:Box_Plot_LLM_5020LG_F1_Real_Dynamic}
    \end{subfigure}
    ~
    \begin{subfigure}[t]{0.23\textwidth}
        \centering
        \includegraphics[width=0.99\textwidth]{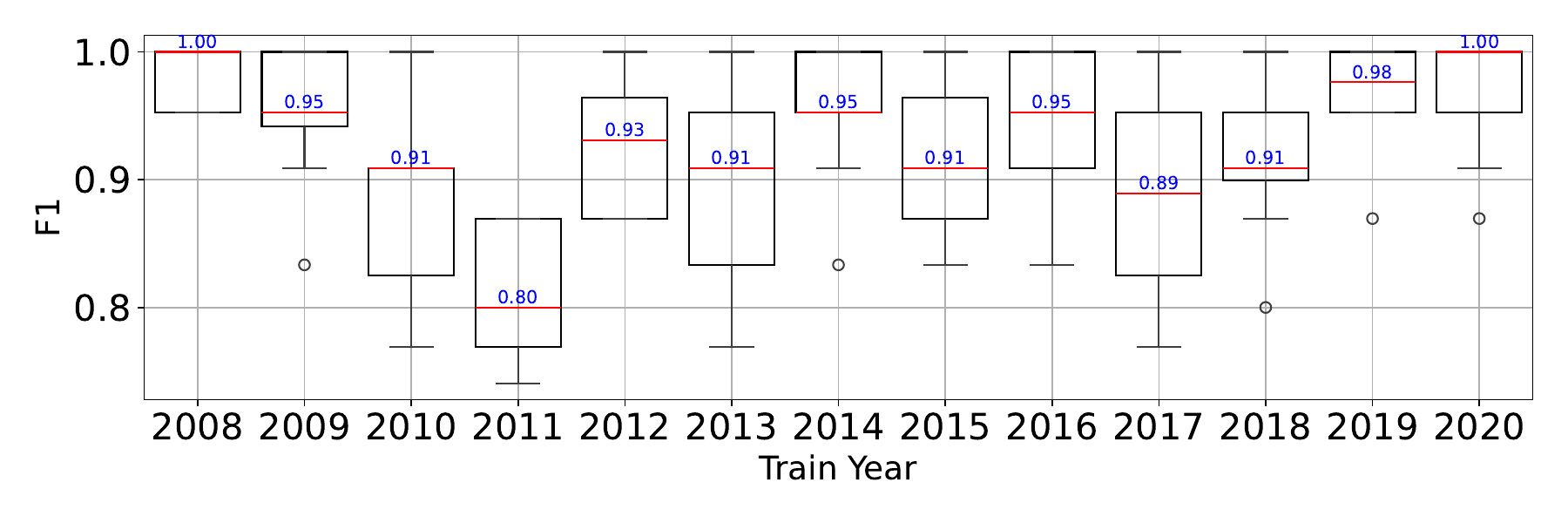}\vspace{-2mm}
        \caption{\normalfont Hybrid}
        \label{fig:fig:Box_Plot_LLM_5020LG_F1_Emu_Hybrid}
    \end{subfigure}
\vspace{-3mm}
    \caption{\normalfont Exaone results on emulator data using static, dynamic, and hybrid features (50 train, 20 test samples).}
    \label{fig:BoxBlot_LLM_LG5020_Hybrid_models}\vspace{-3mm}
\end{figure*}

Under emulator-based settings, Exaone consistently achieved near-perfect median F1 scores ($\approx$1.00) across all years and feature types(\ref{fig:Box_Plot_LLM_LG_F1_Emu_Static}, \ref{fig:Box_Plot_LLM_LG_F1_Emu_Dynamic}, \ref{fig:Box_Plot_LLM_LG_F1_Emu_Hybrid}). To validate this apparent drift resilience, we repeated the experiments with an increased training size (50/20 split). As shown in Figure~\ref{fig:BoxBlot_LLM_LG5020_Hybrid_models}, Exaone maintained high performance overall. However, a closer examination of the lower whiskers and outliers reveals performance degradation in specific years, indicating that the model is not entirely immune to drift, especially in more variable data contexts.

\begin{takeaway}
While LLMs show promise in few-shot malware detection, concept drift remains evident across years and feature types. Moreover, their effectiveness is constrained by token limits, restricting the feature information that can be processed during inference.
\end{takeaway}

\BfPara{Troid Dataset} Two sets of features were selected from Troid: (1) the API call sequences were converted into vectors using TF-IDF and then fed into classifiers, and (2) the hexadecimal data from malware and benign applications, which were used to generate grayscale and color images (RGB) for image-based classification.

\begin{figure*}[t]
    \centering
    \begin{subfigure}[t]{0.15\textwidth} 
        \centering
        \includegraphics[width=0.80\textwidth]{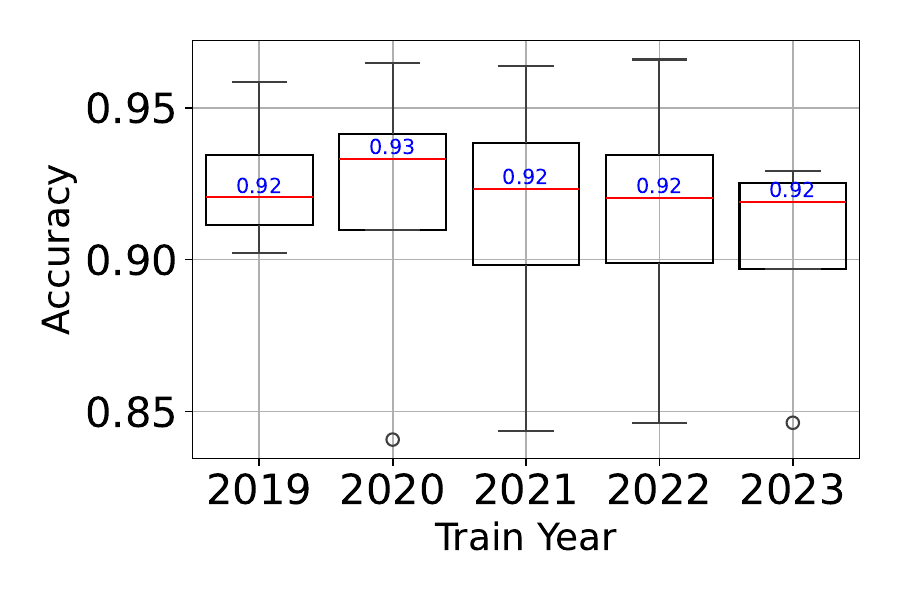}\vspace{-2mm}
        \caption{\normalfont}
        \label{fig:BP_Acc_Pre_ATAT_RF_Troid_NLP}
    \end{subfigure} 
     ~\hspace{-2.5em}
    \begin{subfigure}[t]{0.15\textwidth}  
        \centering
        \includegraphics[width=0.80\textwidth]{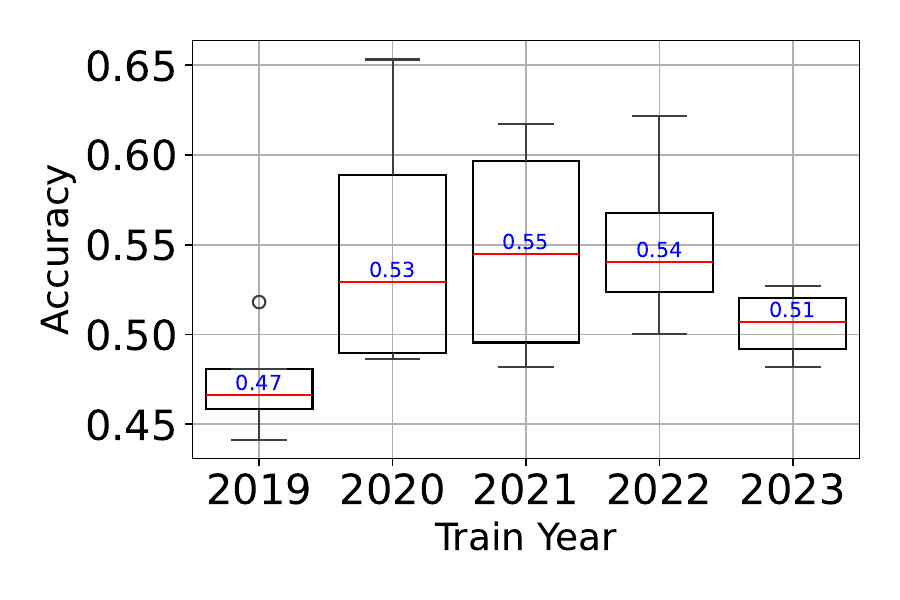}\vspace{-2mm}
        \caption{\normalfont}
        \label{fig:BP_Acc_Post_ATAT_RF_Troid_NLP}
    \end{subfigure}
     ~\hspace{-2.5em}
    \begin{subfigure}[t]{0.15\textwidth}
        \centering
        \includegraphics[width=0.80\textwidth]{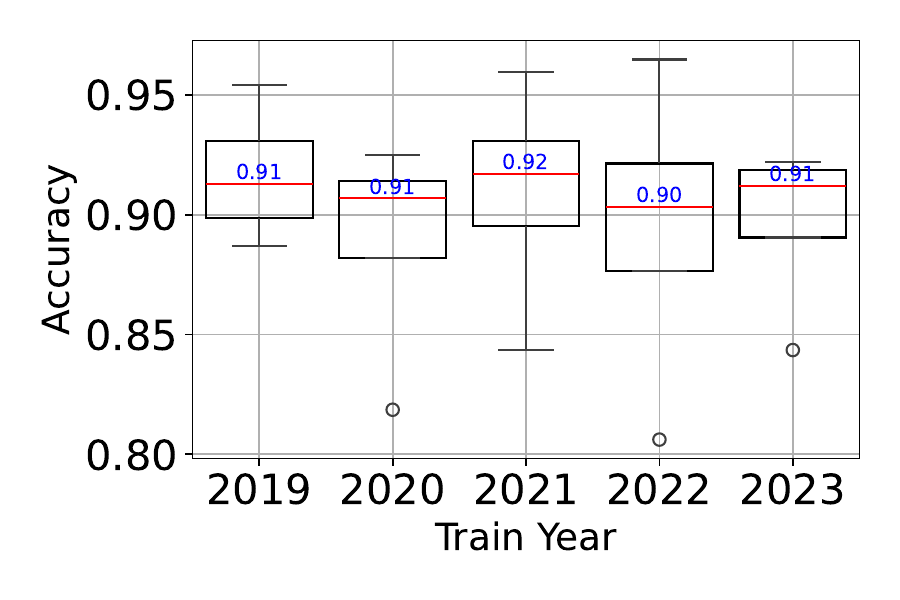}\vspace{-2mm}
        \caption{\normalfont}
        \label{fig:BP_Acc_Pre_ATAT_GB_Troid_NLP}
 
    \end{subfigure}
     ~\hspace{-2.5em}
    \begin{subfigure}[t]{0.15\textwidth}
        \centering
        \includegraphics[width=0.80\textwidth]{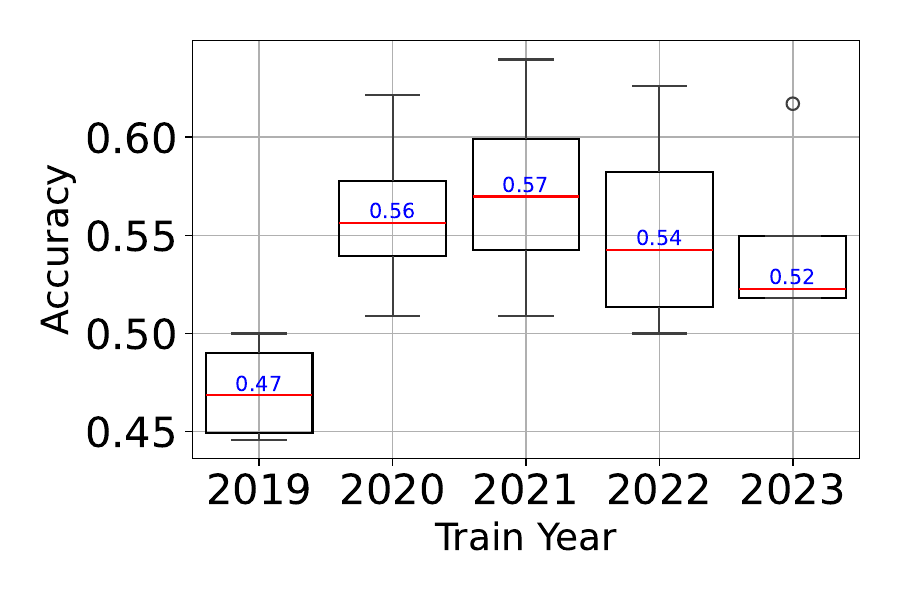}\vspace{-2mm}
        \caption{\normalfont}
        \label{fig:BP_Acc_Post_ATAT_GB_Troid_NLP}
    \end{subfigure} ~\hspace{-2.4em}
       \begin{subfigure}[t]{0.15\textwidth} 
        \centering
        \includegraphics[width=0.80\textwidth]{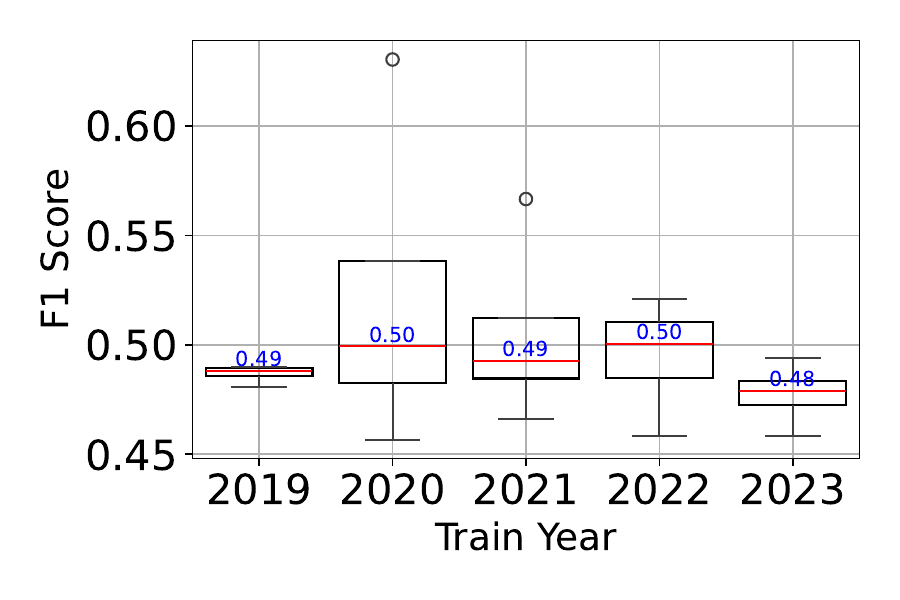}\vspace{-2mm}
        \caption{\normalfont}
        \label{fig:BP_F1_Pre_ATAT_RF_Troid_NLP}
    \end{subfigure}
      ~\hspace{-2.5em}
    \begin{subfigure}[t]{0.15\textwidth}  
        \centering
        \includegraphics[width=0.80\textwidth]{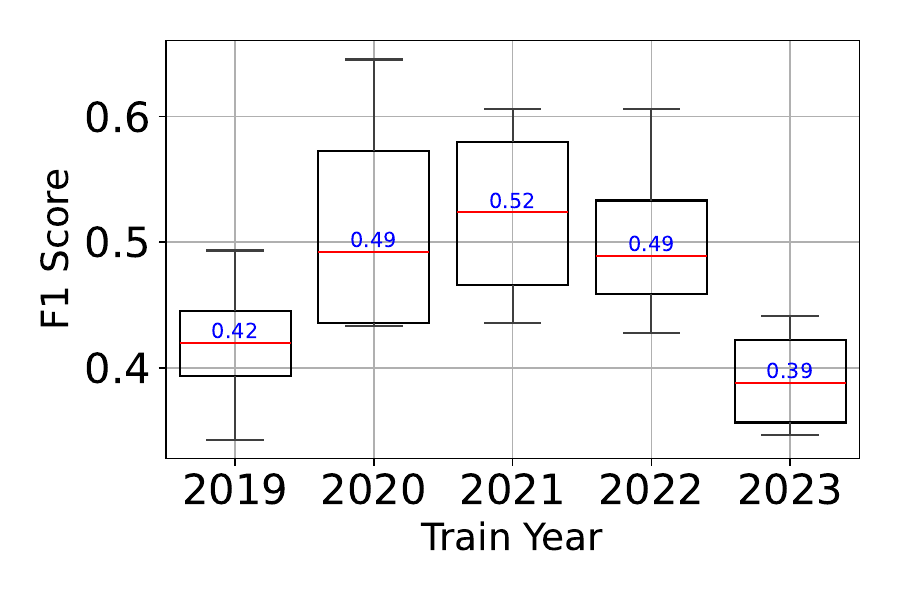}\vspace{-2mm}
        \caption{\normalfont}
        \label{fig:BP_F1_Post_ATAT_RF_Troid_NLP}
    \end{subfigure}
      ~\hspace{-2.5em}
    \begin{subfigure}[t]{0.15\textwidth}
        \centering
        \includegraphics[width=0.80\textwidth]{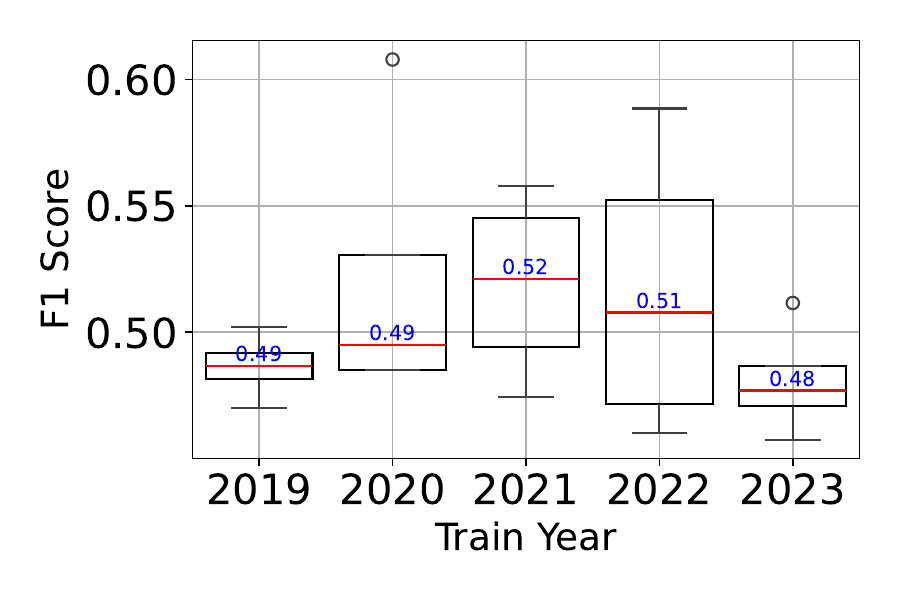}\vspace{-2mm}
        \caption{\normalfont}
        \label{fig:BP_F1_Pre_ATAT_GB_Troid_NLP}
    \end{subfigure}
     ~\hspace{-2.5em}
    \begin{subfigure}[t]{0.15\textwidth}
        \centering
        \includegraphics[width=0.80\textwidth]{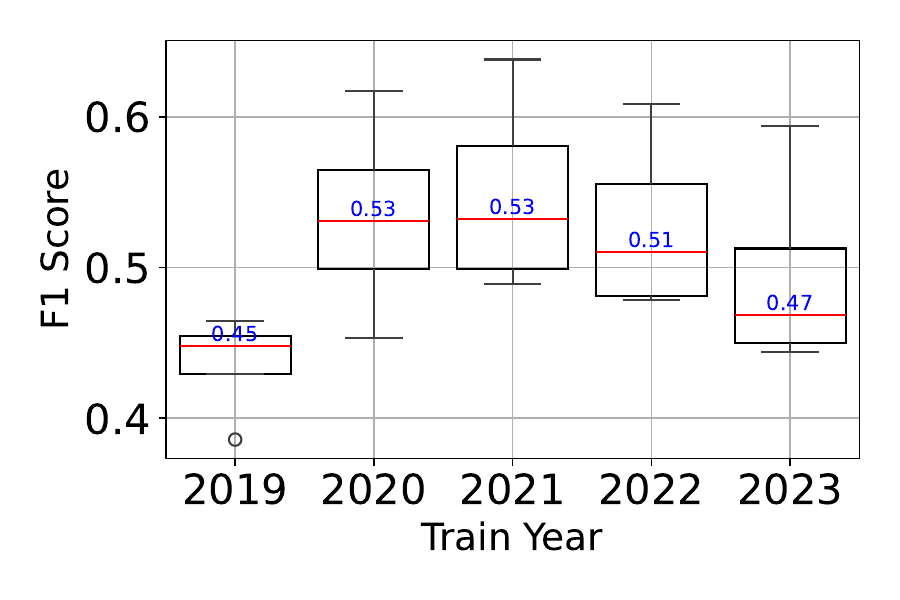}\vspace{-2mm}
        \caption{\normalfont}
        \label{fig:BP_F1_Post_ATAT_GB_Troid_NLP}
    \end{subfigure} \vspace{-2mm}

    \caption{\normalfont The performance of the RF and GB models pre- and post-balancing with API calls sequence features.(a) Pre-RF-Acc, (b) Post-RF-Acc, (c) Pre-GB-Acc, (d) Post-GB-Acc, (e) Pre-RF-F1, (f) Post-RF-F1, (g) Pre-GB-F1., (h) Post-GB-F1. }
    \label{fig:Box_Plot_NLP_RF_GB_TDS_Cross}\vspace{-5mm}
\end{figure*}

\EMP{API Calls}  
Experimental results demonstrate the detection performance of RF and GB models over time using API call sequences, both before and after applying a balancing algorithm. \autoref{fig:Box_Plot_NLP_RF_GB_TDS_Cross} shows that classification performance is influenced by concept drift and class imbalance. Before balancing, both models maintained consistent performance due to the dominance of benign samples. After balancing, accuracy fluctuated, indicating that while class imbalance was addressed, instability was introduced. The F1 score, a more comprehensive metric, exhibited greater fluctuation, highlighting the algorithm's impact on correctly identifying both malware and benign samples. Post-balancing box plots reveal a wider range, particularly for the F1 score, suggesting increased variability in model performance. Additionally, median accuracy and F1 score slightly declined, indicating a potential trade-off between mitigating class imbalance and maintaining overall performance.

\begin{takeaway}
Contrary to expectations based on prior results, balancing API call features exacerbates the concept drift issue, resulting in a decline in both F1 score and accuracy, as evidenced by their average and variance.\end{takeaway}

\begin{figure*}[t]
    \centering
    \begin{subfigure}[t]{0.15\textwidth} 
        \centering
        \includegraphics[width=0.80\textwidth]{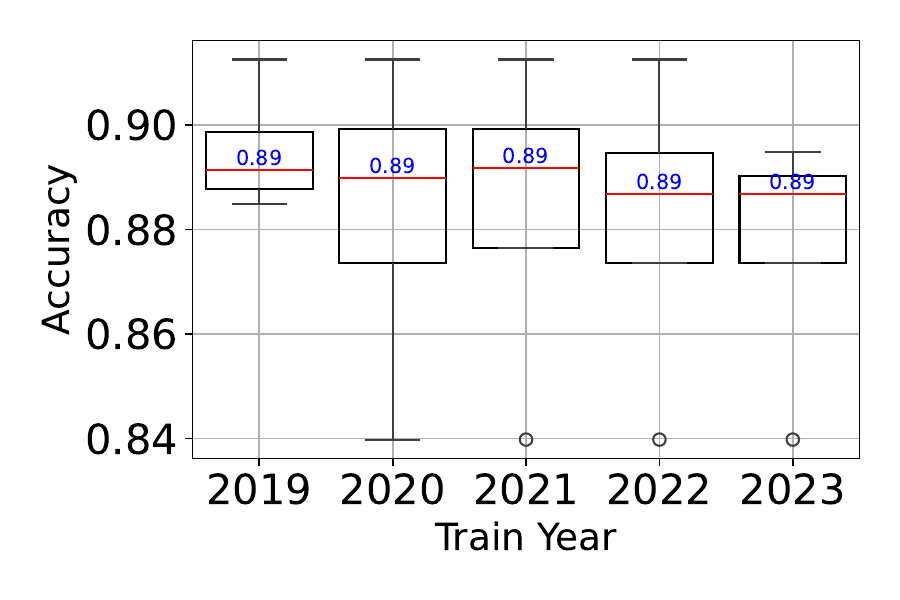}\vspace{-2mm}
        \caption{\normalfont }
        \label{fig:BP_Acc_Pre_ATAT_CNN_Troid_ImageGS}
    \end{subfigure} 
    ~\hspace{-2.5em}
    \begin{subfigure}[t]{0.15\textwidth}  
        \centering
        \includegraphics[width=0.80\textwidth]{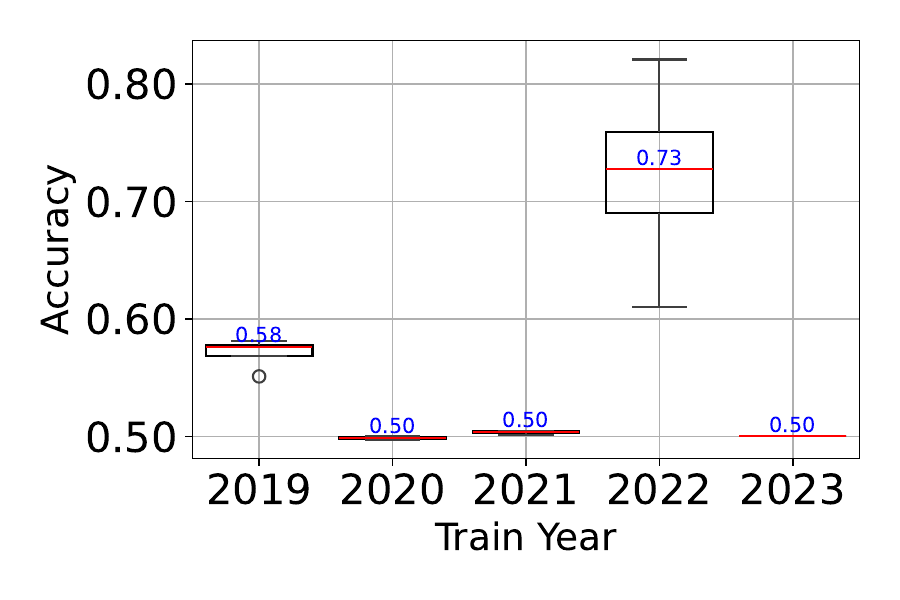}\vspace{-2mm}
        \caption{\normalfont }
        \label{fig:BP_Acc_Post_ATAT_CNN_Troid_ImageGS}
    \end{subfigure}
     ~\hspace{-2.5em}
    \begin{subfigure}[t]{0.16\textwidth}
        \centering
        \includegraphics[width=0.80\textwidth]{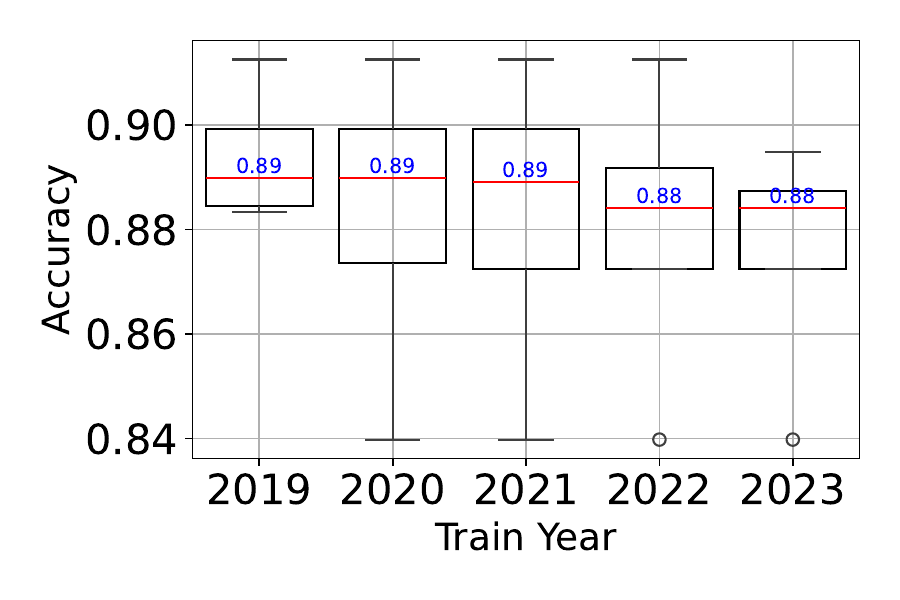}\vspace{-2mm}
        \caption{\normalfont }
        \label{fig:BP_Acc_Pre_ATAT_CNN_Troid_ImageRGB}
    \end{subfigure}
     ~\hspace{-2.5em}
    \begin{subfigure}[t]{0.15\textwidth}
        \centering
        \includegraphics[width=0.80\textwidth]{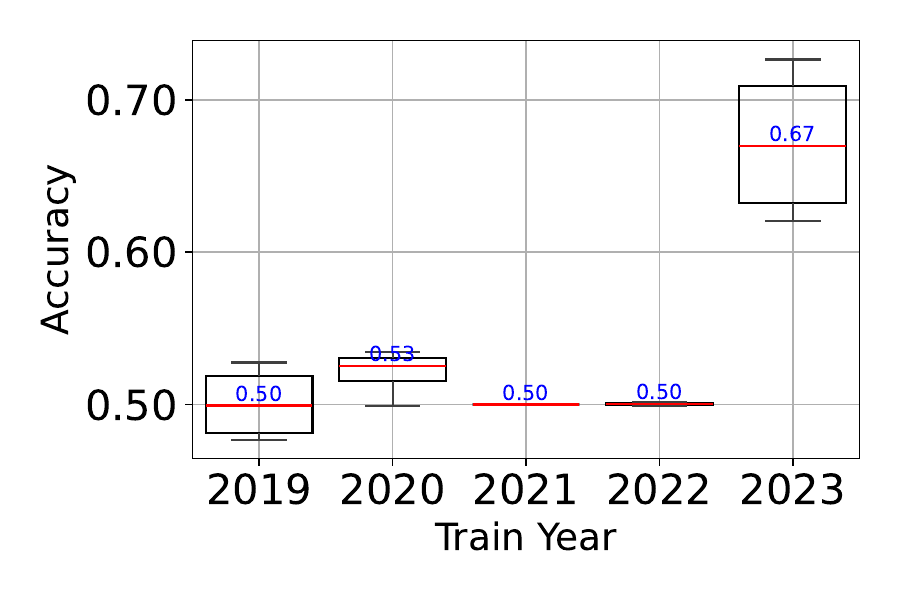}\vspace{-2mm}
        \caption{\normalfont }
        \label{fig:BP_Acc_Post_ATAT_CNN_Troid_ImageRGB}
    \end{subfigure} 
    ~\hspace{-2.5em}
\begin{subfigure}[t]{0.15\textwidth} 
        \centering
        \includegraphics[width=0.80\textwidth]{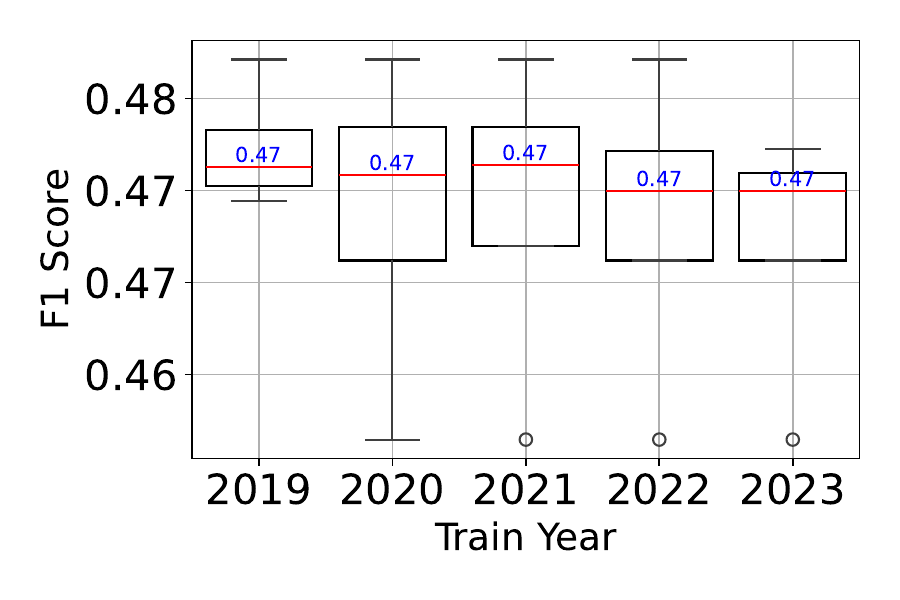}\vspace{-2mm}
        \caption{\normalfont }
        
        \label{fig:BP_F1_Pre_ATAT_CNN_Troid_ImageGS}
    \end{subfigure}
    ~\hspace{-2.5em}
    \begin{subfigure}[t]{0.15\textwidth}  
        \centering
        \includegraphics[width=0.80\textwidth]{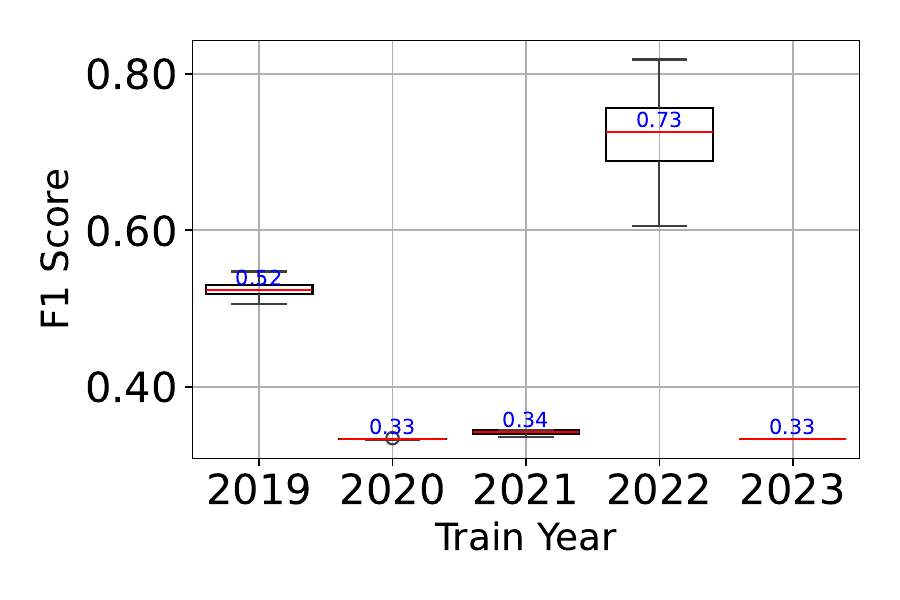}\vspace{-2mm}
        \caption{\normalfont }
        
        \label{fig:BP_F1_Post_ATAT_CNN_Troid_ImageGS}
    \end{subfigure}
    ~\hspace{-2.5em}
    \begin{subfigure}[t]{0.15\textwidth}
        \centering
        \includegraphics[width=0.80\textwidth]{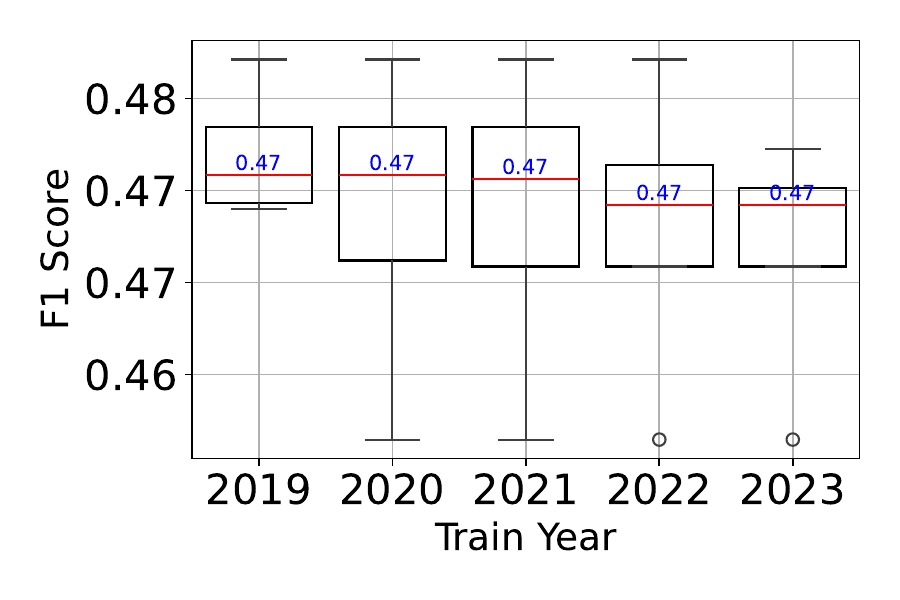}\vspace{-2mm}
        \caption{\normalfont }
        
        \label{fig:BP_F1_Pre_ATAT_CNN_Troid_ImageRGB}
    \end{subfigure}
     ~\hspace{-2.5em}
    \begin{subfigure}[t]{0.15\textwidth}
        \centering
        \includegraphics[width=0.80\textwidth]{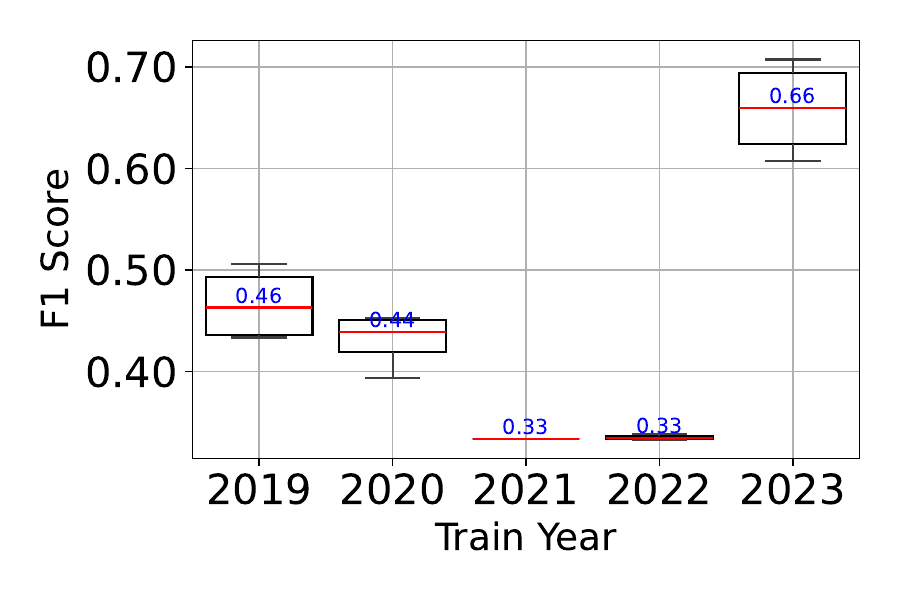}\vspace{-2mm}
        \caption{\normalfont }
        
        \label{fig:BP_F1_Post_ATAT_CNN_Troid_ImageRGB}
    \end{subfigure} \vspace{-2mm}

    \caption{\normalfont The detection performance for CNN  with  RGB and grayscale (GS), pre and post-balancing. (a) Pre-CNN-GS-Acc, (b) Post-CNN-GS-Acc, (c) Pre-CNN-RGB-Acc, (d) Post-CNN-RGB-Acc, (e) Pre-CNN-GS-F1, (f) Post-CNN-GS-F1, (g) Pre-CNN-RGB-F1, (h) Post-CNN-RGB-F1.}
    \label{fig:Box_Plot_RGB_GS_CNN_TDS_Cross}\vspace{-5mm}
\end{figure*}

\EMP{Image Features}  
This approach explores image-based detection to address concept drift, utilizing images derived from hexadecimal values in both grayscale (GS) and RGB formats. The CNN model's performance before and after applying a balancing algorithm is evaluated, with \autoref{fig:Box_Plot_RGB_GS_CNN_TDS_Cross} providing a detailed statistical comparison through box plots.  

Pre-balancing accuracy and F1 scores for grayscale and RGB images, shown in \autoref{fig:BP_Acc_Pre_ATAT_CNN_Troid_ImageGS} and ~\autoref{fig:BP_F1_Post_ATAT_CNN_Troid_ImageGS}, exhibit greater variability and lower median values than post-balancing. The observed difference in accuracy and F1 scores suggests model bias toward the majority class due to imbalance, evident in a parallel improvement pattern where higher accuracy coincides with a decline in F1 scores. This confirms improved performance while reducing its variability.  

Another key observation in the image-based approach is the model's tendency to enter a "forgetting state," where accuracy increases over time while previously learned data is lost. This phenomenon is particularly relevant in the malware landscape, where older malware or its components are repurposed in new attacks, impacting the model's effectiveness in detecting threats. This behavior is further illustrated in \autoref{fig:BP_Acc_Post_ATAT_CNN_Troid_ImageRGB}–\autoref{fig:BP_F1_Post_ATAT_CNN_Troid_ImageRGB}.

\begin{takeaway}
There is a slight correlation between algorithm type and concept drift, regardless of whether models use shallow learning or deep learning.
\end{takeaway}

\subsection{Incremental Strategy}
\label{Incremental_Strategy}
To represent the real-world scenario of how data is collected and used in ML-based detection pipelines, we implement an incremental (or cumulative) strategy for model training and testing. In this strategy, the model is trained on data from a set of consecutive years and tested on the subsequent years, one at a time. For instance, a model was trained with data obtained from 2010 to 2015 (as a single fold) and tested individually with 2016, 2017, and so forth. This is analogous to training a model with data up to 2015 and operating it on incoming data streams for the subsequent years. 

\BfPara{KronoDroid Dataset}
All algorithms displayed the same pattern with differences in accuracy based on the feature types. This indicates that the overall performance is influenced by the features' nature rather than the used algorithm.

\begin{figure*}[t]
    \centering
    \begin{subfigure}[t]{0.235\textwidth} 
        \centering
        \includegraphics[width=0.99\textwidth]{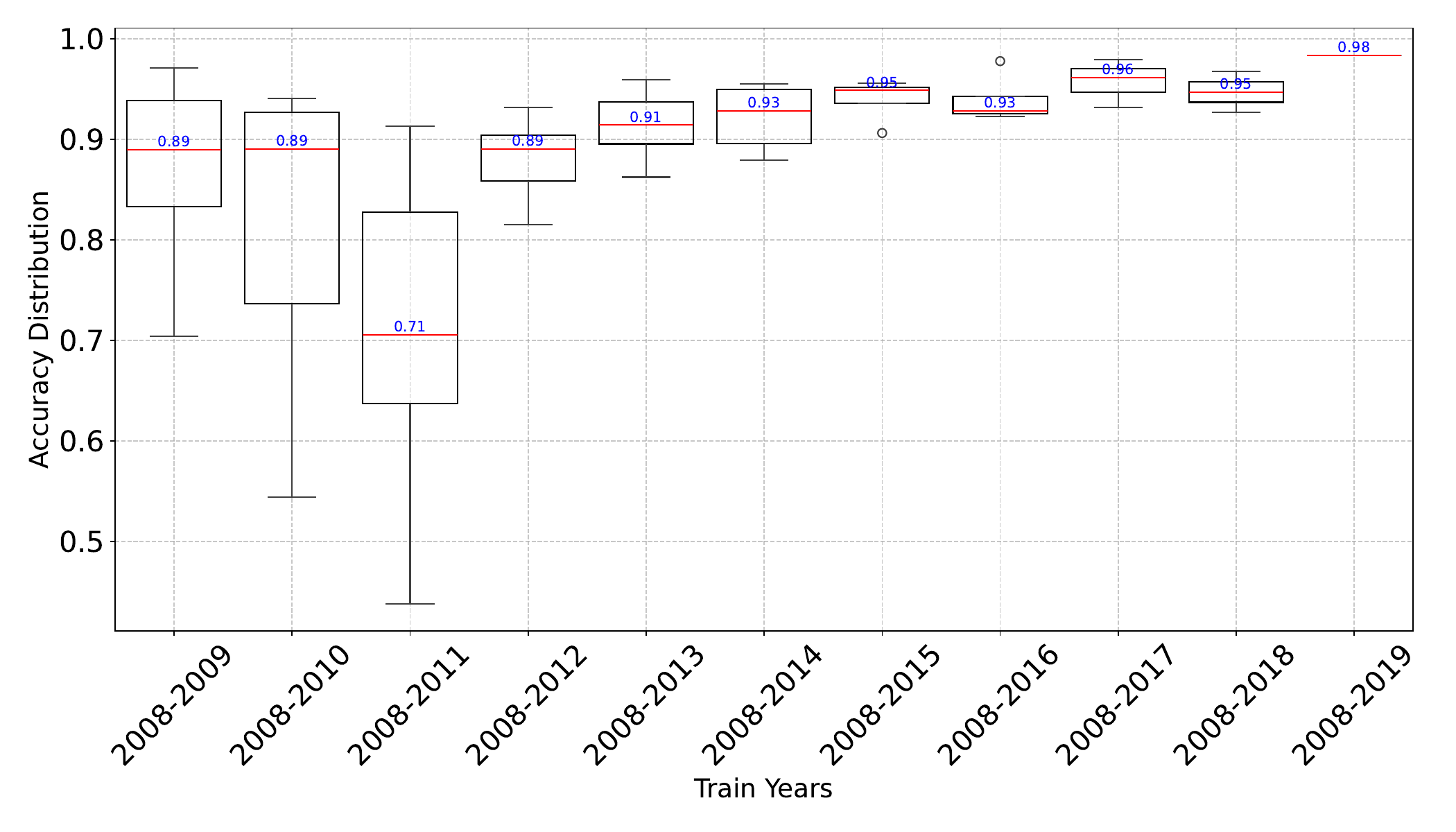}\vspace{-2mm}
        \caption{\normalfont Pre-Real-Acc.}
        \label{fig:BP_Acc_Pre_ATAT_RF_V2_Real_Static_Incremental}
    \end{subfigure}
    ~
    \begin{subfigure}[t]{0.235\textwidth}  
        \centering
        \includegraphics[width=0.99\textwidth]{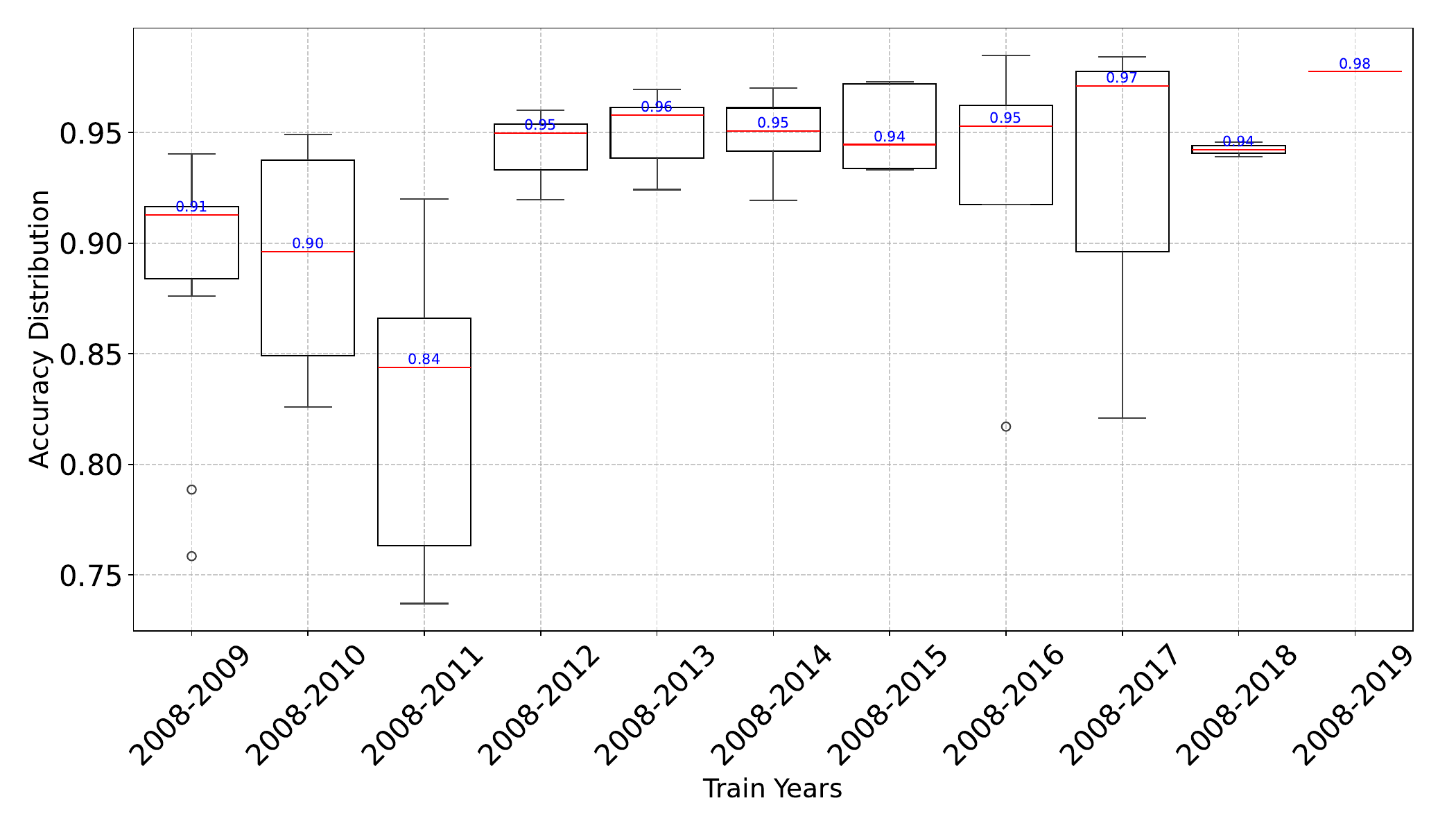}\vspace{-2mm}
        \caption{\normalfont Post-Real-Acc.}
        \label{fig:BP_Acc_Post_ATAT_RF_V2_Real_Static_Incremental}
    \end{subfigure}
    ~
    \begin{subfigure}[t]{0.235\textwidth}
        \centering
        \includegraphics[width=0.99\textwidth]{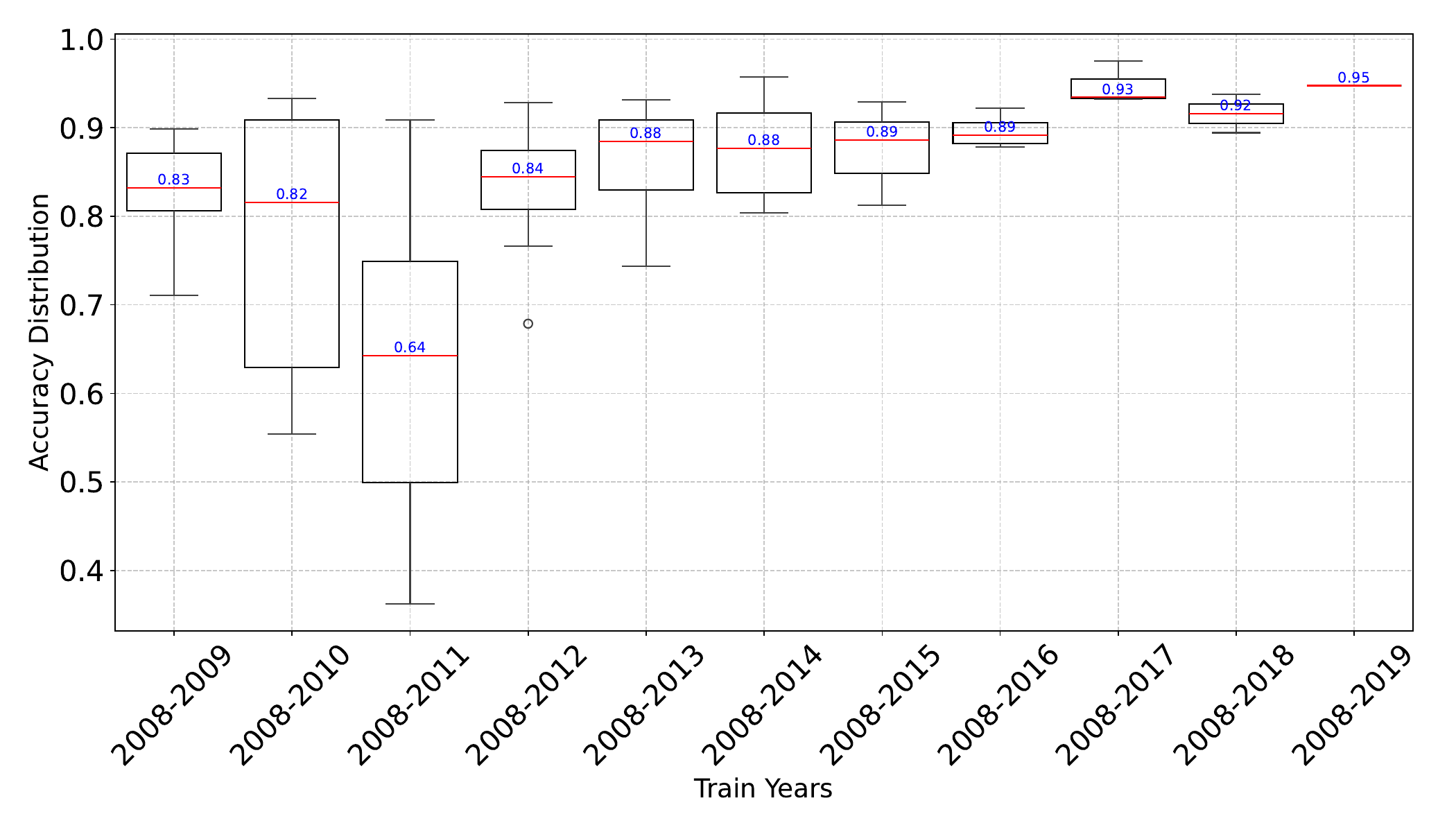}\vspace{-2mm}
        \caption{\normalfont Pre-Emu-Acc.}
        \label{fig:BP_Acc_Pre_ATAT_RF_V2_Emu_Static_Incremental}
    \end{subfigure}
    ~
    \begin{subfigure}[t]{0.2352\textwidth}
        \centering
        \includegraphics[width=0.99\textwidth]{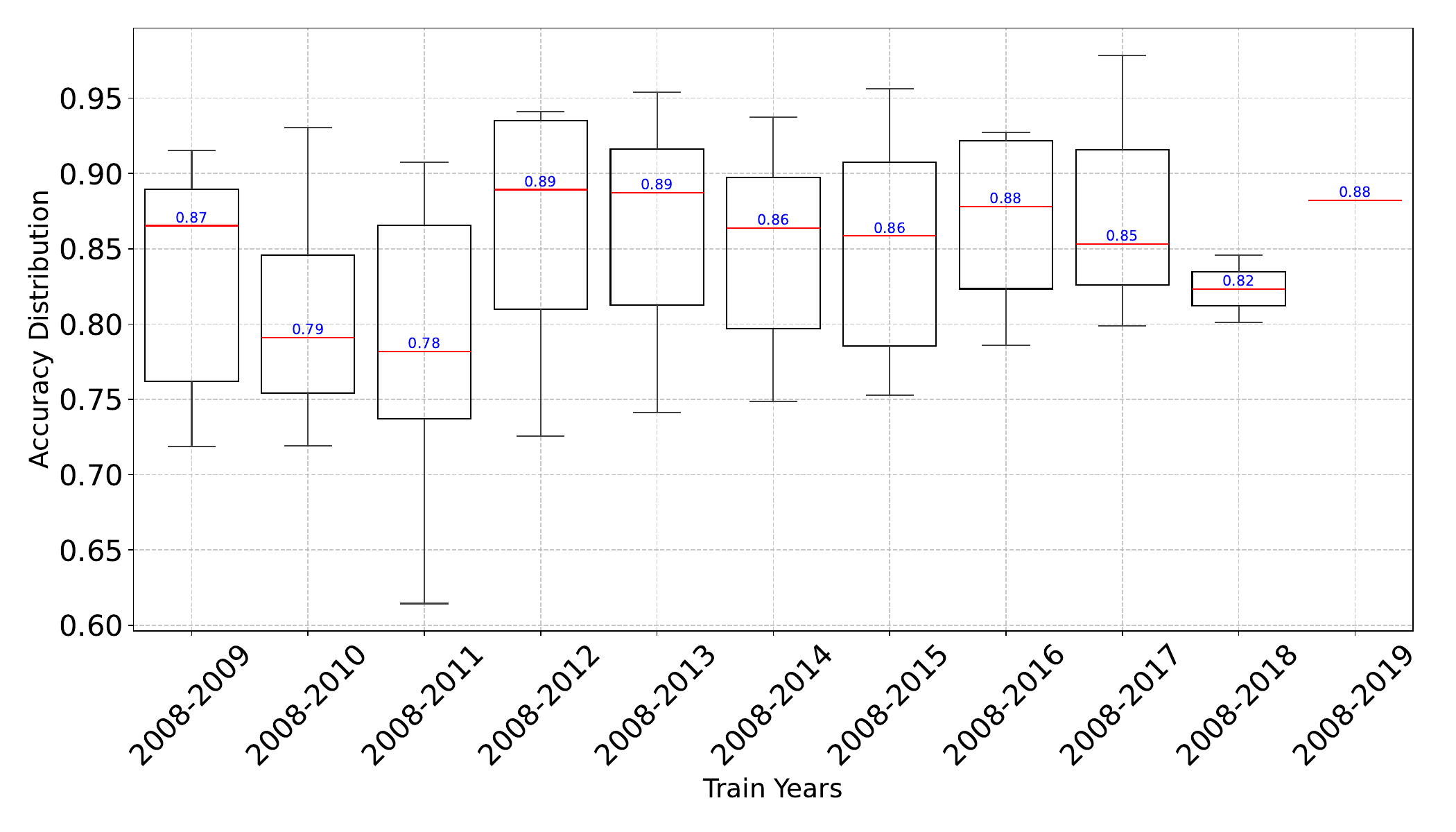}\vspace{-2mm}
        \caption{\normalfont Post-Emu-Acc.}
        \label{fig:BP_Acc_Post_ATAT_RF_V2_Emu_Static_Incremental}
    \end{subfigure}

\vspace{-2mm}
    \caption{RF accuracy pre- and post-balancing on real and emulator data with static features. (F1 in Appendix; \autoref{fig:BoxPlot_RF_Static_Incremental_Emu_Real_F1})}\vspace{-3mm}
    \label{fig:BoxPlot_RF_Static_Incremental_Emu_Real}
\end{figure*}

\EMP{Static Features}  
Besides the model drift over time, additional insights related to the real device and emulator data are shown. These findings suggest that feature selection significantly impacts both accuracy and F1 score. While some differences exist, emulator data can reasonably approximate real-world performance. The results emphasize the importance of feature engineering, addressing the class imbalance, and accounting for emulator limitations when developing and deploying time-dependent detection models. The outcomes of this approach are presented in \autoref{fig:BoxPlot_RF_Static_Incremental_Emu_Real}.

\begin{figure*}[t]
    \centering
    \begin{subfigure}[t]{0.235\textwidth} 
        \centering
        \includegraphics[width=0.99\textwidth]{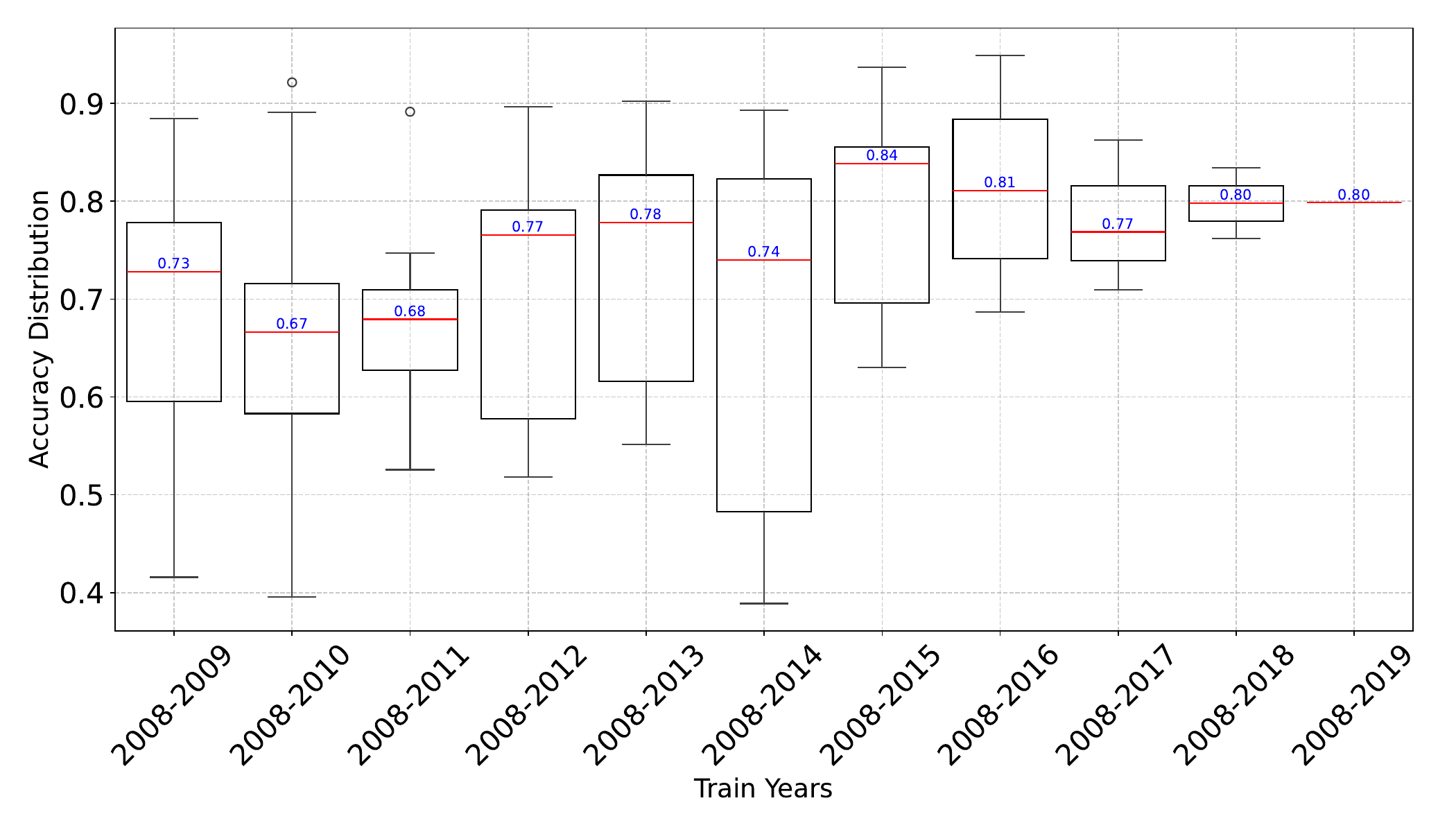}\vspace{-2mm}
        \caption{\normalfont Pre-Real-Acc.}
        \label{fig:BP_Acc_Pre_ATAT_GRU_V2_Real_Dynamic_Incremental}
    \end{subfigure}
    ~
    \begin{subfigure}[t]{0.235\textwidth}  
        \centering
        \includegraphics[width=0.99\textwidth]{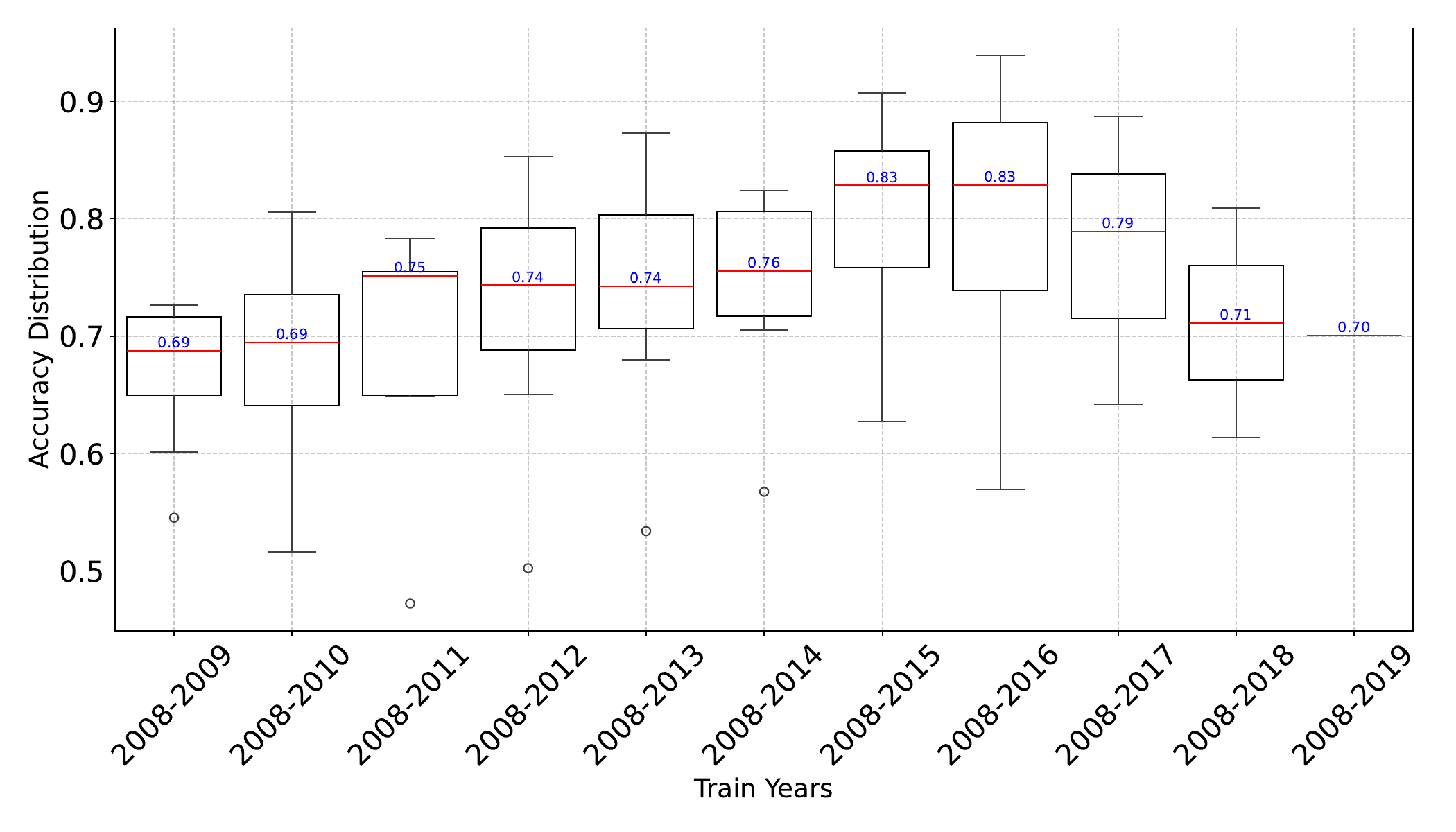}\vspace{-2mm}
        \caption{\normalfont Post-Real-Acc.}
        \label{fig:BP_Acc_Post_ATAT_GRU_V2_Real_Dynamic_Incremental}
    \end{subfigure}
    ~
    \begin{subfigure}[t]{0.235\textwidth}
        \centering
        \includegraphics[width=0.99\textwidth]{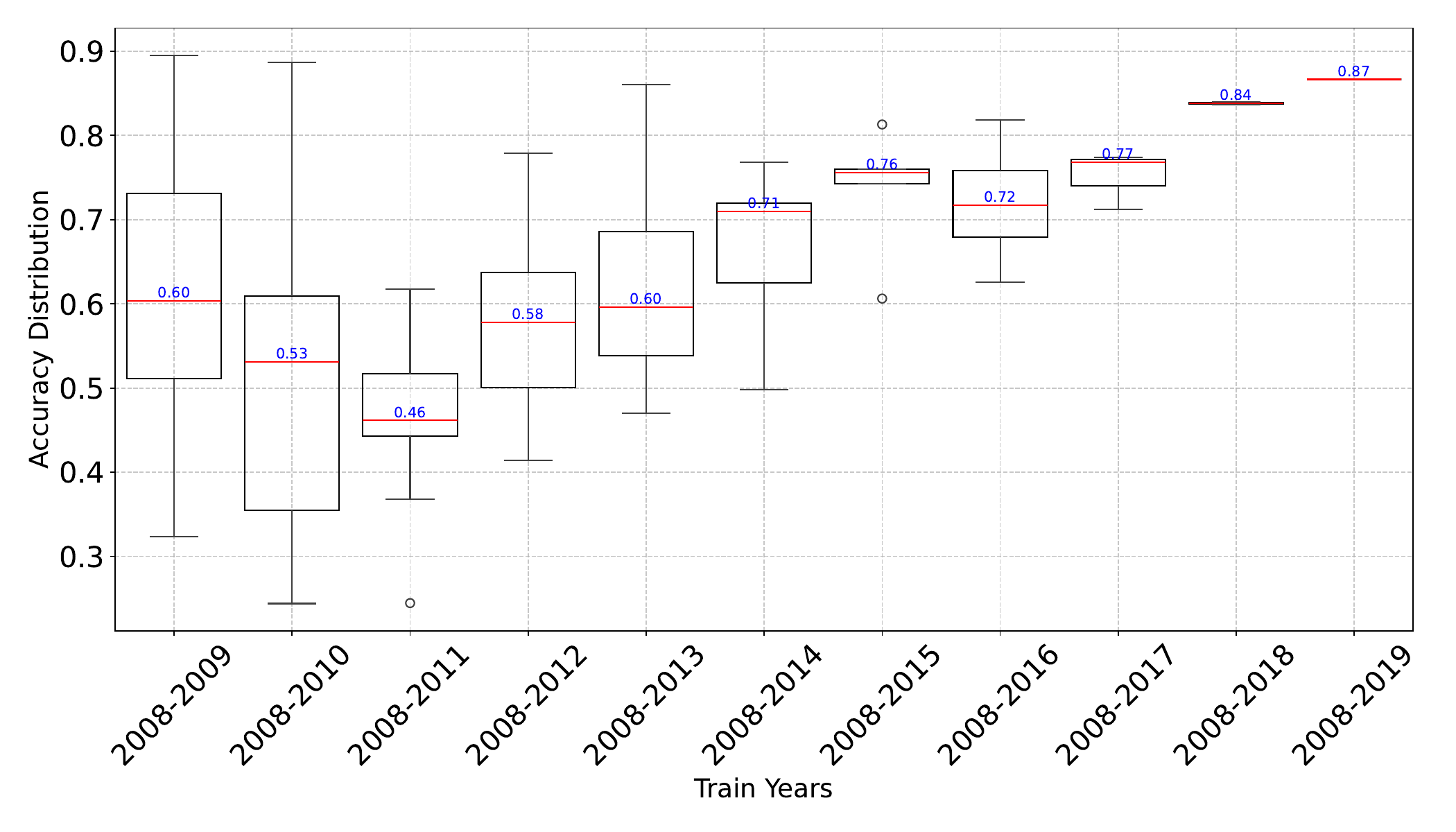}\vspace{-2mm}
        \caption{\normalfont Pre-Emu-Acc.}
        \label{fig:BP_Acc_Pre_ATAT_GRU_V2_Emu_Dynamic_Incremental}
    \end{subfigure}
    ~
    \begin{subfigure}[t]{0.235\textwidth}
        \centering
        \includegraphics[width=0.99\textwidth]{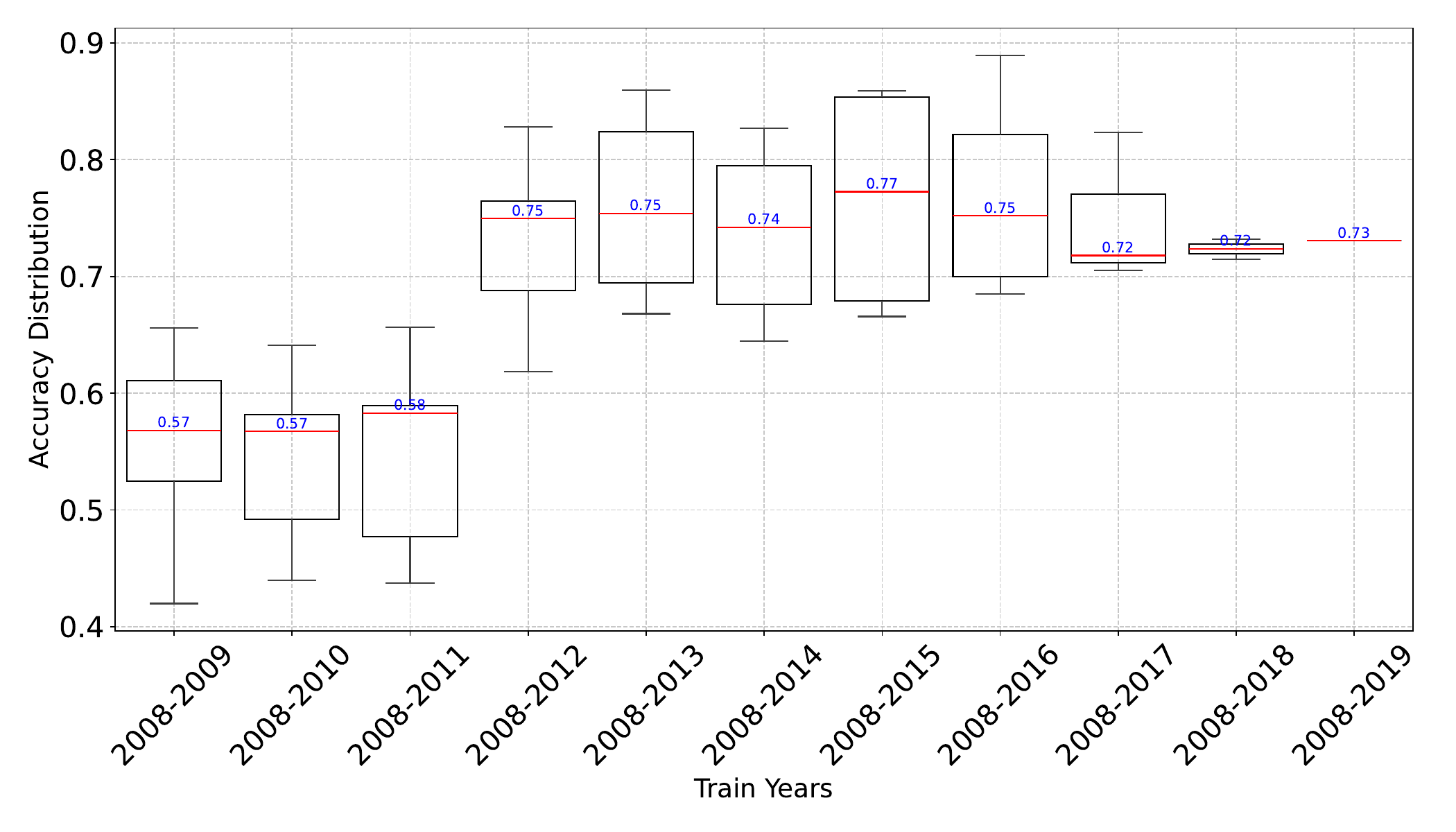}\vspace{-2mm}
        \caption{\normalfont Post-Emu-Acc.}
        \label{fig:/BP_Acc_Post_ATAT_GRU_V2_Emu_Dynamic_Incremental}
    \end{subfigure}

    \vspace{-2mm}

    \caption{\normalfont GRU's accuracy pre- and post-balancing on real/emulator data and dynamic features.  (F1 in Appendix; Fig.~\ref{fig:BoxPlot_RF_Incremental_Emu_Real_F1})}
    \label{fig:BoxPlot_RF_Incremental_Emu_Real}\vspace{-3mm}
\end{figure*}

\EMP{Dynamic Features}  
The results in \autoref{fig:BoxPlot_RF_Incremental_Emu_Real} indicate the presence of concept drift over time. The RF model, utilizing dynamic features, is analyzed before and after applying a balancing algorithm. In the real dataset, pre-balancing results show significant fluctuations in accuracy and F1 score, with inconsistencies as new data samples are introduced. Post-balancing, these variations decrease, suggesting that the balancing algorithm mitigates concept drift, resulting in more stable and reliable performance metrics.  Additionally, emulator data and pre-balancing results exhibit considerable variations in accuracy and F1 score, showing the challenge of maintaining consistent model performance as the data evolves, further confirming concept drift impact.

\begin{figure*}[t]
    \centering
    \begin{subfigure}[t]{0.235\textwidth} 
        \centering
        \includegraphics[width=0.99\textwidth]{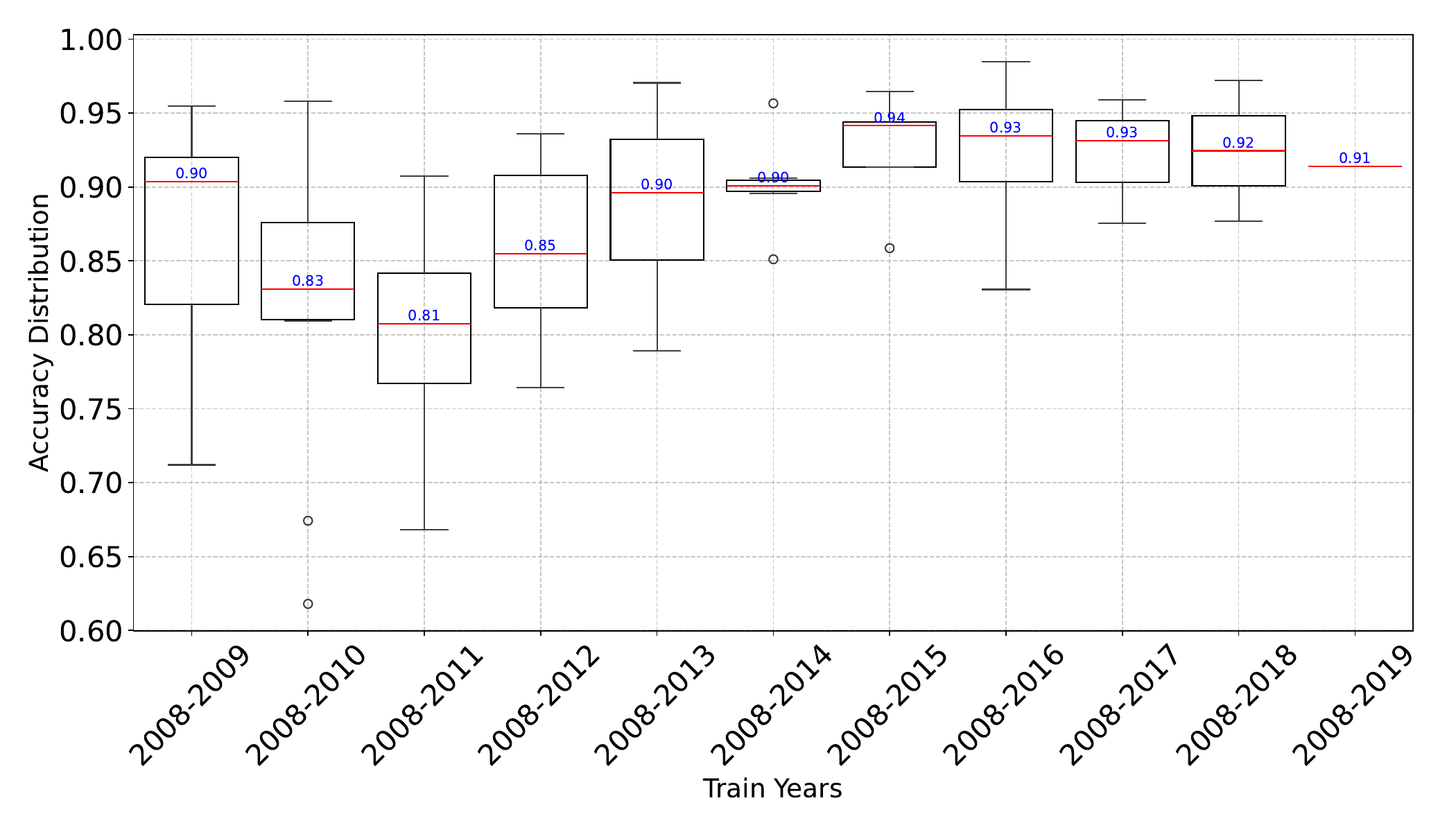}\vspace{-2mm}
        \caption{\normalfont Pre-Real-Acc.}
        \label{fig:BP_Acc_Pre_ATAT_RNN_V2_Real_Hybrid_Incremental}
    \end{subfigure}
    ~
    \begin{subfigure}[t]{0.235\textwidth}  
        \centering
        \includegraphics[width=0.99\textwidth]{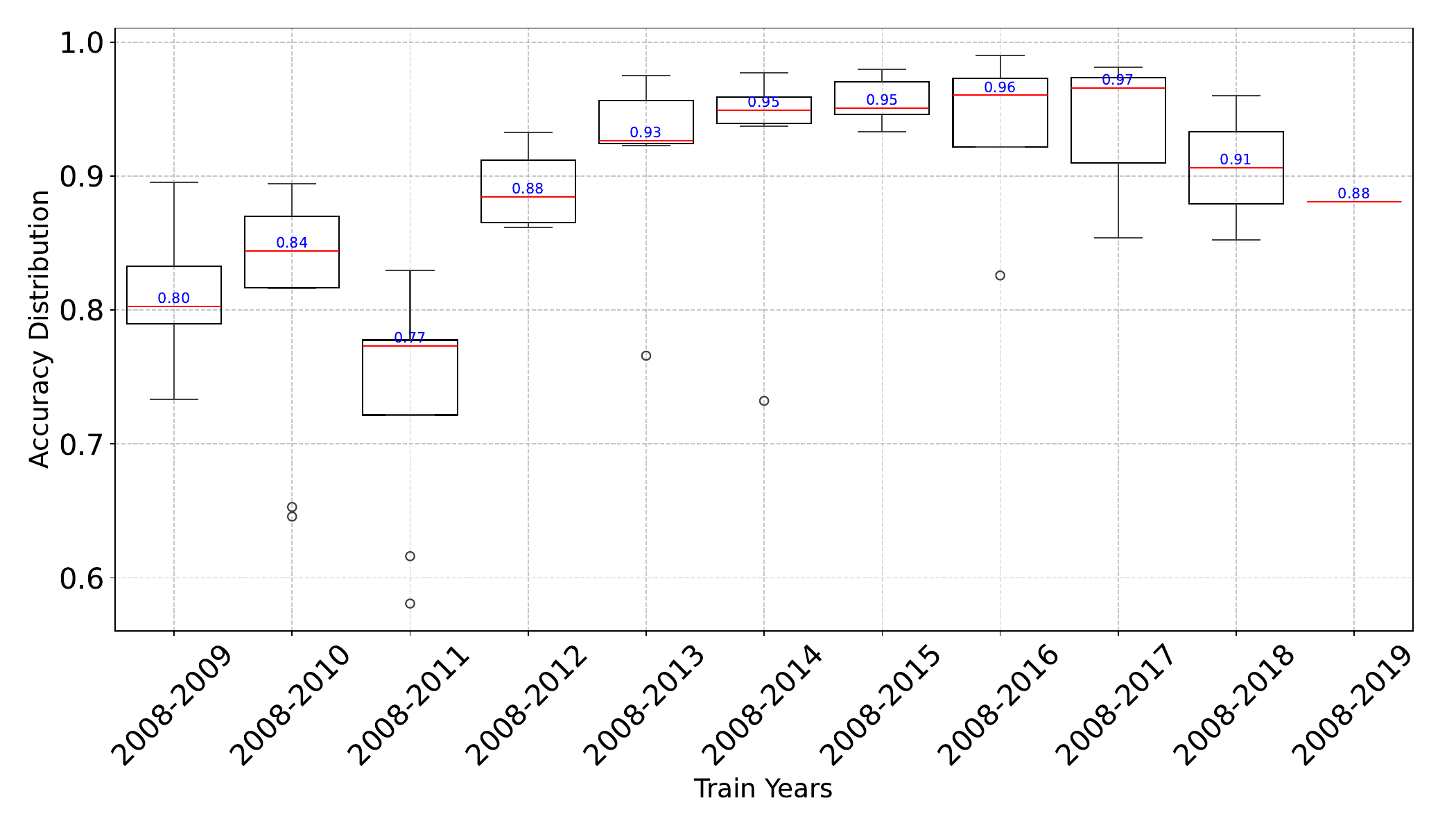}\vspace{-2mm}
        \caption{\normalfont Post-Real-Acc.}
        \label{fig:BP_Acc_Post_ATAT_RNN_V2_Real_Hybrid_Incremental}
    \end{subfigure}
    ~
    \begin{subfigure}[t]{0.235\textwidth}
        \centering
        \includegraphics[width=0.99\textwidth]{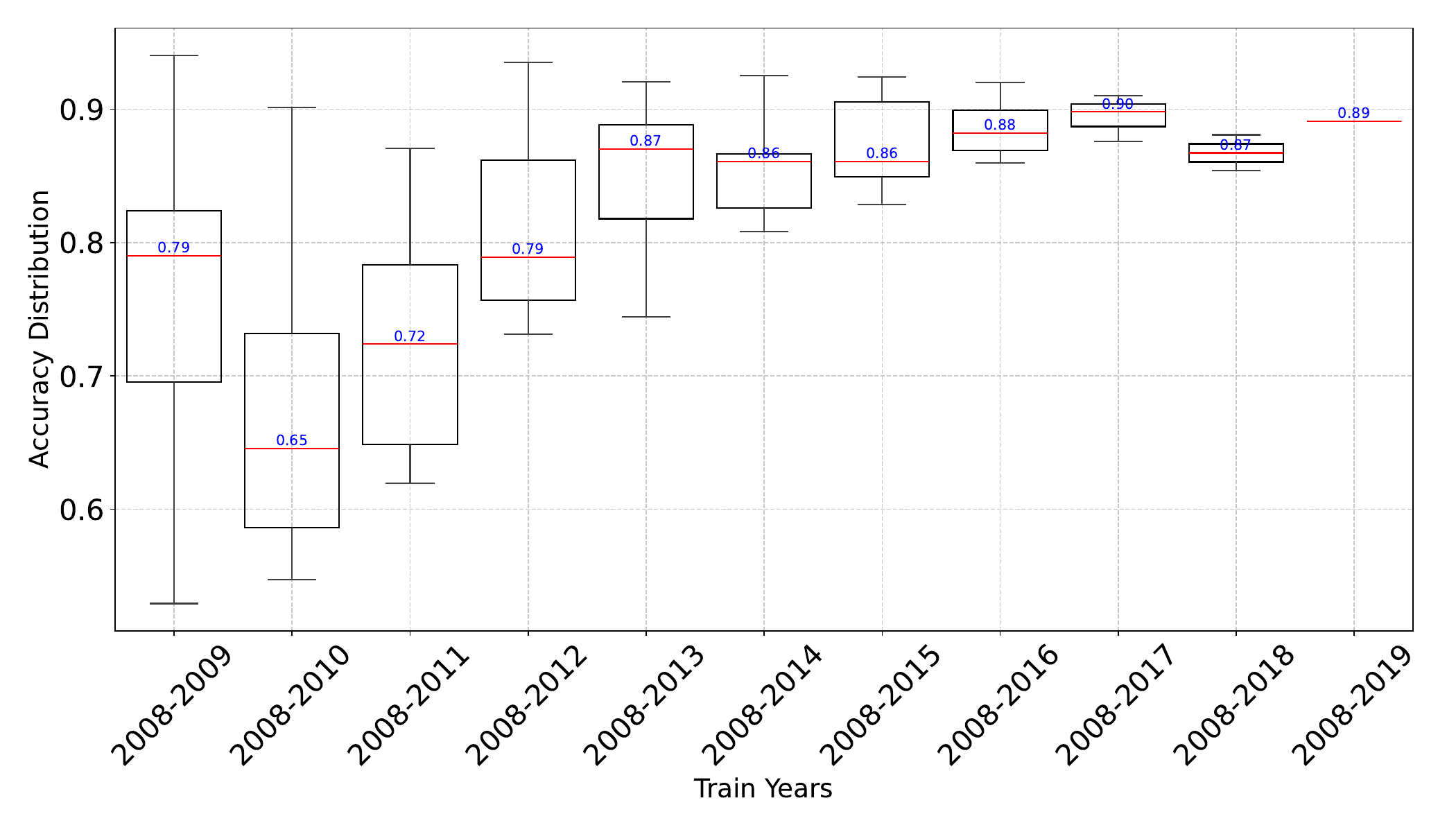}\vspace{-2mm}
        \caption{\normalfont Pre-Emu-Acc.}
        \label{fig:BP_Acc_Pre_ATAT_RNN_V2_Emu_Hybrid_Incremental}
    \end{subfigure}
    ~
    \begin{subfigure}[t]{0.235\textwidth}
        \centering
        \includegraphics[width=0.99\textwidth]{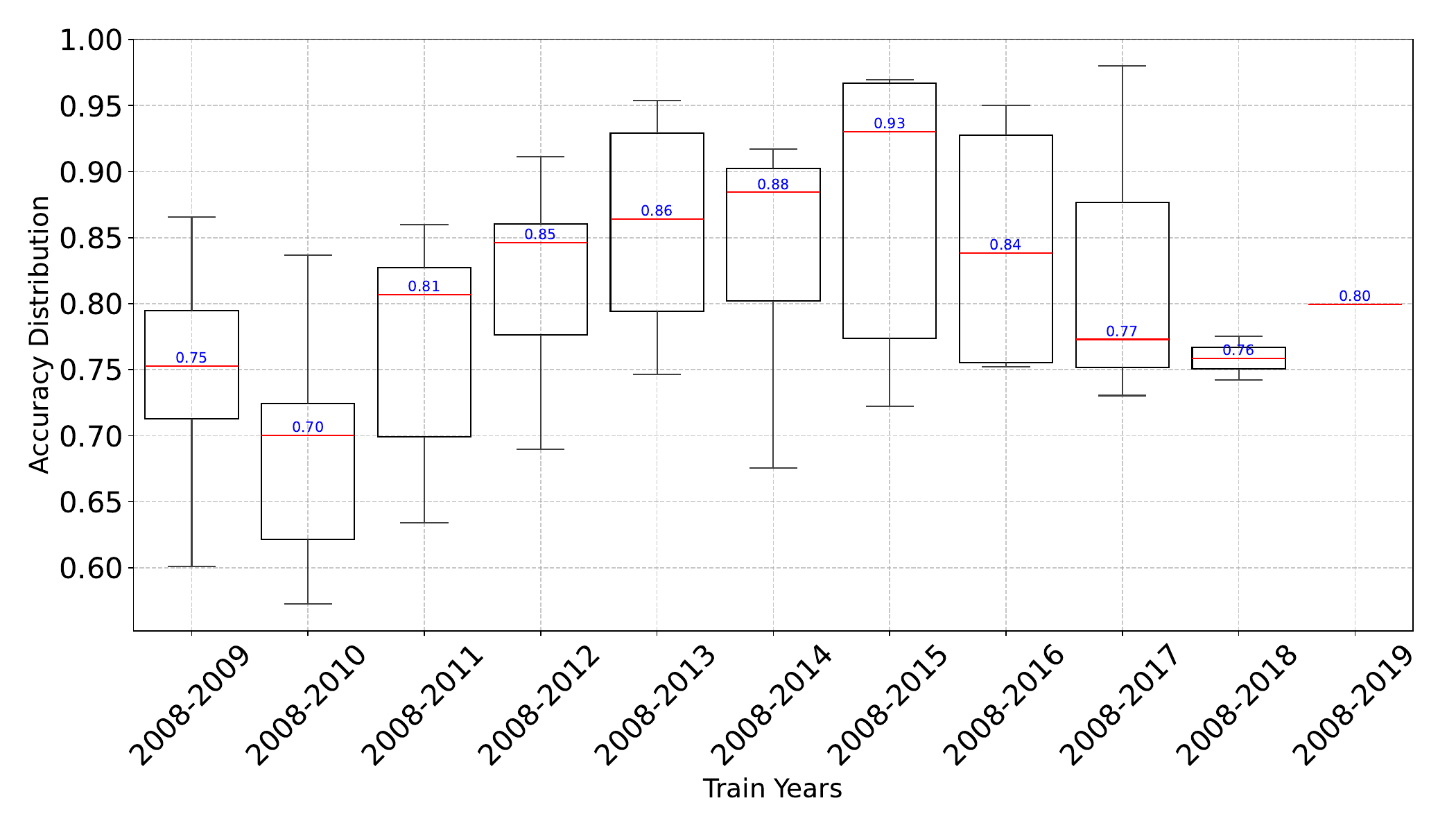}\vspace{-2mm}
        \caption{\normalfont Post-Emu-Acc.}
        \label{fig:BP_Acc_Post_ATAT_RNN_V2_Emu_Hybrid_Incremental}
    \end{subfigure}

    \vspace{-2mm}
    \caption{RNN accuracy with incremental training, real/emulator data, and hybrid features. (F1 in Appendix; Fig.~\ref{fig:BoxPlot_RNN_Incremental_Emu_Real_F1})}
    \label{fig:BoxPlot_RNN_Incremental_Emu_Real} \vspace{-4mm}
\end{figure*}

\EMP{Hybrid Features}  
The hybrid features outperform static and dynamic features, achieving an accuracy of 0.99 (per \autoref{tab:IgrnoreYearsRealDevice}). \autoref{fig:BoxPlot_RNN_Incremental_Emu_Real} offers insights into concept drift over time by illustrating how model performance evolves as training data from different years is incrementally added. Notably, it reveals fluctuations across training years, particularly in earlier periods, highlighting performance instability and reinforcing the presence of concept drift.

\begin{takeaway}
The incremental strategy for malware detection, reflecting real-world scenarios, clearly shows the presence of concept drift and its impact on performance.
\end{takeaway}

\BfPara{Troid Dataset}
The results, presented in \autoref{tab:TDS_GB_NLP_Pre_Incremental}–\autoref{tab:TDS_GB_RGB_Post_Incremental}, demonstrate the effectiveness of incremental strategies using the GB model across different approaches, including semantic and image-based methods (grayscale and RGB). Both pre- and post-balancing results reveal concept drift under various conditions.   Among the evaluated methods, image-based approaches exhibit the most stable performance post-balancing, with accuracy and F1 scores approaching 0.8 in 2023, outperforming the semantic approach. This suggests that RGB-based features may be more resistant to drift and imbalance when combined with ML models. Conversely, while the  approach also stabilizes after balancing, it experiences greater fluctuations and a notable accuracy drop from 0.96 to 0.55 (\autoref{tab:TDS_GB_NLP_Pre_Incremental}, \autoref{tab:TDS_GB_NLP_Post_Incremental}), indicating that textual features may be more susceptible to drift.

\begin{takeaway}
RGB-based features are more resistant to drift and imbalance, whereas the textual features appear to be more susceptible to drift.
\end{takeaway}

\begin{table*}[t]
\begin{minipage}{.33\textwidth} %
   \centering
    \caption{\normalfont Incremental strategy using GB and TF-IDF approach pre balancing.} \vspace{-2mm}
    \scalebox{0.75}{
    \begin{tabular}{l ccc ccc ccc}
\toprule
Test Year & \multicolumn{2}{c}{2021} & \multicolumn{2}{c}{2022} & \multicolumn{2}{c}{2023} \\
Train Years & A & F1 & A & F1 & A & F1 \\
\midrule
2019-2020 & 0.93 & 0.60 & 0.93 & 0.49 & 0.96 & 0.49 \\
2019-2021 & - & - & 0.93 & 0.49 & 0.96 & 0.54 \\
2019-2022 & - & - & - & - & 0.96 & 0.49 \\
\bottomrule
\end{tabular}}
\label{tab:TDS_GB_NLP_Pre_Incremental}
\end{minipage}\hspace{2mm}~
\begin{minipage}{.33\textwidth} %
    \centering
    \caption{\normalfont Incremental strategy using GB  and TF-IDF approach post balancing.} \vspace{-2mm}
    \scalebox{0.75}{
\begin{tabular}{l ccc ccc ccc}
\toprule
Test Year & \multicolumn{2}{c}{2021} & \multicolumn{2}{c}{2022} & \multicolumn{2}{c}{2023} \\
Train Years & A & F1 & A & F1 & A & F1 \\
\midrule
2019-2020 & 0.63 & 0.62 & 0.52 & 0.46 & 0.49 & 0.44 \\
2019-2021 & - & - & 0.55 & 0.53 & 0.51 & 0.48 \\
2019-2022 & - & - & - & - & 0.55 & 0.51 \\
\bottomrule
\end{tabular}
    }
    \label{tab:TDS_GB_NLP_Post_Incremental}
\end{minipage} \hspace{2mm}~
\begin{minipage}{.33\textwidth} %
    \centering
    \caption{\normalfont Incremental strategy using GB and gray-scale image  pre balancing.} \vspace{-2mm}
    \scalebox{0.75}{
\begin{tabular}{l ccc ccc ccc}
\toprule
Test Year & \multicolumn{2}{c}{2021} & \multicolumn{2}{c}{2022} & \multicolumn{2}{c}{2023} \\
 Train Years& A & F1 & A & F1 & A & F1 \\
\midrule
2019-2020 & 0.88 & 0.58 & 0.89 & 0.57 & 0.91 & 0.66 \\
2019-2021 & - & - & 0.90 & 0.66 & 0.93 & 0.75 \\
2019-2022 & - & - & - & - & 0.94 & 0.77 \\
\bottomrule
\end{tabular}
    }
\label{tab:TDS_GB_GS_Pre_Incremental}
\end{minipage} \hspace{2mm}
\end{table*}

\begin{table*}[t]
\begin{minipage}{.33\textwidth} %
   \centering
    \caption{\normalfont Incremental strategy using GB and gray-scale image post balancing.}\vspace{-2mm}
    \scalebox{0.70}{
    \begin{tabular}{l ccc ccc ccc}
\toprule
Test Year & \multicolumn{2}{c}{2021} & \multicolumn{2}{c}{2022} & \multicolumn{2}{c}{2023} \\
Train Years & A & F1 & A & F1 & A & F1 \\
\midrule
2019-2020 & 0.59 & 0.52 & 0.54 & 0.42 & 0.57 & 0.47 \\
2019-2021 & - & - & 0.66 & 0.62 & 0.73 & 0.71 \\
2019-2022 & - & - & - & - & 0.80 & 0.80 \\
\bottomrule
\end{tabular}}
\label{tab:TDS_GB_GS_Post_Incremental}
\end{minipage} \hspace{2mm}~
\begin{minipage}{.33\textwidth} %

    \centering
    \caption{\normalfont Incremental strategy using GB and RGB image pre balancing.} \vspace{-2mm}
    \scalebox{0.75}{
\begin{tabular}{l ccc ccc ccc}
\toprule
Test Year & \multicolumn{2}{c}{2021} & \multicolumn{2}{c}{2022} & \multicolumn{2}{c}{2023} \\
Train Years & A & F1 & A & F1 & A & F1 \\
\midrule
2019-2020 & 0.89 & 0.62 & 0.89 & 0.62 & 0.92 & 0.69 \\
2019-2021 & - & - & 0.91 & 0.66 & 0.94 & 0.75 \\
2019-2022 & - & - & - & - & 0.94 & 0.76 \\
\bottomrule
\end{tabular}
    }
    \label{tab:TDS_GB_RGB_Pre_Incremental}
\end{minipage}~\hspace{2mm}
\begin{minipage}{.33\textwidth} %
    \centering
    \caption{\normalfont Incremental strategy using GB and RGB image post balancing.} \vspace{-2mm}
    \scalebox{0.75}{
\begin{tabular}{l ccc ccc ccc}
\toprule
Test Year & \multicolumn{2}{c}{2021} & \multicolumn{2}{c}{2022} & \multicolumn{2}{c}{2023} \\
 Train Years& A & F1 & A & F1 & A & F1 \\
\midrule
2019-2020 & 0.62 & 0.56 & 0.62 & 0.56 & 0.67 & 0.62 \\
2019-2021 & - & - & 0.68 & 0.65 & 0.77 & 0.76 \\
2019-2022 & - & - & - & - & 0.81 & 0.80 \\
\bottomrule
\end{tabular}
    }
\label{tab:TDS_GB_RGB_Post_Incremental}
\end{minipage}\vspace{-5mm}
\end{table*}

\subsection{Grouping Years Strategy} 
To classify malware into their respective families, we selected only malware samples from the KronoDroid dataset. The dataset was then divided into three groups based on the distribution of malware applications over the years: Group 1 (2008–2012), Group 2 (2013–2016), and Group 3 (2017–2020). Classification models were trained on one group and tested on the others. The top ten malware families, selected from both real devices and emulators, accounted for 0.75 and 0.71 of the samples, respectively.  

\begin{figure*}[t]
    \centering
    \begin{subfigure}[t]{0.19\textwidth} 
        \centering
        \includegraphics[width=0.85\textwidth]{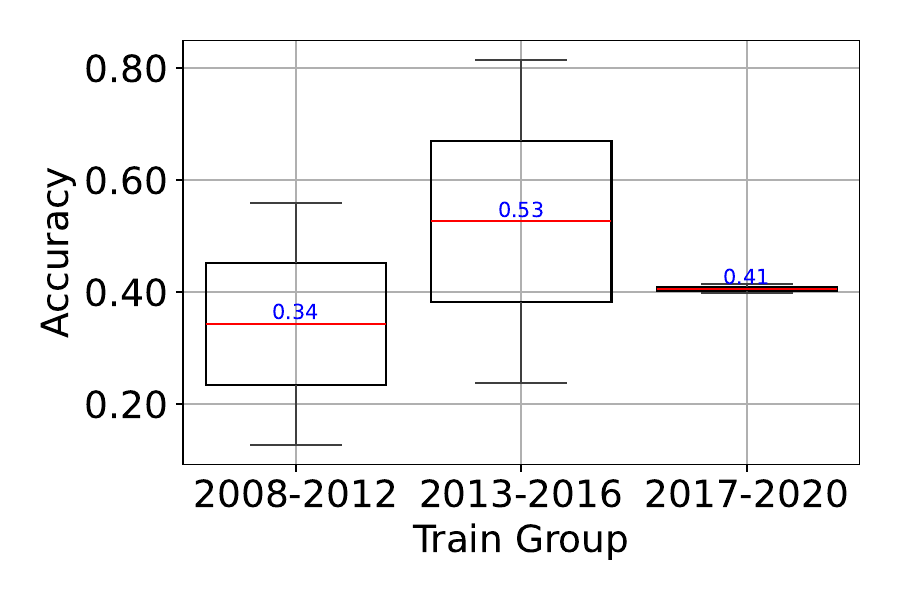}\vspace{-2mm}
        \caption{ Pre-R-S-Acc.}
    \label{fig:BP_Acc_Pre_Family_ATAT_RF_Real_Static}
    \end{subfigure} 
    ~\hspace{-2.2em}
    \begin{subfigure}[t]{0.19\textwidth}  
        \centering
        \includegraphics[width=0.85\textwidth]{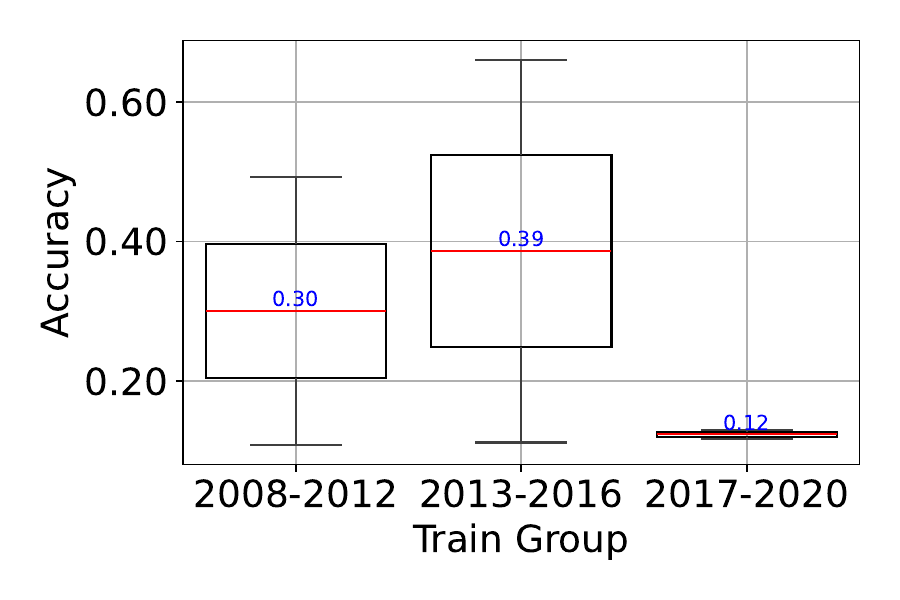}\vspace{-2mm}
        \caption{\normalfont Pre-R-D-Acc.}
        \label{fig:BP_Acc_Pre_Family_ATAT_RF_Real_Dynamic}
    \end{subfigure}
    ~\hspace{-2.2em}
    \begin{subfigure}[t]{0.19\textwidth}
        \centering
        \includegraphics[width=0.85\textwidth]{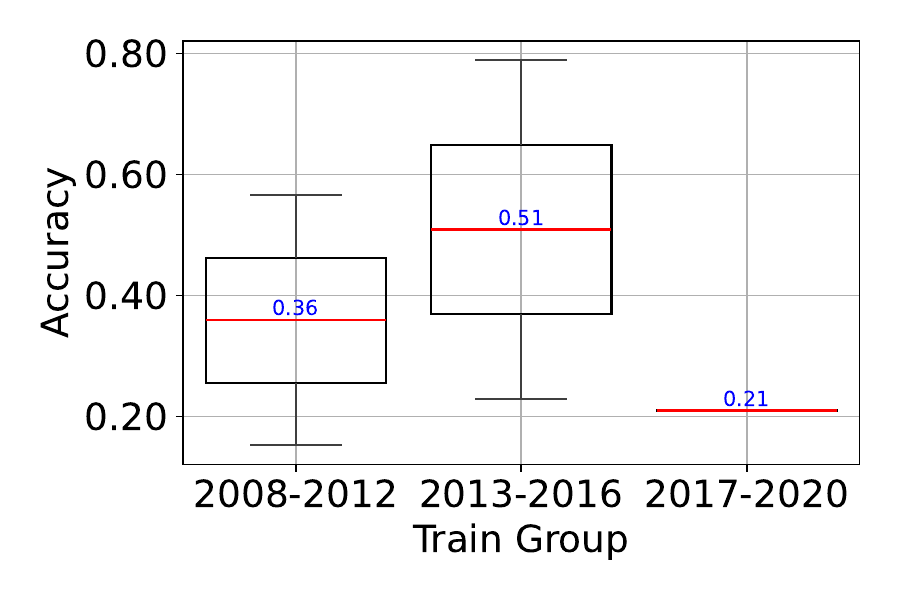}\vspace{-2mm}
        \caption{\normalfont Pre-R-H-Acc.}
        \label{fig:BP_Acc_Pre_Family_ATAT_RF_Real_Hybrid}
    \end{subfigure}
    ~\hspace{-2.2em}
    \begin{subfigure}[t]{0.19\textwidth}
        \centering
        \includegraphics[width=0.85\textwidth]{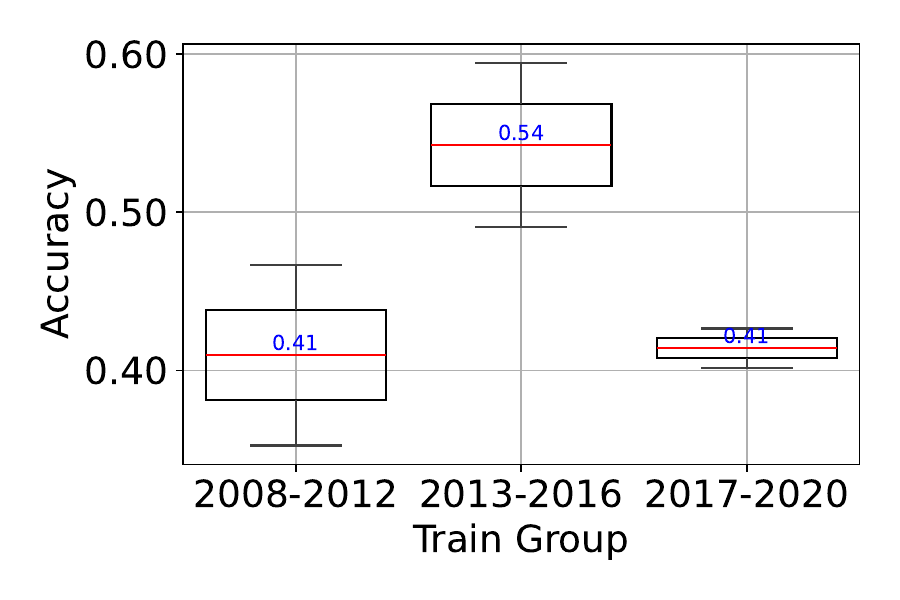}\vspace{-2mm}
        \caption{\normalfont Post-R-S-Acc.}
        \label{fig:BP_Acc_Post_Family_ATAT_RF_Real_Static}
    \end{subfigure} ~\hspace{-2em}
       \begin{subfigure}[t]{0.19\textwidth} 
        \centering
        \includegraphics[width=0.85\textwidth]{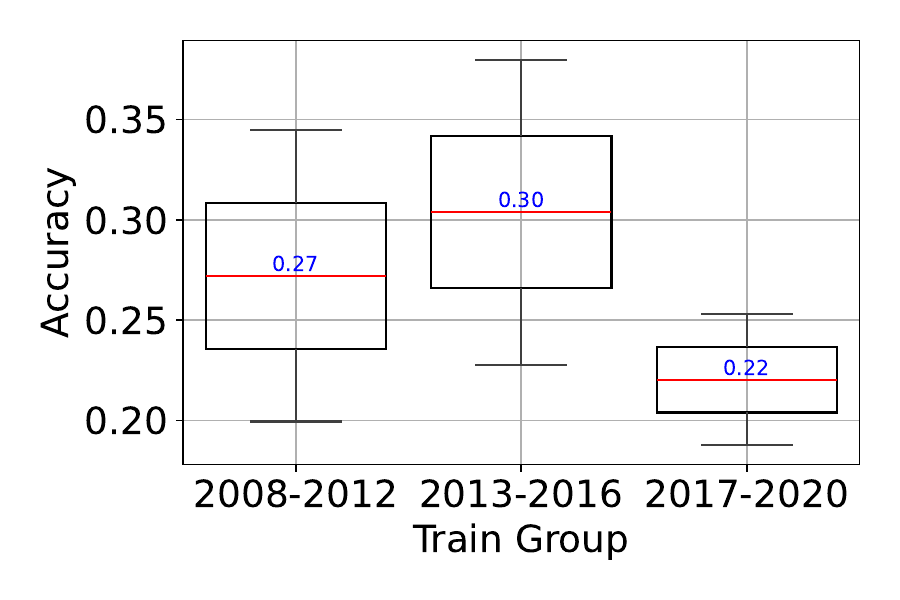}\vspace{-2mm}
        \caption{\normalfont Post-R-D-Acc.}
        \label{fig:BP_Acc_Post_Family_ATAT_RF_Real_Dynamic}
    \end{subfigure}
     ~\hspace{-2.2em}
    \begin{subfigure}[t]{0.19\textwidth}  
        \centering
        \includegraphics[width=0.85\textwidth]{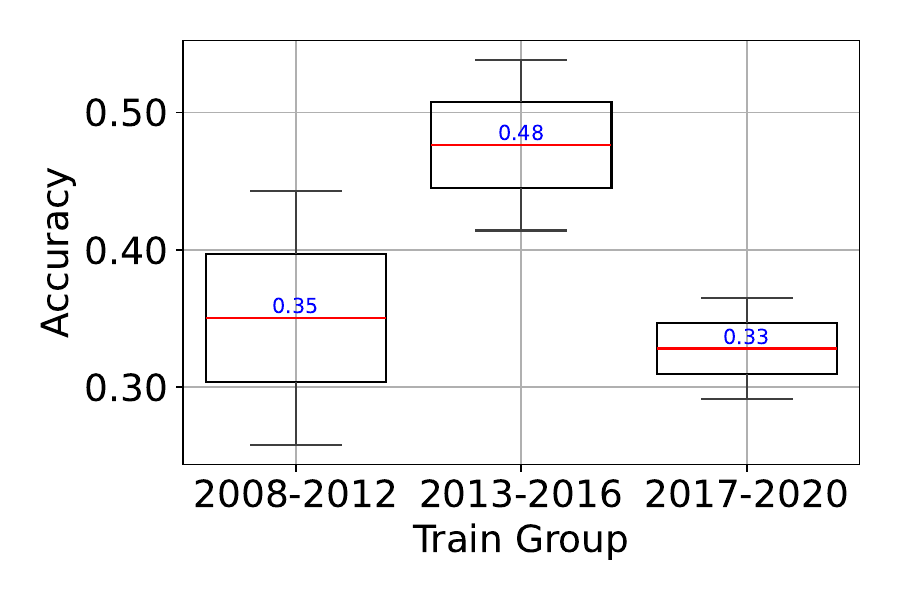}\vspace{-2mm}
        \caption{\normalfont Post-R-H-Acc.}
        \label{fig:BP_Acc_Post_Family_ATAT_RF_Real_Hybrid}
    \end{subfigure}
     ~\hspace{-2.2em}
     
     \begin{subfigure}[t]{0.19\textwidth} 
        \centering
        \includegraphics[width=0.85\textwidth]{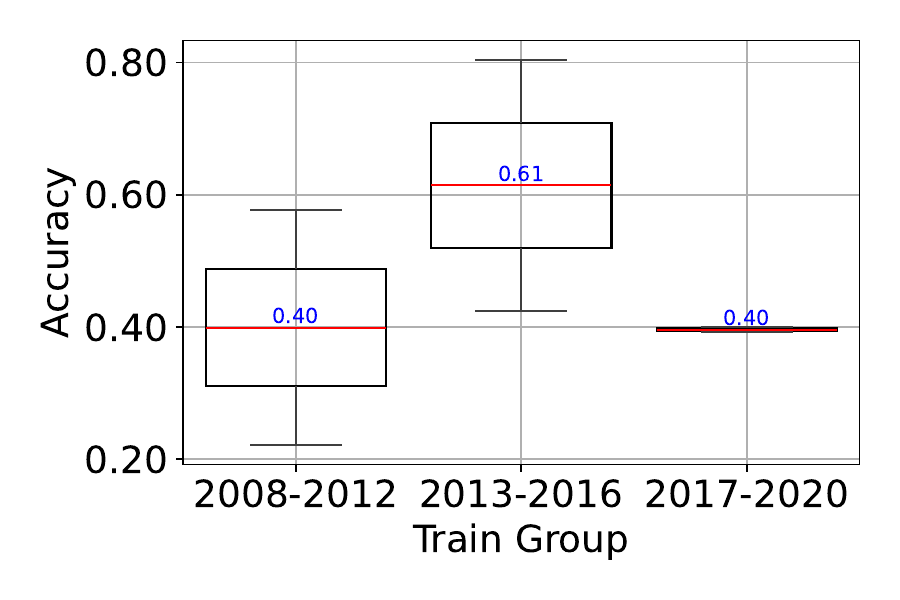}\vspace{-2mm}
        \caption{\normalfont Pre-E-S-Acc.}
    \label{fig:BP_Acc_Pre_Family_ATAT_RF_Emu_Static}
    \end{subfigure} 
    ~\hspace{-2.2em}
    \begin{subfigure}[t]{0.19\textwidth}  
        \centering
        \includegraphics[width=0.85\textwidth]{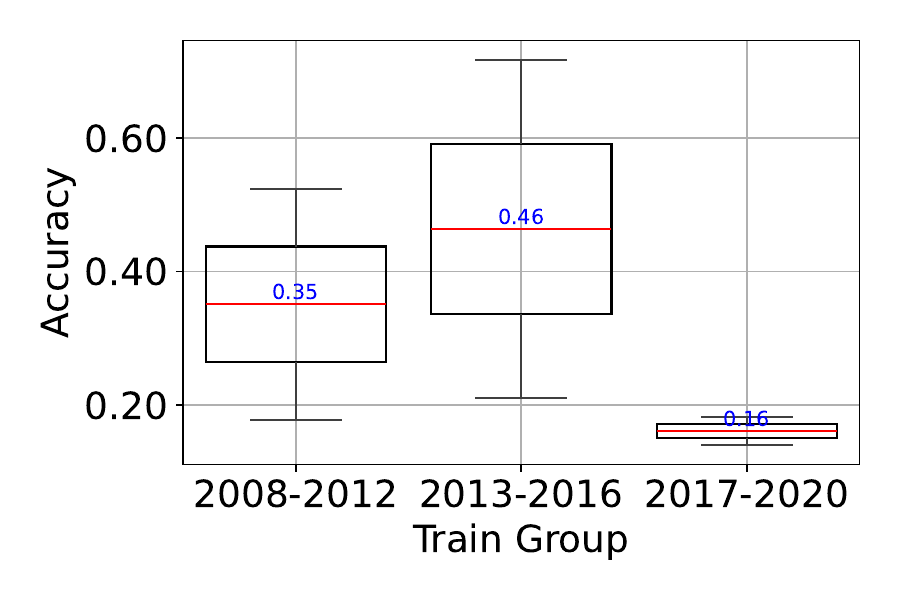}\vspace{-2mm}
        \caption{\normalfont Pre-E-D-Acc.}
        \label{fig:BP_Acc_Pre_Family_ATAT_RF_Emu_Dynamic}
    \end{subfigure}
    ~\hspace{-2.2em}
    \begin{subfigure}[t]{0.19\textwidth}
        \centering
        \includegraphics[width=0.85\textwidth]{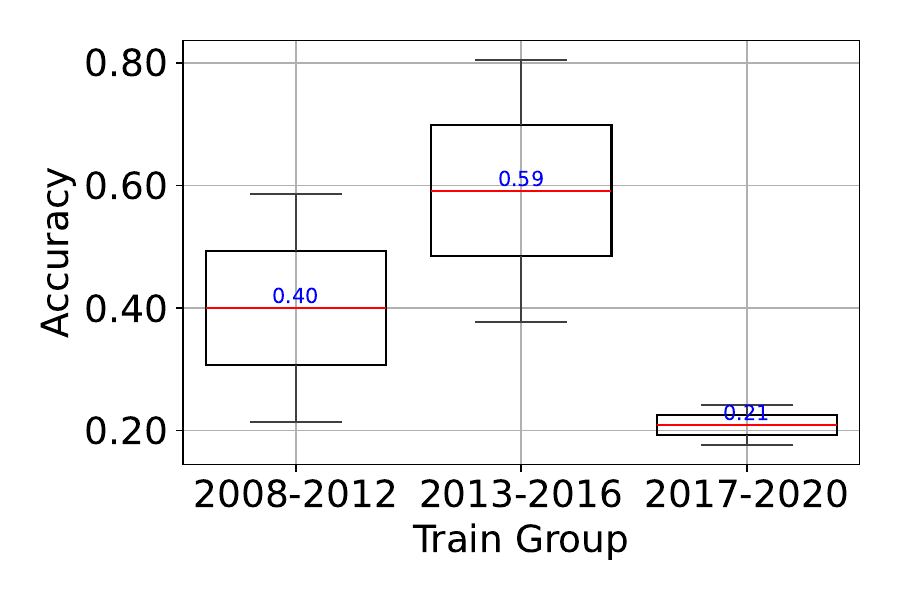}\vspace{-2mm}
        \caption{\normalfont Pre-E-H-Acc.}
        \label{fig:BP_Acc_Pre_Family_ATAT_RF_Emu_Hybrid}
    \end{subfigure}
    ~\hspace{-2.2em}
    \begin{subfigure}[t]{0.19\textwidth}
        \centering
        \includegraphics[width=0.85\textwidth]{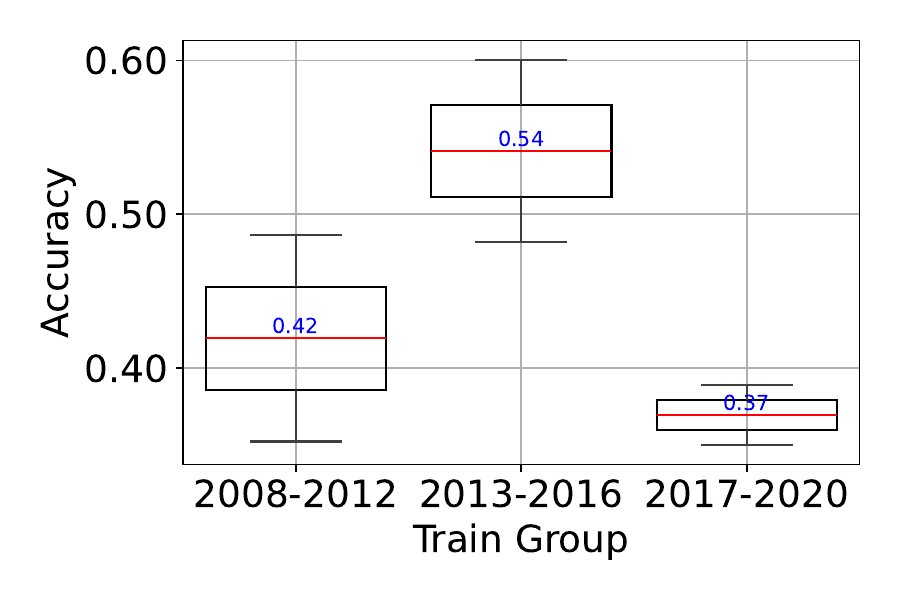}\vspace{-2mm}
        \caption{\normalfont Post-E-S-Acc.}
        \label{fig:BP_Acc_Post_Family_ATAT_RF_Emu_Static}
    \end{subfigure} ~\hspace{-2.2em}
       \begin{subfigure}[t]{0.19\textwidth} 
        \centering
        \includegraphics[width=0.85\textwidth]{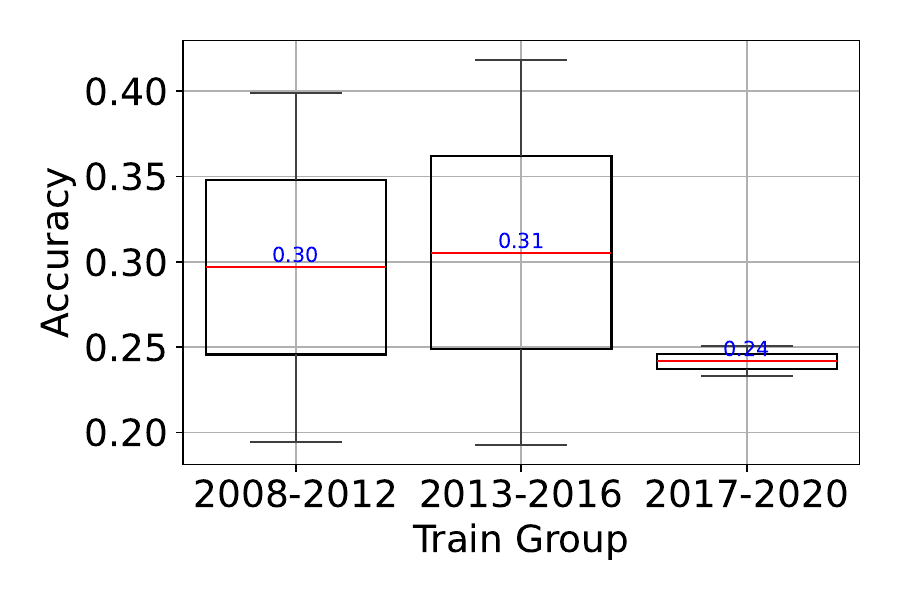}\vspace{-2mm}
        \caption{\normalfont Post-E-D-Acc.}
        \label{fig:BP_Acc_Post_Family_ATAT_RF_Emu_Dynamic}
    \end{subfigure}
     ~\hspace{-2.4em}
    \begin{subfigure}[t]{0.19\textwidth}  
        \centering
        \includegraphics[width=0.85\textwidth]{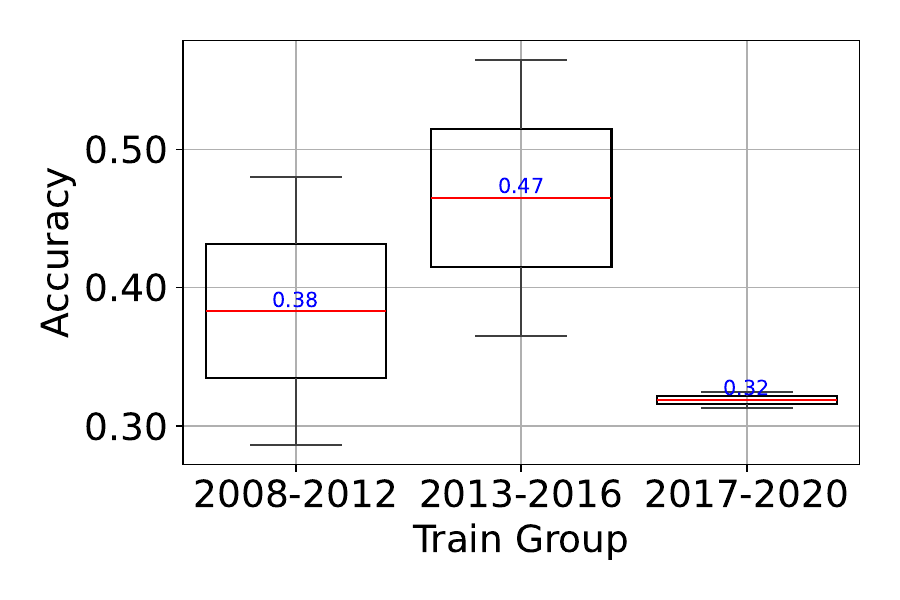}\vspace{-2mm}
        \caption{\normalfont Post-E-H-Acc.}
        \label{fig:BP_Acc_Post_Family_ATAT_RF_Emu_Hybrid}
    \end{subfigure}
     ~\hspace{-2.2em}
     
      \begin{subfigure}[t]{0.19\textwidth} 
        \centering
        \includegraphics[width=0.85\textwidth]{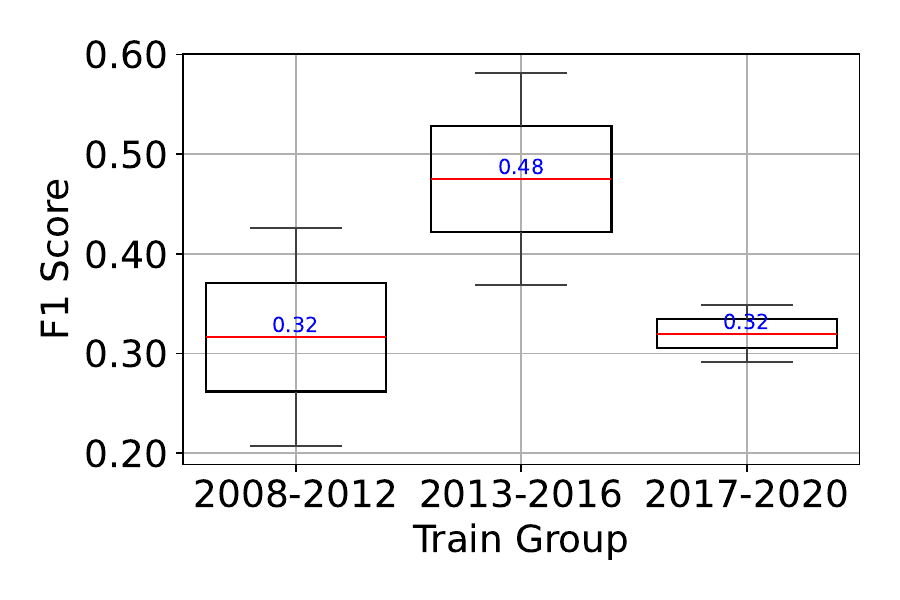}\vspace{-2mm}
        \caption{\normalfont Pre-R-S-F1.}
    \label{fig:BP_F1_Pre_Family_ATAT_RF_Real_Static}
    \end{subfigure} 
    ~\hspace{-2.2em}
    \begin{subfigure}[t]{0.19\textwidth}  
        \centering
        \includegraphics[width=0.85\textwidth]{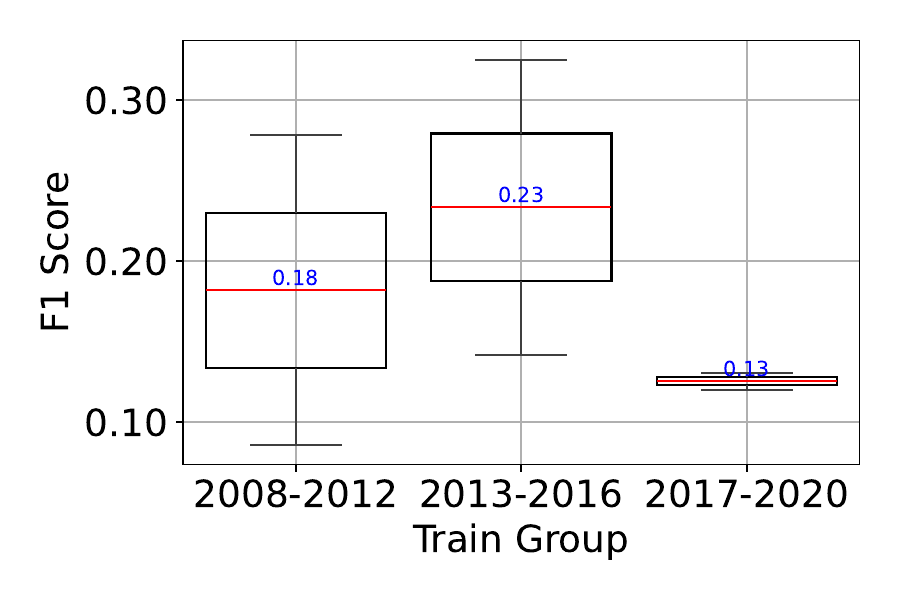}\vspace{-2mm}
        \caption{\normalfont Pre-R-D-F1.}
        \label{fig:BP_F1_Pre_Family_ATAT_RF_Real_Dynamic}
    \end{subfigure}
    ~\hspace{-2.2em}
    \begin{subfigure}[t]{0.19\textwidth}
        \centering
        \includegraphics[width=0.85\textwidth]{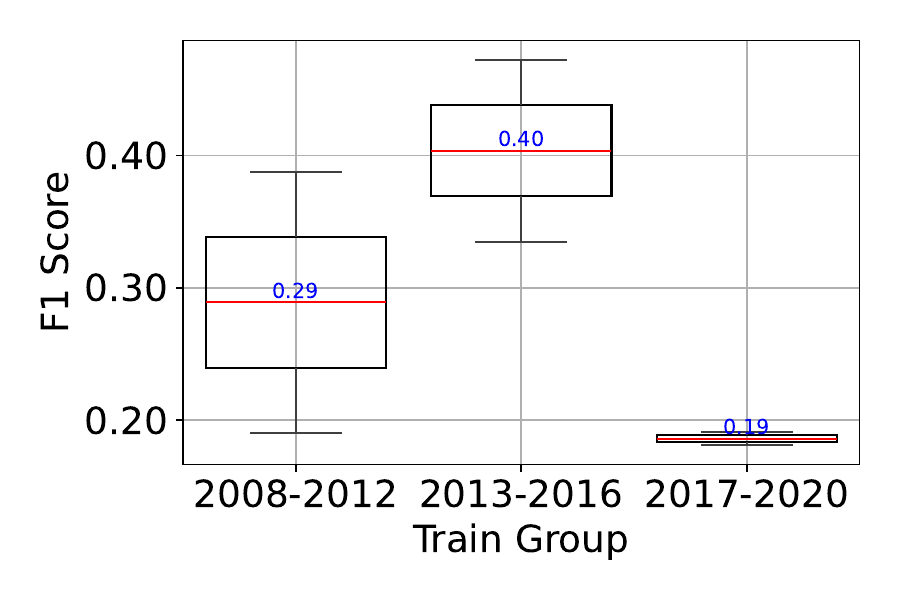}\vspace{-2mm}
        \caption{\normalfont Pre-R-H-F1.}
        \label{fig:BP_F1_Pre_Family_ATAT_RF_Real_Hybrid}
    \end{subfigure}
    ~\hspace{-2.2em}
    \begin{subfigure}[t]{0.19\textwidth}
        \centering
        \includegraphics[width=0.85\textwidth]{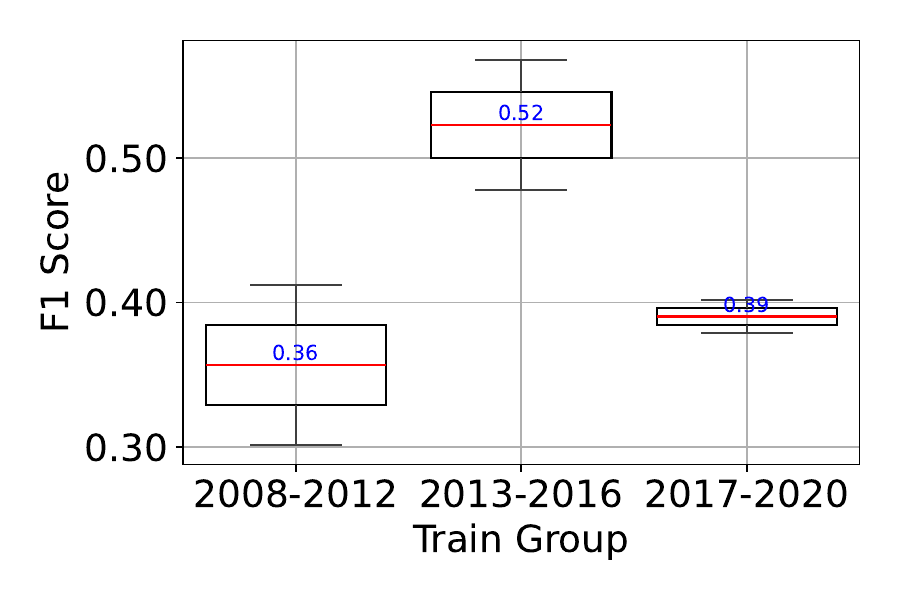}\vspace{-2mm}
        \caption{\normalfont Post-R-S-F1.}
        \label{fig:BP_F1_Post_Family_ATAT_RF_Real_Static}
    \end{subfigure} ~\hspace{-2.2em}
       \begin{subfigure}[t]{0.19\textwidth} 
        \centering
        \includegraphics[width=0.85\textwidth]{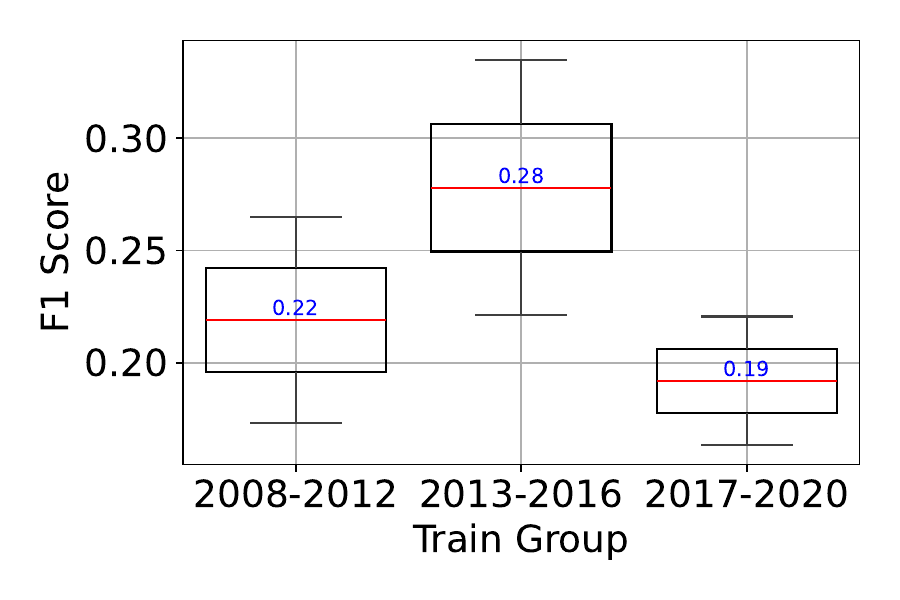}\vspace{-2mm}
        \caption{\normalfont Post-R-D-F1.}
        \label{fig:BP_F1_Post_Family_ATAT_RF_Real_Dynamic}
    \end{subfigure}
     ~\hspace{-2.2em}
    \begin{subfigure}[t]{0.19\textwidth}  
        \centering
        \includegraphics[width=0.85\textwidth]{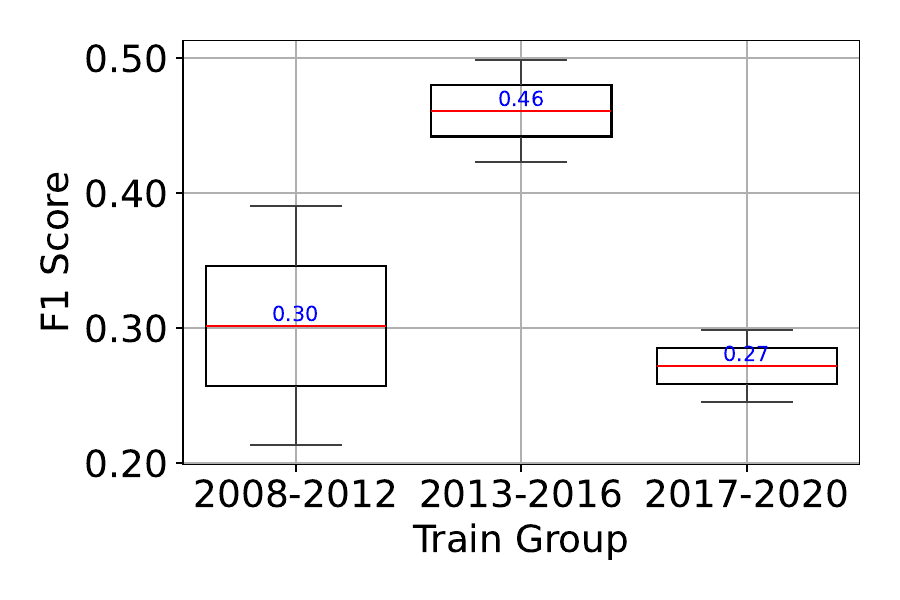}\vspace{-2mm}
        \caption{\normalfont Post-R-H-F1.}
        \label{fig:BP_F1_Post_Family_ATAT_RF_Real_Hybrid}
    \end{subfigure}
     ~\hspace{-2.2em}
     
     \begin{subfigure}[t]{0.19\textwidth} 
        \centering
        \includegraphics[width=0.85\textwidth]{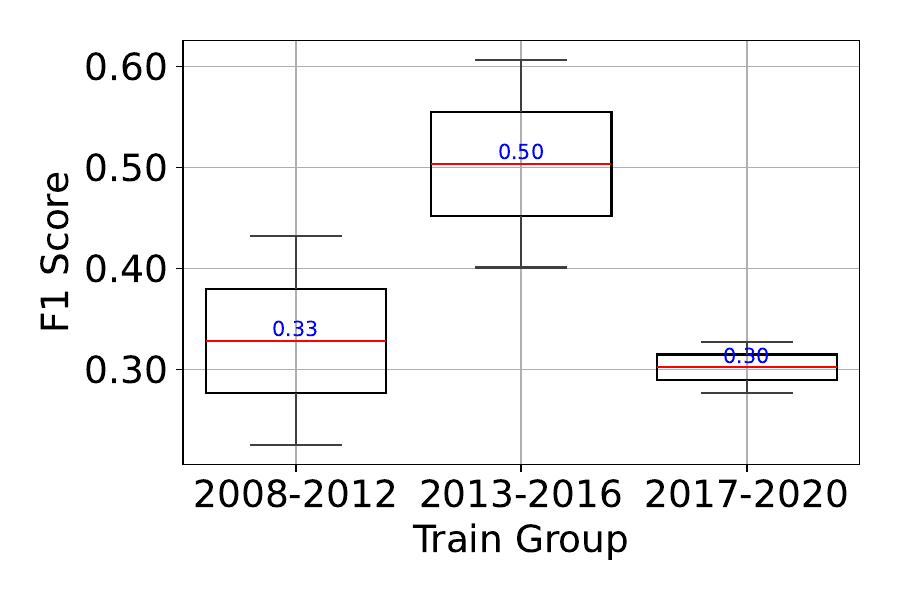}\vspace{-2mm}
        \caption{\normalfont Pre-E-S-F1.}
    \label{fig:BP_F1_Pre_Family_ATAT_RF_Emu_Static}
    \end{subfigure} 
    ~\hspace{-2.2em}
    \begin{subfigure}[t]{0.19\textwidth}  
        \centering
        \includegraphics[width=0.85\textwidth]{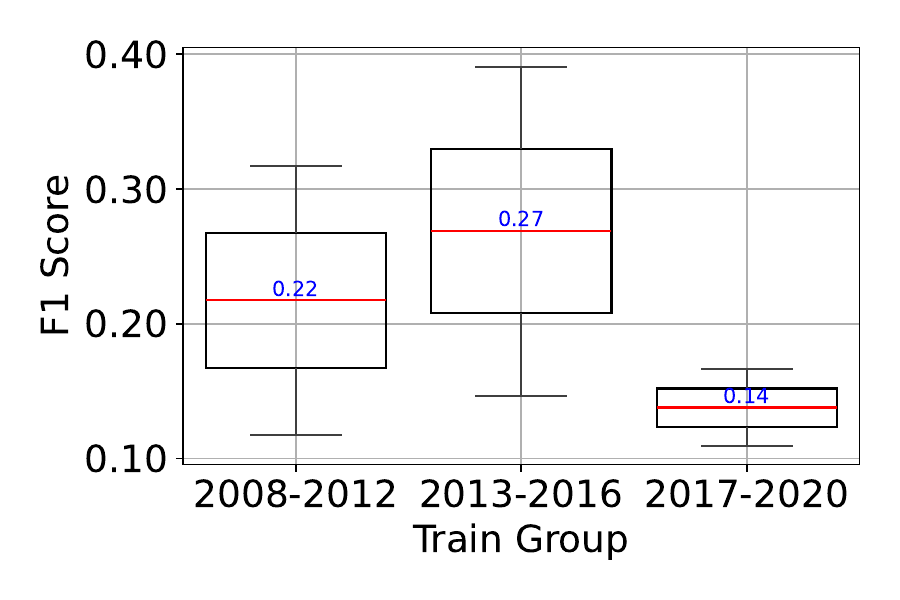}\vspace{-2mm}
        \caption{\normalfont Pre-E-D-F1.}
        \label{fig:BP_F1_Pre_Family_ATAT_RF_Emu_Dynamic}
    \end{subfigure}
    ~\hspace{-2.2em}
    \begin{subfigure}[t]{0.19\textwidth}
        \centering
        \includegraphics[width=0.85\textwidth]{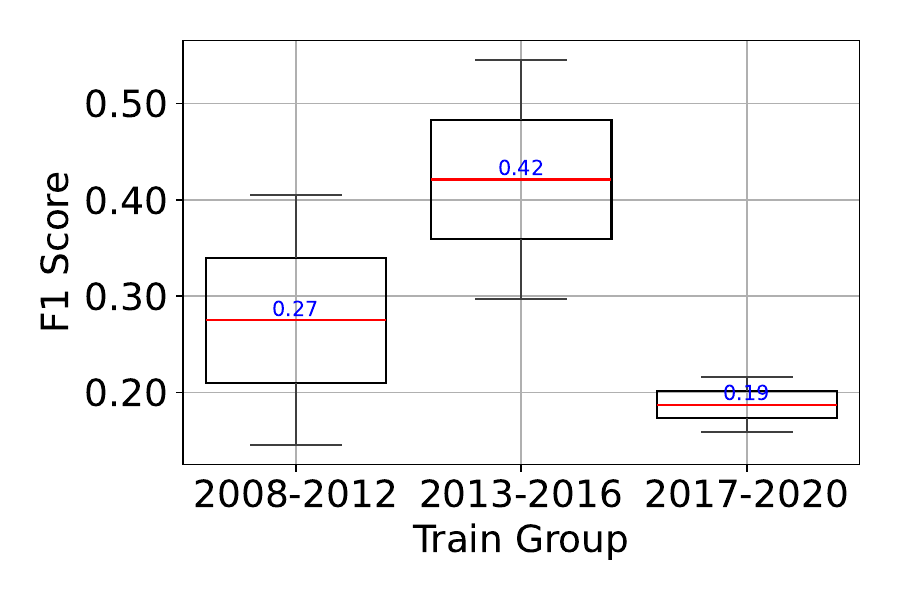}\vspace{-2mm}
        \caption{\normalfont Pre-E-H-F1.}
        \label{fig:BP_F1_Pre_Family_ATAT_RF_Emu_Hybrid}
    \end{subfigure}
    ~\hspace{-2.2em}
    \begin{subfigure}[t]{0.19\textwidth}
        \centering
        \includegraphics[width=0.85\textwidth]{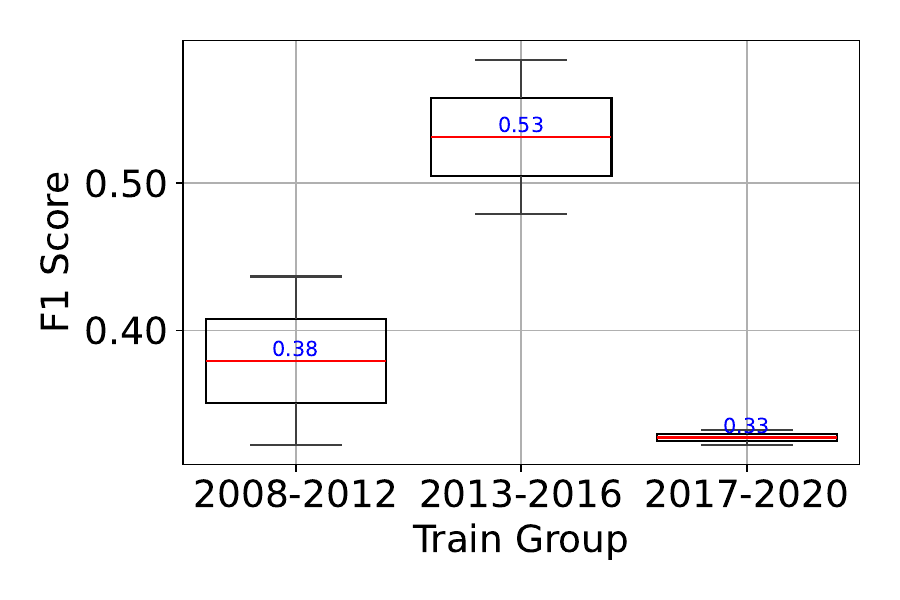}\vspace{-2mm}
        \caption{\normalfont Post-E-S-F1.}
        \label{fig:BP_F1_Post_Family_ATAT_RF_Emu_Static}
    \end{subfigure} ~\hspace{-2.2em}
       \begin{subfigure}[t]{0.19\textwidth} 
        \centering
        \includegraphics[width=0.85\textwidth]{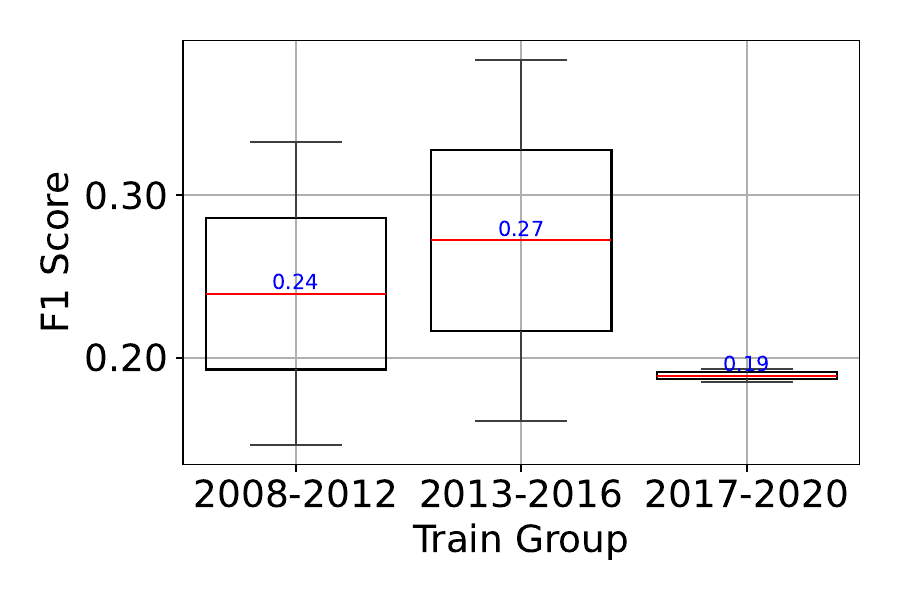}\vspace{-2mm}
        \caption{\normalfont Post-E-D-F1.}
        \label{fig:BP_F1_Post_Family_ATAT_RF_Emu_Dynamic}
    \end{subfigure}
     ~\hspace{-2.2em}
    \begin{subfigure}[t]{0.19\textwidth}  
        \centering
        \includegraphics[width=0.85\textwidth]{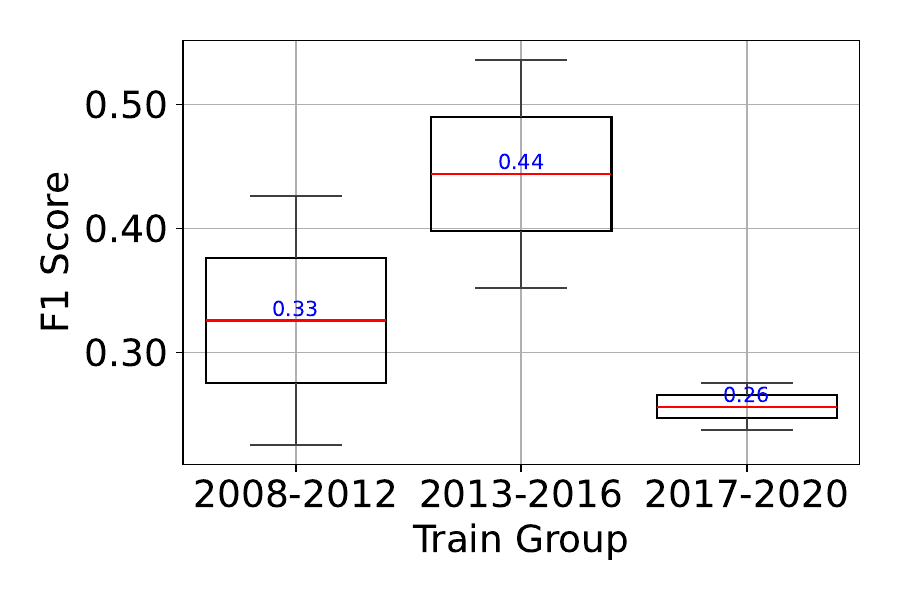}\vspace{-2mm}
        \caption{\normalfont Post-E-H-F1.}
        \label{fig:BP_F1_Post_Family_ATAT_RF_Emu_Hybrid}
    \end{subfigure}
     ~\hspace{-2.2em}
\vspace{-2mm}
    \caption{{Grouping strategy performance of RF pre- and post-balancing for real device (R) and emulator (E), and static (S), dynamic (D), hybrid (H) features.}}
    \label{fig:Box_Plot_RF_Real_Emu_Grouping}\vspace{-5mm}
\end{figure*}

The results indicate temporal drift across all models (RF, CNN, and RNN). Models trained on older datasets, e.g., Group 1 (2008–2012), performed poorly when tested on newer data (2017–2020), with noticeable declines in both accuracy and F1 score, as shown in \autoref{fig:Box_Plot_RF_Real_Emu_Grouping}. Even models trained on intermediate data (2013–2016) failed to maintain performance, showing the continuous evolution of malware.  

Static features generally outperformed dynamic features, but hybrid features yielded further improvements. For Group 1 (2008–2012), accuracy increased from 0.37 (static) to 0.47 (hybrid). Similarly, for Group 2 (2013–2016), accuracy improved from 0.46 to 0.54. However, for Group 3 (2017–2020), accuracy declined, indicating model drift.  
 
Regarding the class imbalance, we notice that balancing improved the F1 score, as illustrated in \autoref{fig:BP_F1_Post_Family_ATAT_RF_Real_Static}. However, the accuracy improvements were less pronounced and, in some cases, declined post-balancing, as shown in \autoref{fig:BP_Acc_Post_Family_ATAT_RF_Emu_Dynamic}. This is expected in imbalanced multi-class settings, where balancing improves minority class predictions, which may not be reflected in the overall accuracy. Since accuracy can be misleading in such settings, the F1 score is a more appropriate performance metric. A key insight emerges from comparing real device and emulator data. Surprisingly, models trained on emulator data sometimes outperformed those trained on real device data in adapting to new malware samples. This contrasts with previous binary classification results, where real device data was more resistant to concept drift than emulator data. This divergence underscores the importance of selecting an appropriate data source for a given task to mitigate drift.

\begin{takeaway}
The data source is crucial, and aligning it with the task objective enhances model robustness.
\end{takeaway}

Among the classification models, the differences in addressing temporal concept drift appear small but provide valuable insights into their specific strengths. RNN showed a slight edge in adapting to concept drift, particularly when working with dynamic and hybrid features. CNN also showed competitive performance, such as with hybrid features, highlighting its ability to model complex features effectively. Despite RF being a simpler algorithm, it outperformed them in many scenarios and remained a solid baseline, especially when using static features.

\begin{takeaway}
The small differences in performance across classification algorithms propose that factors like feature selection, data balancing, model adapting, and retraining to address concept drift may be more crucial to attaining strong performance than only selecting an algorithm
\end{takeaway}

\subsection{Discussion and Results Summary}
The empirical evidence driven by our results strongly indicates that concept drift is very prevalent in Android malware detection across various feature types (static, dynamic, hybrid) and different representations such as image, textual, or numeric data. Two datasets were evaluated using seven ML algorithms and two LLMs to address research questions. By this result, we answer {\bf RQ1} positively.

To answer {\bf RQ1.1}, we summarize our findings as follows: 

\BfPara{1) Feature Types} Static, dynamic, and hybrid feature types were analyzed. Concept drift was evident across all types under both cross-year and incremental strategies. Dynamic features were more susceptible to drift, as they capture malware behavior and runtime characteristics, which evolve over time. In contrast, static features remained more stable. As shown in \autoref{fig:BP_Acc_Post_ATAT_RF_V2_Real_Static_Incremental}, models trained on static data from 2008–2014 achieved 0.95 accuracy, whereas the RF model trained on dynamic features during the same period achieved only 0.76. For malware family classification, static features remained more stable but were less responsive to malware behavior changes. In contrast, dynamic and hybrid features captured evolving patterns more effectively but were more vulnerable to concept drift. On the other hand, when using LLMs, hybrid features proved to be the most robust--especially when combined with Exaone and emulator data.

\BfPara{2) Data Collection Environments} Data were collected from both real and emulated devices. Performance on real device data was generally comparable to emulator data when time was not considered. However, when accounting for temporal variations, real device data demonstrated greater resilience and exhibited less concept drift. In contrast, for malware family classification, emulator data performed slightly better than real device data, indicating its potential for improved adaptability in family classification. For LLMs trained with emulator data, they generally outperformed those using real device data.  

To answer {\bf RQ1.2} we found the following:

\BfPara{3) Classifiers} Seven ML and deep learning models were used to assess the impact of algorithms on concept drift. While some influence was observed, it was not substantial enough to be the primary driver of concept drift, as measured by performance metrics across both the KronoDroid and Troid datasets. For the malware family, the impact of drift varied based on feature type. RF performed best with static features, while RNN excelled with hybrid features. Both LLMs, LLaMA and Exaone, showed performance variation over time, indicating sensitivity to concept drift.

\BfPara{4) Detection Approach} Various Android malware detection strategies were examined for concept drift. Numeric-based approaches (e.g., binary values for requested permissions and intents) were classified as static data. The dynamic approach analyzed system call frequency using the Kronodroid dataset. For the semantic approach, TF-IDF techniques were applied to API calls, while the image-based method examined grayscale and RGB images.

All methods were tested with nine classifiers to assess the extent of concept drift. Results showed discrepancies across approaches, with static data being the most resistant, followed by dynamic data. RGB-based image methods exhibited similar resistance to grayscale methods in cross-year analysis but were more resilient under incremental strategies. The semantic-based method was the most susceptible to concept drift, although it outperformed image-based methods in general. However, a significant gap remained between its baseline performance and other strategies. For multi-class, we found that malware family classification is highly susceptible to concept drift due to the increasing complexity of distinguishing evolving families. Both LLMs showed promising results using the few-shot approach, though their performance was affected by concept drift, highlighting the need for more investigation.
 
To answer {\bf RQ1.3}, we found that the effect of balanced data generally improved the model's reliability. Where no significant differences between accuracy and F1 scores were observed with the Kronodroid dataset before balancing, the accuracy of the models improved after balancing the data. The Troid dataset, which originally had high accuracy and a low F1 score before balancing, was more consistent with the F1 score, indicating an overall improvement in model reliability. This suggests that the balancing algorithm effectively addresses class imbalance, improving the results. These findings emphasize the importance of balancing techniques, although they do not solve the issue of concept drift.

\section{Limitations and Future Work} \label{sec:limitations}
\BfPara{Malware Evolution and Concept Drift}
Concept drift can be categorized into several types based on how the data distribution changes over time~\cite{XiangZCW23}. \textbf{Incremental drift} occurs when the malware feature space evolves gradually---for example, as new Android APIs or permissions are introduced and slowly adopted by malware authors. In contrast, \textbf{gradual drift} is observed when old and new behaviors coexist temporarily, such as when malware developers experiment with both traditional and obfuscated payloads before fully shifting to the newer technique. \textbf{Abrupt drift} results from sudden changes, such as when Google deprecates key permissions (e.g., \texttt{WRITE\_SMS}, \texttt{CALL\_LOG} in Android 9), immediately affecting feature availability and rendering older model assumptions invalid. Finally, \textbf{recurring drift} arises when older malware behaviors re-emerge, such as the reappearance of SMS-based Trojans that had previously declined. A fully fledged line of analysis remains an open direction for future work to comprehensively evaluate how platform-level changes, particularly deprecated and restricted permissions, contribute to concept drift.

\BfPara{Recommendation} While this study focuses on understanding the factors that lead to concept drift, the results suggest that some approaches, such as LLMs, show promising potential for addressing it. Several studies have explored transfer learning~\cite{FuDG21,GarciaDC23}, online learning~\cite{ChenZKYCPPCW23}, active learning~\cite{AlamFMR24}, and unsupervised learning~\cite{TangSS14}, which are suitable directions for handling evolving data distributions. Additionally, it is worth investigating MLLMs and vision transformers for a deeper analysis of feature representations and drift-resilient detection. These approaches will be further explored in future work using our experimental strategies.

\BfPara{Threats to Validity}
Despite the promising results, several limitations must be addressed in future research.

\begin{description}
    \item[External Validity] One limitation of this study is the limited size of the Troid dataset, particularly in terms of malware samples. The dataset includes 4,146 benign apps for API call features and only 358 malware samples. For the hex dump features, there are 4,457 benign samples and 566 malware samples. This imbalance may limit the generalizability of our findings to real-world malware detection scenarios in the context of TF-IDF and image-based approaches. To mitigate this issue, we applied balancing techniques and relied on the F1 score to better reflect performance on unbalanced data. However, the relatively small malware sample may still impact the model's ability to generalize.
    \item [Internal Validity] Another potential limitation of this study is the lack of hyperparameter tuning for the learning models. Instead of optimizing hyperparameters for each algorithm, we used default configurations, which may introduce bias in the results. Some models may inherently perform better under default settings, while others might require fine-tuning to achieve better performance. This could impact the comparative evaluation of the models, leading to unintended advantages or disadvantages for specific algorithms. To mitigate this, we included a diverse set of nine models--three ML algorithms, four deep learning models, and two LLMs--to observe consistent trends across different approaches. However, future work could explore the impact of hyperparameter optimization on model stability and effectiveness in addressing concept drift.
\end{description}

\section{Conclusion}\label{sec:conclusion}
In this work, we examined the causes of concept drift in Android malware detection using KronoDroid and Troid across diverse detection approaches--static, dynamic, hybrid, textual (TF-IDF), and image-based. The results show that concept drift consistently degrades performance across all models, independent of algorithm or feature type. Models trained on older data perform poorly on newer samples, primarily due to distribution shifts rather than algorithmic flaws. While data balancing offers partial relief, it is often inadequate or even harmful. LLMs with few-shot learning show promise but fall short of fully addressing drift. Additionally, multi-class classification, which requires distinguishing malware families, exacerbates drift effects compared to binary classification, increasing model sensitivity to feature shifts.


\vspace{-10mm}
\begin{IEEEbiography}[{\includegraphics[width=1in,height=1.25in,clip,keepaspectratio]{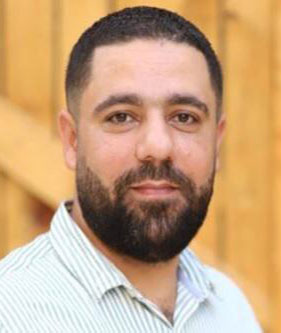}}]{Ahmed Sabbah }
received a Bachelor's degree in computer science from An-Najah National University, Palestine, in 2008 and a Master's degree in software engineering from Birzeit University in 2021. He is currently working toward a Ph.D. degree with the Department of Computer Science, Birzeit University. His research interests include security, ML, software engineering, and mobile malware analysis.
\end{IEEEbiography}

\vspace{-10mm}
\begin{IEEEbiography}[{\includegraphics[width=1in,height=1.25in,clip,keepaspectratio]{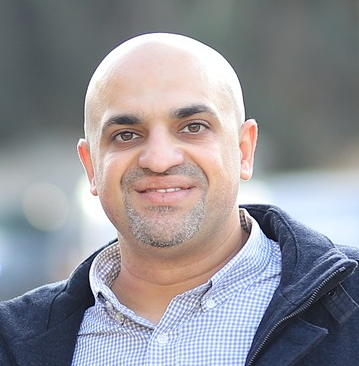}}]{Radi Jarrar}
obtained his B.Sc. in Computer Information Technology from the Arab American University in 2007 and a Ph.D. in Computer Science from Monash University in 2012. Since 2015, he has been an assistant professor in the Department of Computer Science at Birzeit University, Ramallah, Palestine. His research interests include ML, computer vision, and data science, with applications in computer security.
\end{IEEEbiography}

\vspace{-10mm}
\begin{IEEEbiography}[{\includegraphics[width=1in,height=1.25in,clip,keepaspectratio]{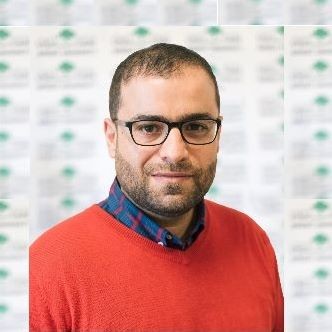}}]{Samer Zein} is an Associate Professor in the Computer Science Department at Birzeit University. He has over 20 years of academic experience and has also contributed as a software engineer to several management information system (MIS) projects since 2000. His primary area of expertise is mobile application software engineering. He received the M.Sc. degree in Software Engineering from Northumbria University, United Kingdom, in 2004, and the Ph.D. degree in Mobile Software Engineering from the International Islamic University Malaysia (IIUM) in 2016. His research interests include mobile app software engineering, empirical software engineering, model-driven software development, and systematic literature reviews (SLRs). He has also conducted multiple qualitative studies involving contemporary industrial case studies.
\end{IEEEbiography}

\vspace{-10mm}
\begin{IEEEbiography}[{\includegraphics[width=1in,height=1.25in,clip,keepaspectratio]{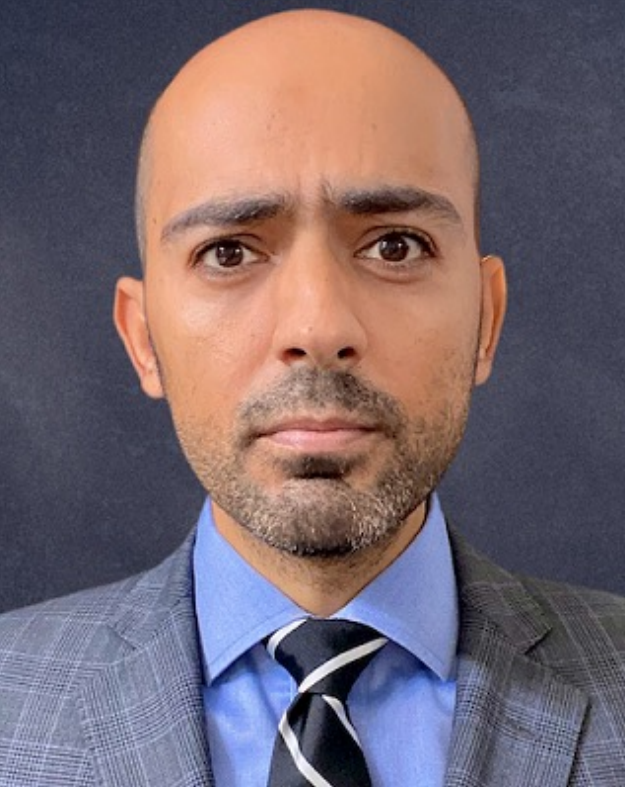}}]{David Mohaisen}
(Senior Member, IEEE) received the MSc and PhD degrees from the University of Minnesota in 2012. He is currently a full professor at the University of Central Florida and an affiliated professor at Birzeit University, where he directs the Security and Analytics Lab. From 2015 to 2017, he was an assistant professor at SUNY Buffalo, and from 2012 to
2015, he was a senior research scientist with Verisign Labs. His research interests span networked systems security, online privacy, and measurements. He has been an associate editor for the IEEE Transactions on Mobile Computing, IEEE Transactions on Cloud Computing, IEEE Transactions on Parallel and Distributed Systems, and IEEE Transactions on Dependable and Secure Computing. He is a senior member of ACM (2018) and IEEE (2015), a distinguished speaker of the ACM, and a distinguished visitor of the IEEE Computer Society.
\end{IEEEbiography}

\vspace{-10mm}
\newpage
\appendices
\section{F1 Score for Cross-Year Strategy} 

\begin{figure}[h]
    \centering
    \begin{subfigure}[t]{0.2352\textwidth} 
        \centering
        \includegraphics[width=0.99\textwidth]{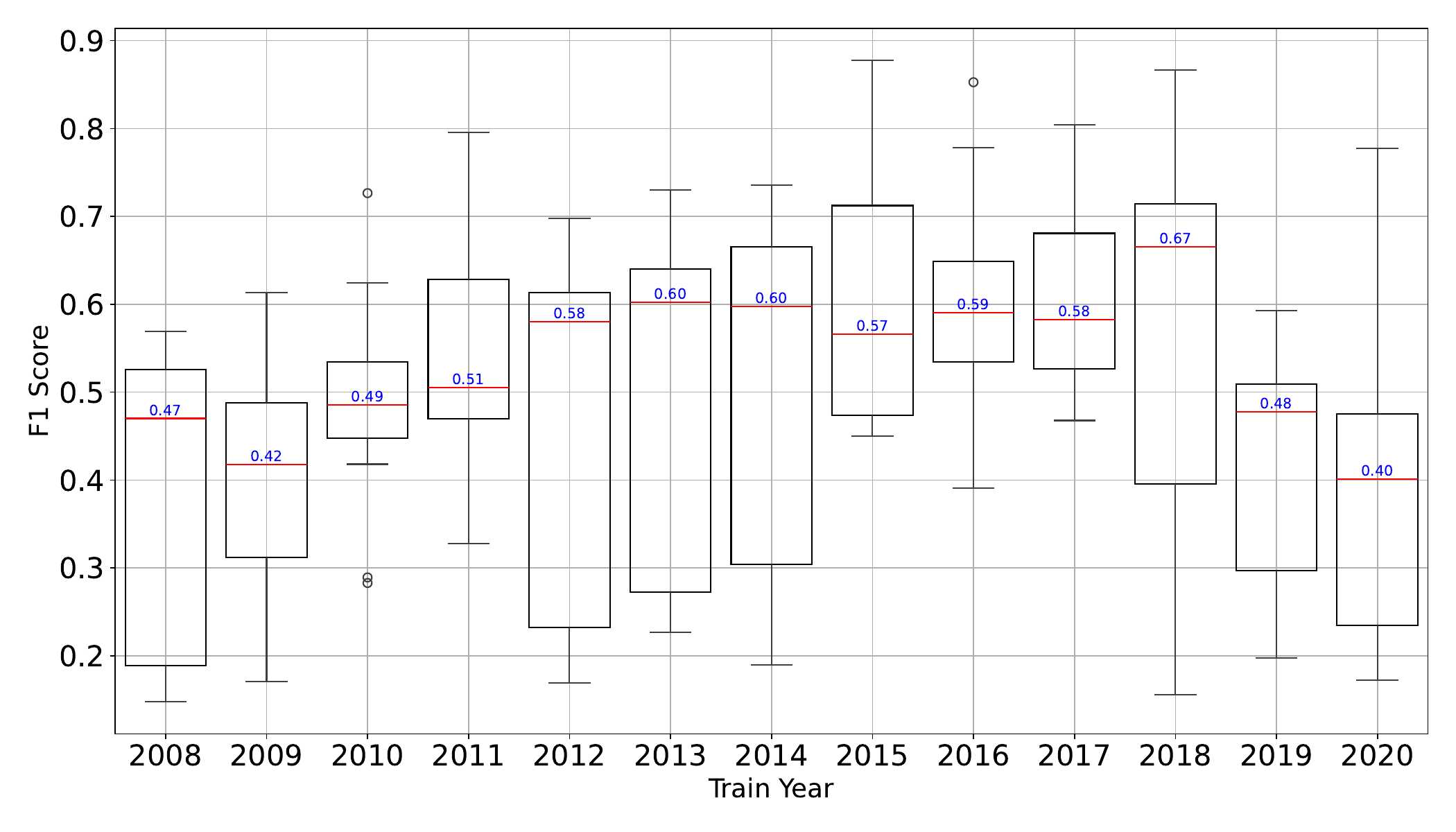}\vspace{-2mm}
        \caption{\normalfont Pre-Real-F1.}
        \label{fig:Box_Plot_GRU_Dynamic_Pre_Real_F1}
    \end{subfigure}
    ~
    \begin{subfigure}[t]{0.2352\textwidth}  
        \centering
        \includegraphics[width=0.99\textwidth]{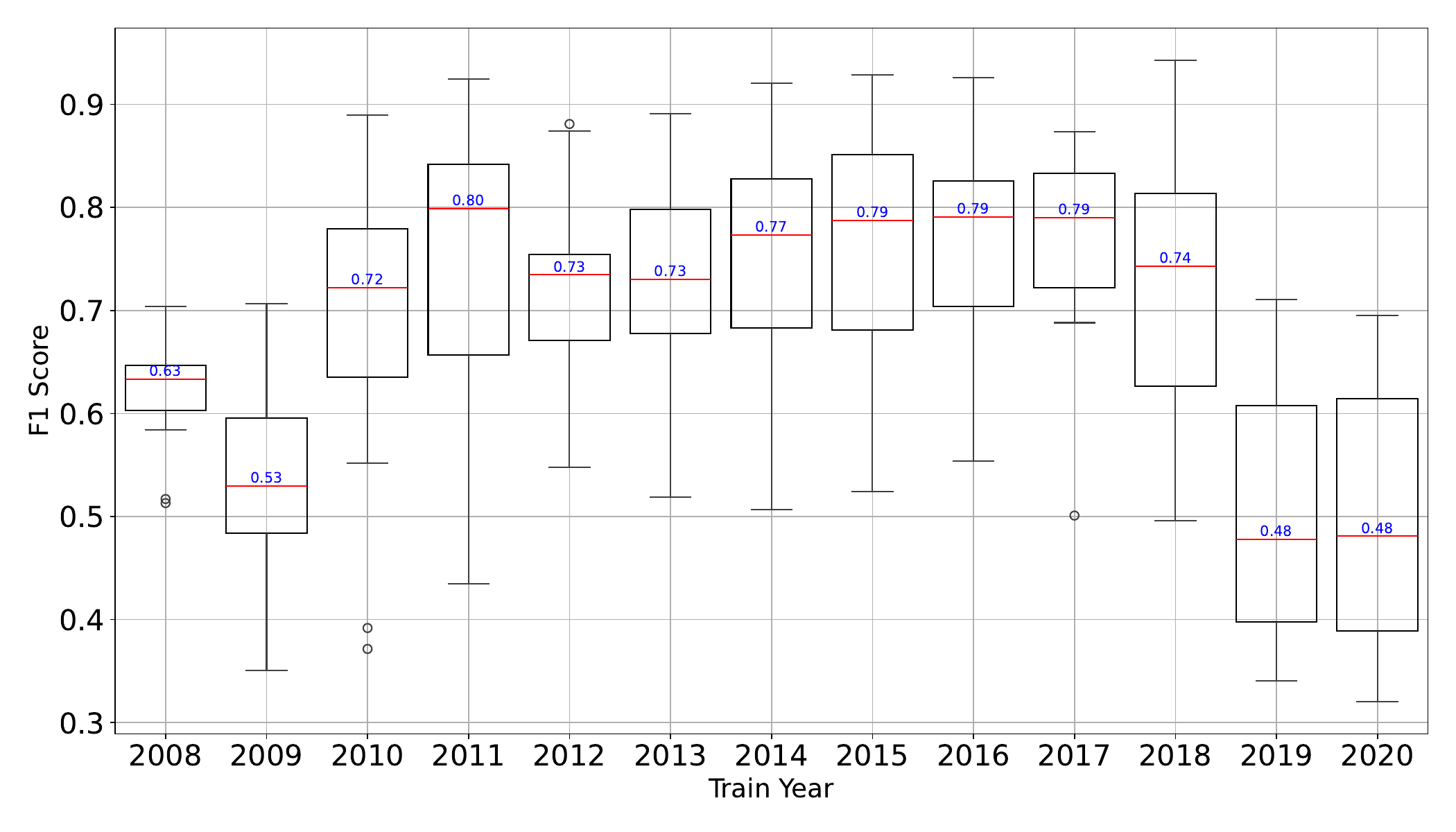}\vspace{-2mm}
        \caption{\normalfont Post-Real-F1.}
        \label{fig:Box_Plot_GRU_Dynamic_Post_Real_F1}
    \end{subfigure}
    ~
    \begin{subfigure}[t]{0.235\textwidth}
        \centering
        \includegraphics[width=0.99\textwidth]{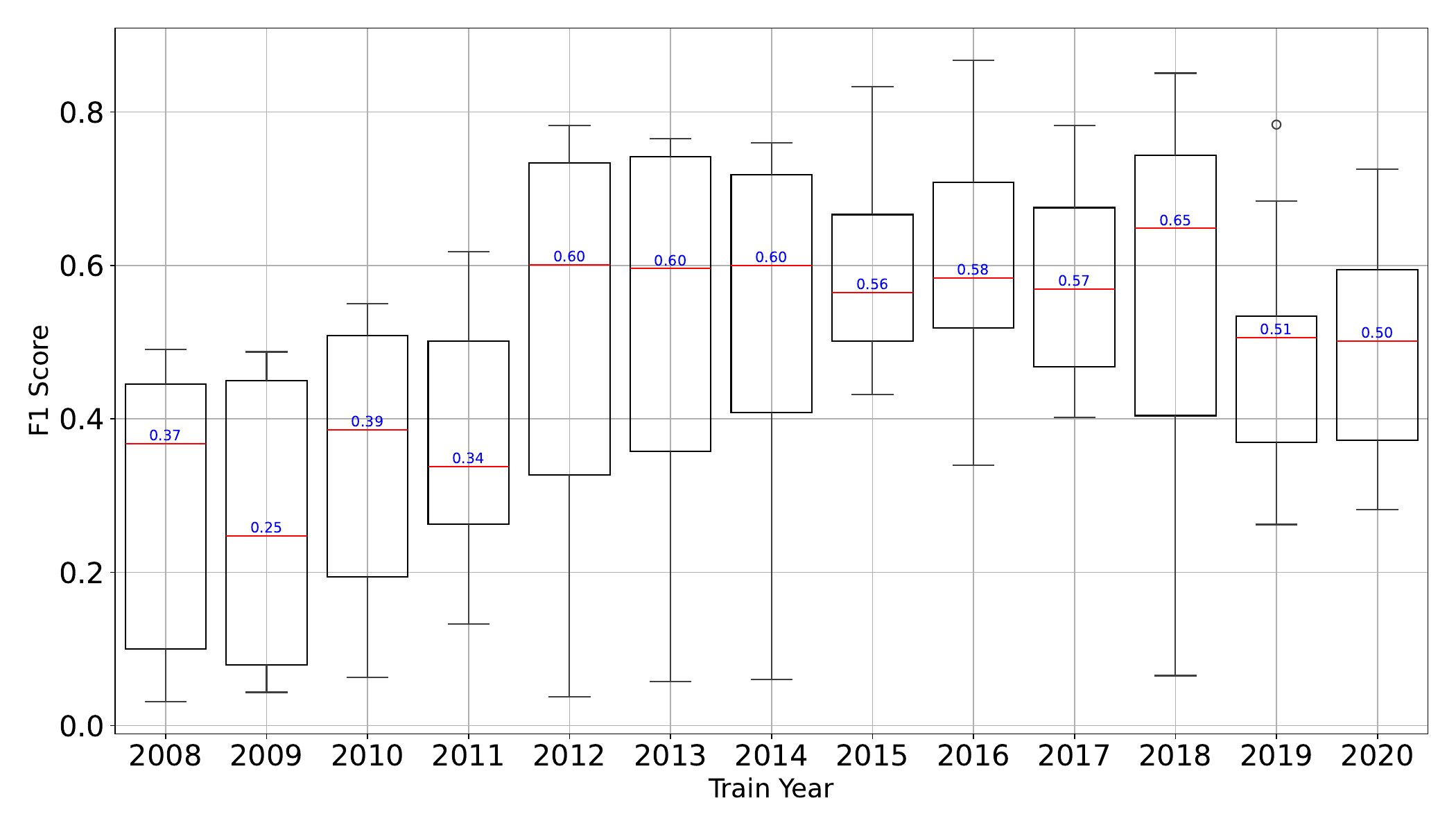}\vspace{-2mm}
        \caption{\normalfont Pre-Emu-F1.}
        \label{fig:Box_Plot_GRU_Dynamic_Pre_Emu_F1}
    \end{subfigure}
    ~
    \begin{subfigure}[t]{0.235\textwidth}
        \centering
        \includegraphics[width=0.99\textwidth]{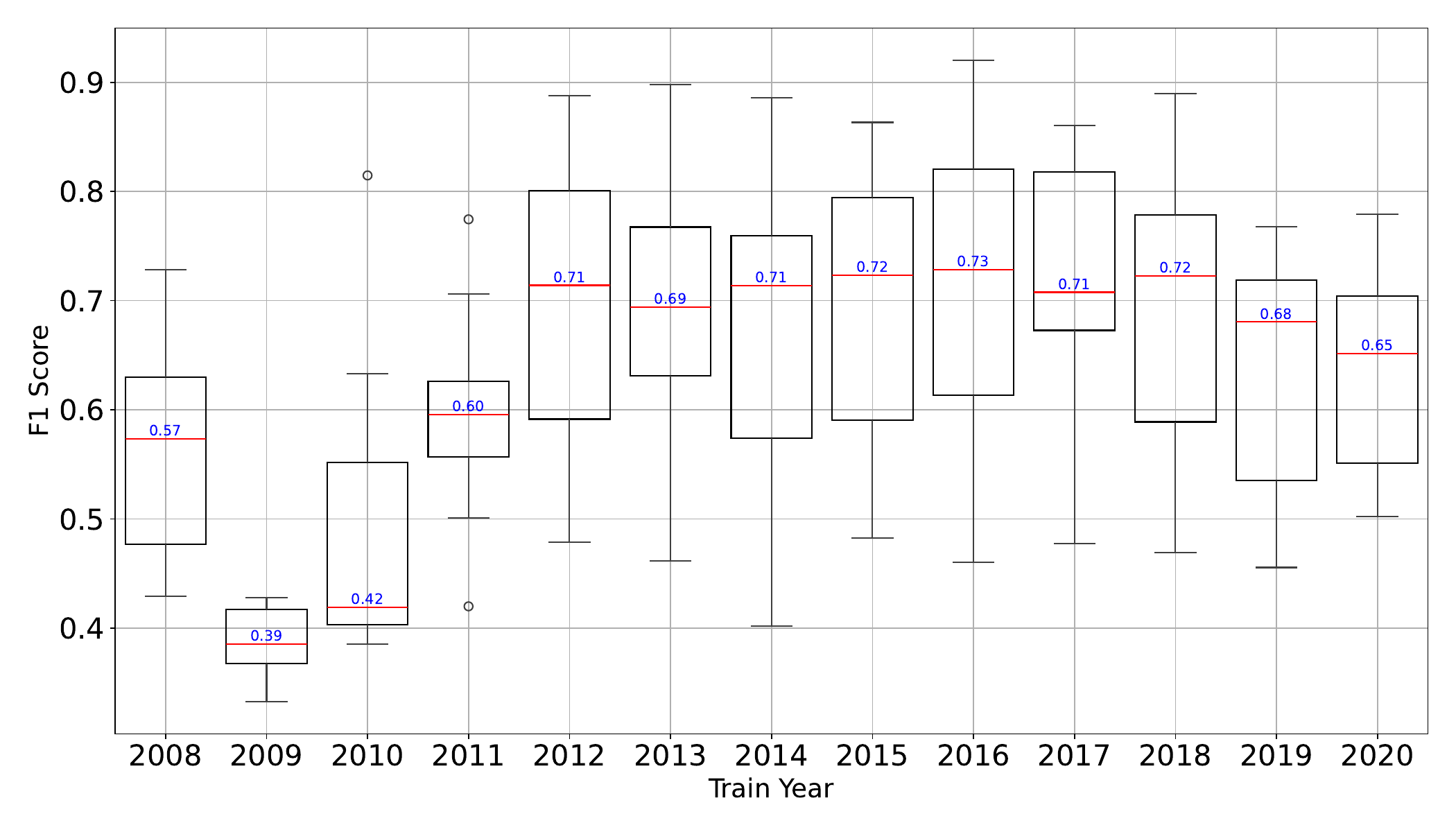}\vspace{-2mm}
        \caption{\normalfont Post-Emu-F1.}
        \label{fig:Box_Plot_GRU_Dynamic_Post_Emu_F1}
    \end{subfigure}
    
    \vspace{-2mm}
\caption{\normalfont Cross-year GRU pre- and post-balancing performance in terms of F1 scores with dynamic features using real and emulator data.}
    
    \label{fig:Boxplot_Gru_Dynamic_Real_Emu_Cross_f1}
\end{figure}

Figure~\ref{fig:Boxplot_Gru_Dynamic_Real_Emu_Cross_f1} presents the cross-year performance of GRU models trained with dynamic features under four configurations—pre- and post-balancing on both real and emulator data—measured by F1 scores across training years. In Fig.~\ref{fig:Boxplot_Gru_Dynamic_Real_Emu_Cross_f1}(a), \textit{Pre-Real-F1}, the F1 scores range approximately from 0.2 to 0.8, with notable fluctuations across training years. Earlier years (2008–2012) yield lower and more dispersed scores, while models trained on mid-range years (2015–2018) tend to show improved and more stable performance. However, high variance and numerous outliers persist throughout, indicating inconsistent generalization. After applying class balancing, Fig.~\ref{fig:Boxplot_Gru_Dynamic_Real_Emu_Cross_f1}(b), \textit{Post-Real-F1}, shows clear improvement: median F1 scores increase, and variance across test years decreases. This stability persists until around 2018, beyond which performance starts to degrade, potentially due to emerging concept drift or data quality issues in recent years.

Fig.~\ref{fig:Boxplot_Gru_Dynamic_Real_Emu_Cross_f1}(c), \textit{Pre-Emu-F1}, exhibits similar trends using emulator data, though with even lower performance in earlier years—suggesting emulator data may suffer more from representation noise. Again, models trained between 2013 and 2017 offer better median F1 scores, with reduced spread. Upon applying balancing, Fig.~\ref{fig:Boxplot_Gru_Dynamic_Real_Emu_Cross_f1}(d), \textit{Post-Emu-F1}, mirrors the improvements seen with emulator data. Median scores rise notably (often exceeding 0.7), and the interquartile ranges shrink (IQR), indicating enhanced consistency. Notably, balancing has a more dramatic impact on emulator data than real data, highlighting the former's sensitivity to class imbalance. Across all settings, the training years 2019 and 2020 remain the most volatile, suggesting these periods may reflect significant data shifts or evolving malware behaviors that challenge temporal generalization. Overall, class balancing significantly improves model robustness across both data sources, with post-balancing emulator performance approaching that of real data, especially for mid-range training years.

\begin{figure}[t]
    \centering
    \begin{subfigure}[t]{0.235\textwidth} 
        \centering
        \includegraphics[width=0.99\textwidth]{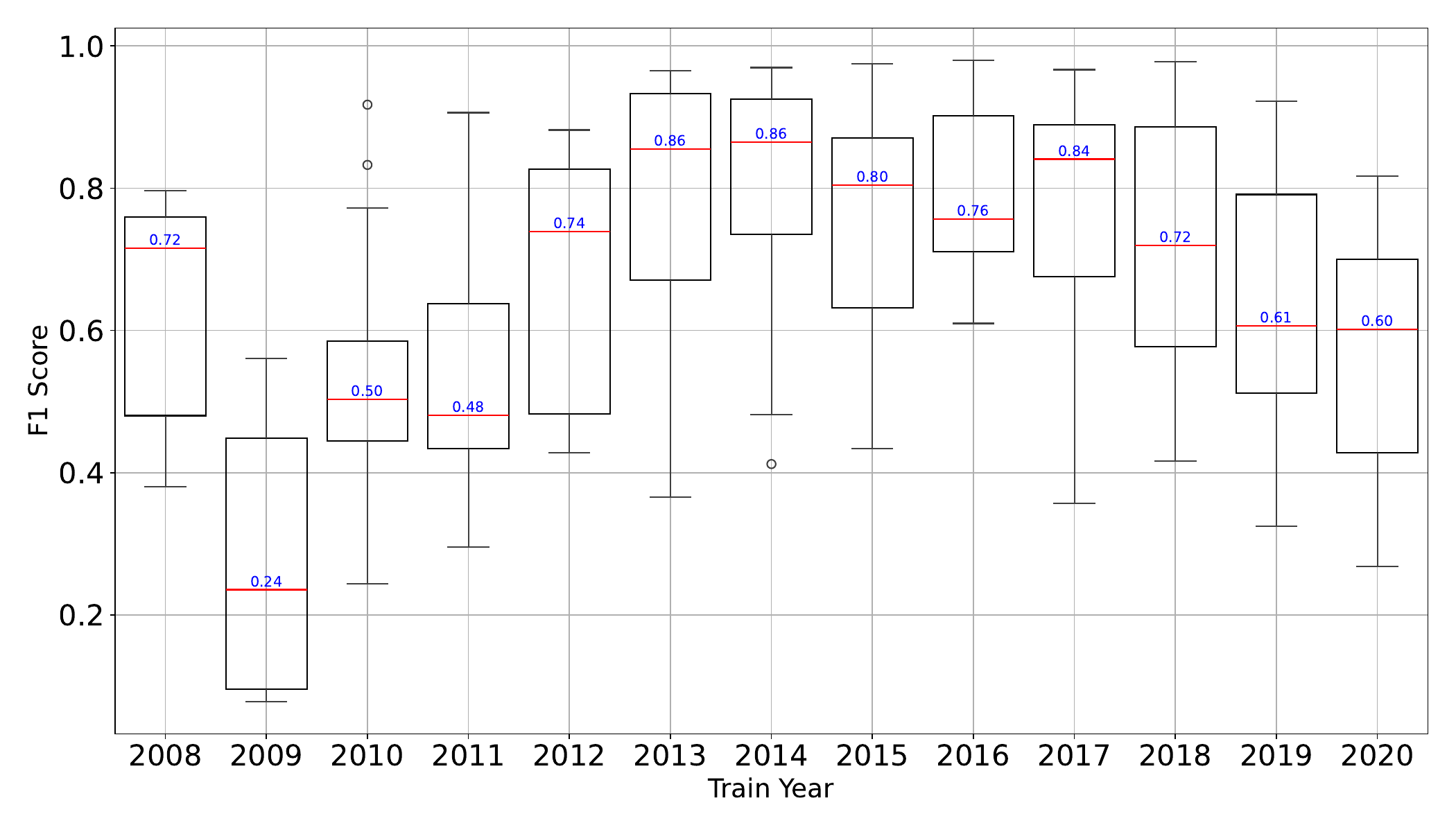}\vspace{-2mm}
        \caption{\normalfont Pre-Real-F1.}
        \label{fig:Box_Plot_RNN_Hybrid_Pre_Real_F1}
    \end{subfigure}
    ~
    \begin{subfigure}[t]{0.235\textwidth}  
        \centering
        \includegraphics[width=0.99\textwidth]{figsr/KDS/Cross/RNN/BP_Acc_Post_ATAT_RNN_Real_Hybrid.pdf}\vspace{-2mm}
        \caption{\normalfont Post-Real-F1.}
        \label{fig:Box_Plot_RNN_Hybrid_Post_Real_F1}
    \end{subfigure}
    ~
    \begin{subfigure}[t]{0.235\textwidth}
        \centering
        \includegraphics[width=0.99\textwidth]{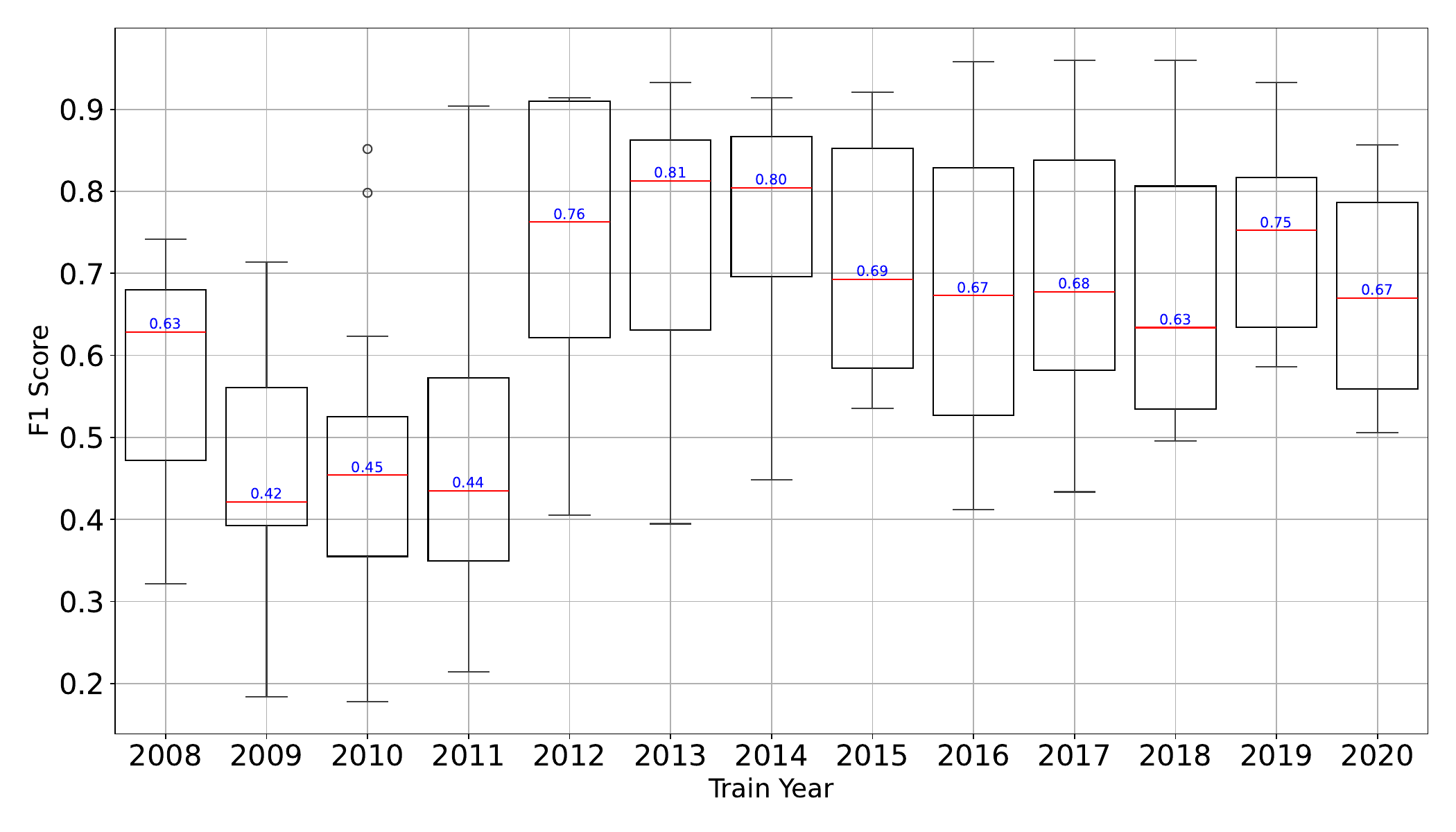}\vspace{-2mm}
        \caption{\normalfont Pre-Emu-F1.}
        \label{fig:Box_Plot_RNN_Hybrid_Pre_Emu_F1}
    \end{subfigure}
    ~
    \begin{subfigure}[t]{0.235\textwidth}
        \centering
        \includegraphics[width=0.99\textwidth]{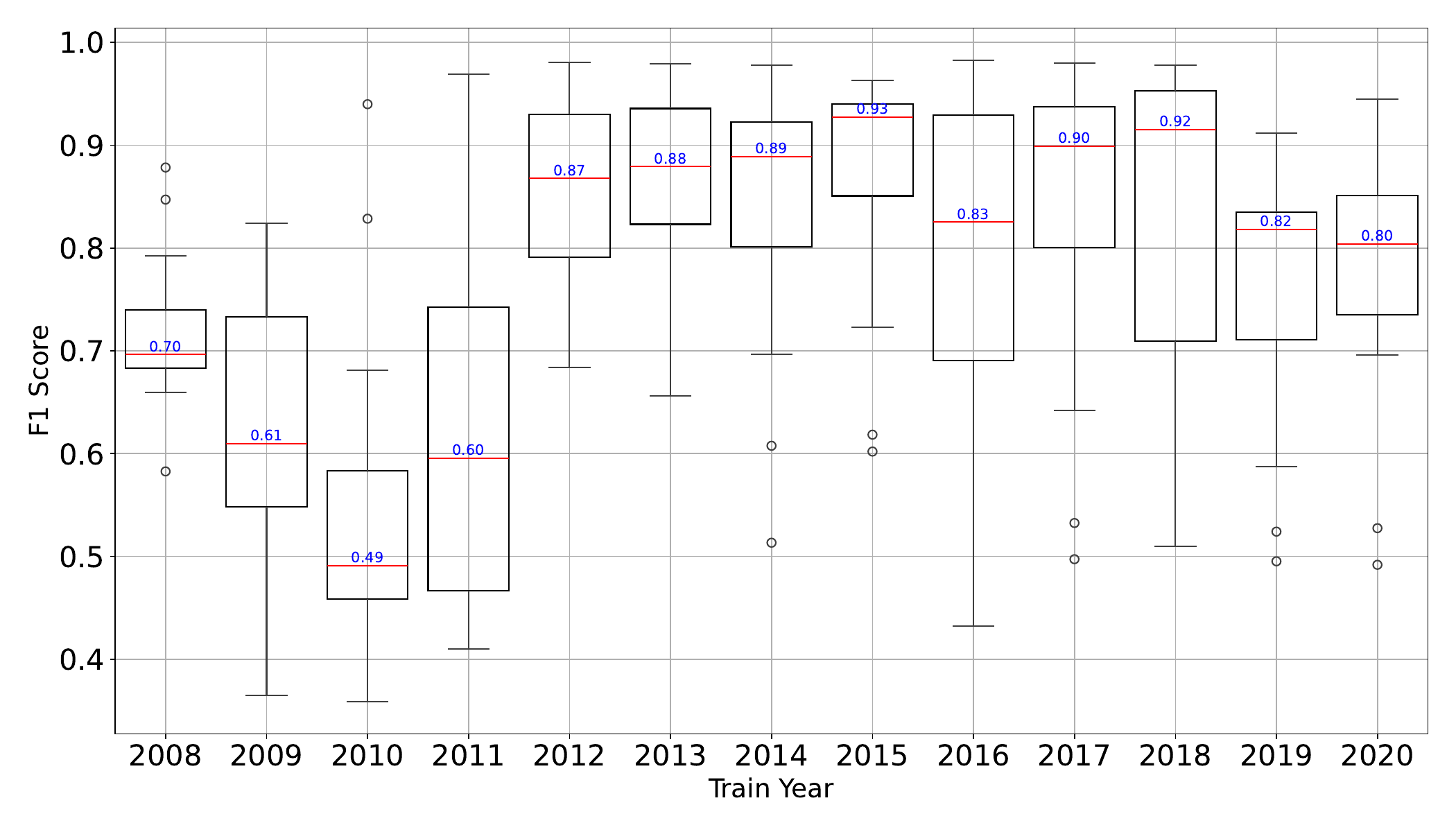}\vspace{-2mm}
        \caption{\normalfont Post-Emu-F1.}
        \label{fig:Box_Plot_RNN_Hybrid_Post_Emu_F1x}
    \end{subfigure}

    \vspace{-2mm}

    \caption{\normalfont Cross-year RNN pre- and post-balancing F1 score results with hybrid features using real and emulator data.}
    \label{fig:BoxBlot_RNN_Hybrid_Real_Emu_F1}
\end{figure}

Figure~\ref{fig:BoxBlot_RNN_Hybrid_Real_Emu_F1} presents the cross-year F1 score performance of RNN-based models trained with hybrid features under four configurations: pre- and post-class balancing using both real and emulator data. In Fig.~\ref{fig:BoxBlot_RNN_Hybrid_Real_Emu_F1}(a), \textit{Pre-Real-F1}, the F1 scores show substantial variability across training years, ranging from as low as 0.2 to as high as 0.9. Earlier years (2008–2011) exhibit poor and inconsistent performance, whereas mid-range years (2013–2017) show higher medians and reduced variance, indicating improved generalization. After balancing, Fig.~\ref{fig:BoxBlot_RNN_Hybrid_Real_Emu_F1}(b), \textit{Post-Real-F1}, demonstrates a marked improvement, with median F1 scores approaching or exceeding 0.9 in many years and significantly reduced IQR. However, a noticeable decline appears in 2019 and 2020, pointing again to potential temporal drift.

Fig.~\ref{fig:BoxBlot_RNN_Hybrid_Real_Emu_F1}(c), \textit{Pre-Emu-F1}, reflects similar dynamics with emulator data: earlier years yield unstable and low performance, while mid-to-late years improve markedly. Post-balancing results in Fig.~\ref{fig:BoxBlot_RNN_Hybrid_Real_Emu_F1}(d), \textit{Post-Emu-F1}, show consistently high F1 scores (often above 0.8) with reduced variance, confirming the benefits of class rebalancing on emulator data. Although performance in recent years (2019–2020) still lags slightly, the RNN appears generally more resilient than in the pre-balancing scenario.

Compared to the GRU-based results in Figure~\ref{fig:Boxplot_Gru_Dynamic_Real_Emu_Cross_f1}, the RNN with hybrid features demonstrates higher peak F1 scores, especially after balancing, and generally tighter distributions. While both architectures benefit from class balancing, RNNs paired with hybrid features tend to achieve better stability and top-end performance across a wider range of training years. Nevertheless, like GRUs, RNNs are not immune to recent-year performance drops, underscoring the persistent challenge of drift-aware generalization.

\section{F1 Score for Incremental Strategy}

\begin{figure}[H]
    \centering
    \begin{subfigure}[t]{0.235\textwidth} 
        \centering
        \includegraphics[width=0.99\textwidth]{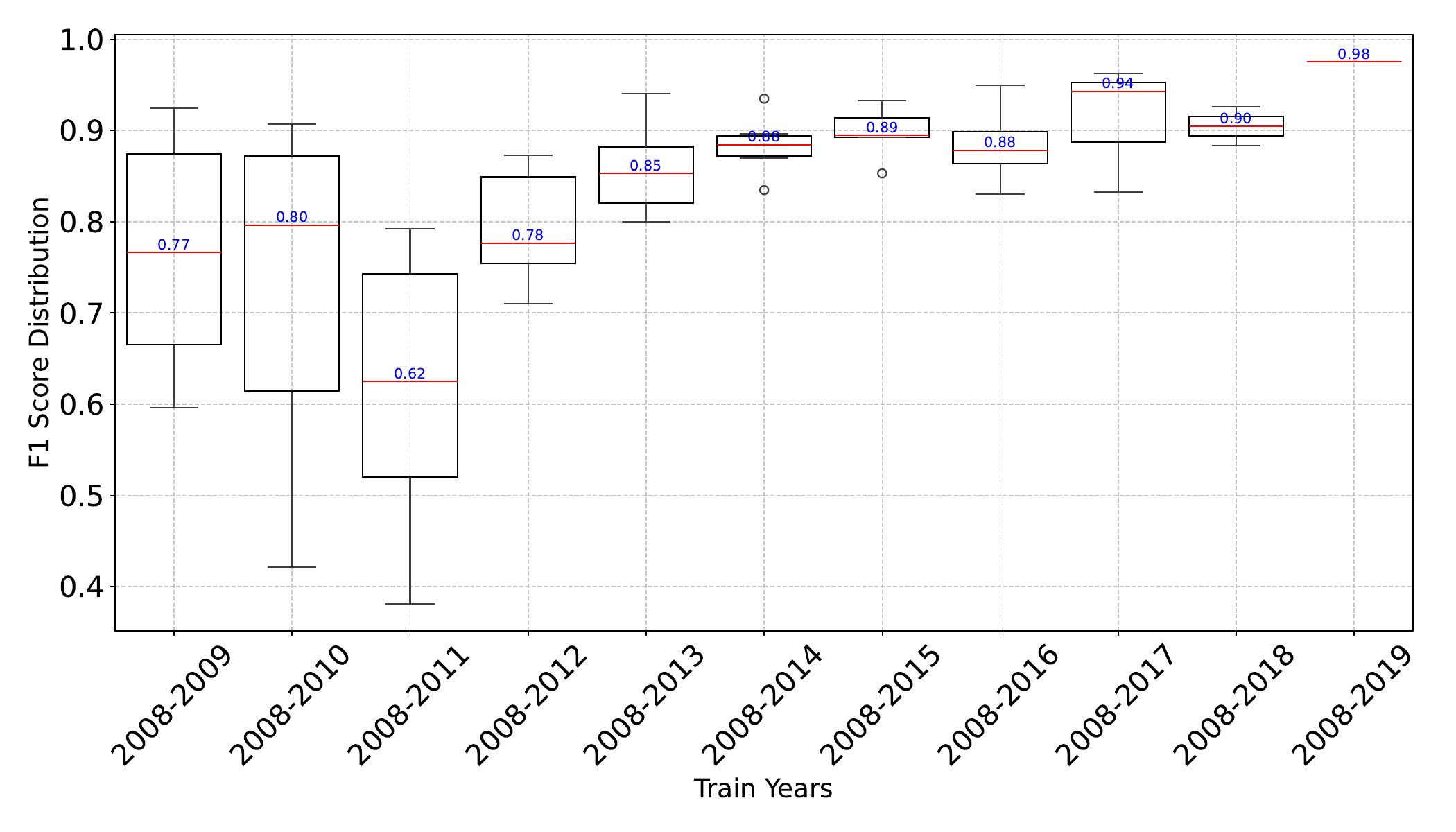}\vspace{-2mm}
        \caption{\normalfont Pre-Real-F1.}
        \label{fig:BP_F1_Pre_ATAT_RF_V2_Real_Static_Incremental}
    \end{subfigure}
    ~
    \begin{subfigure}[t]{0.235\textwidth}  
        \centering
        \includegraphics[width=0.99\textwidth]{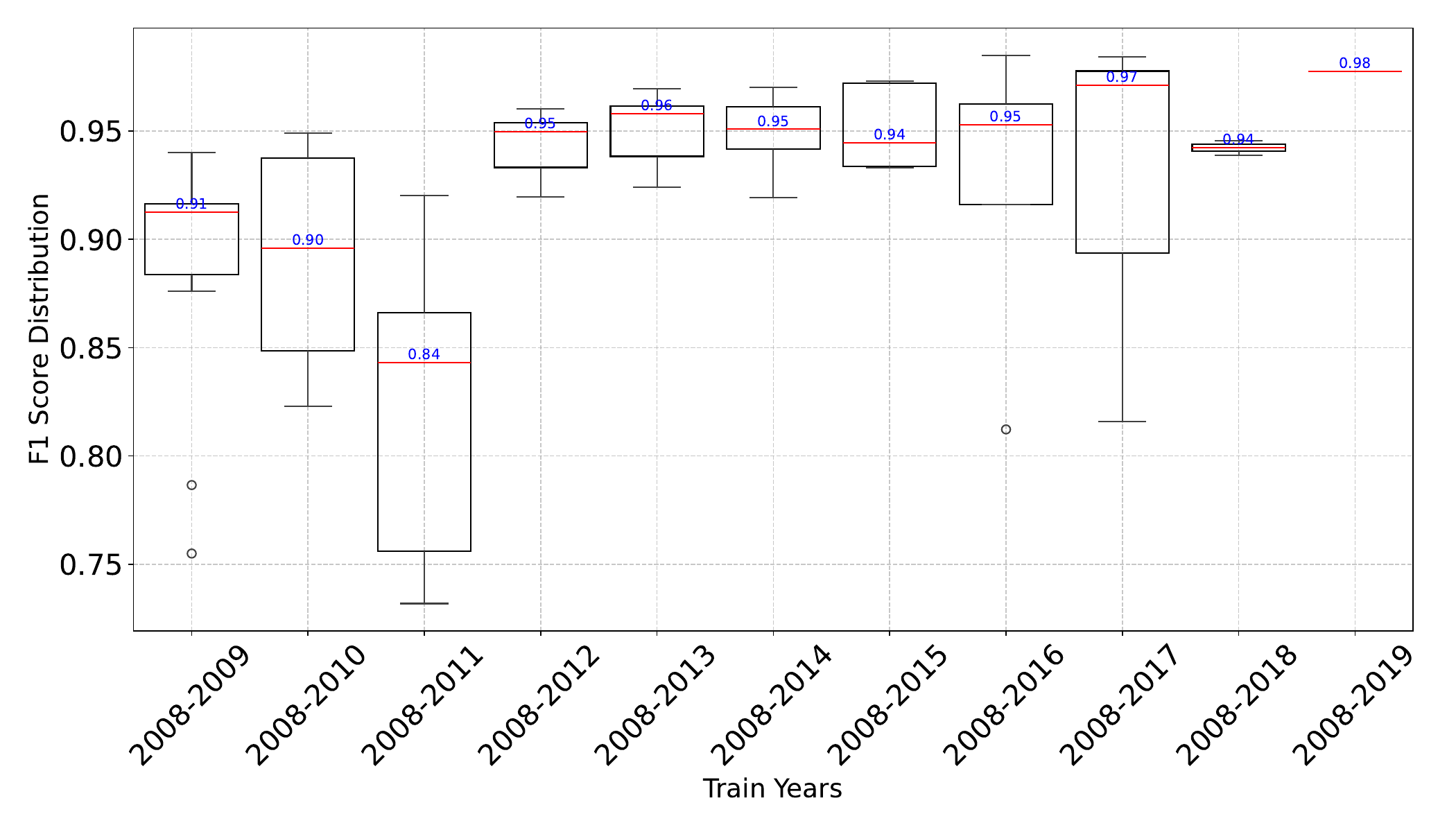}\vspace{-2mm}
        \caption{\normalfont Post-Real-F1.}
        \label{fig:BP_F1_Post_ATAT_RF_V2_Real_Static_Incremental}
    \end{subfigure}
    ~
    \begin{subfigure}[t]{0.235\textwidth}
        \centering
        \includegraphics[width=0.99\textwidth]{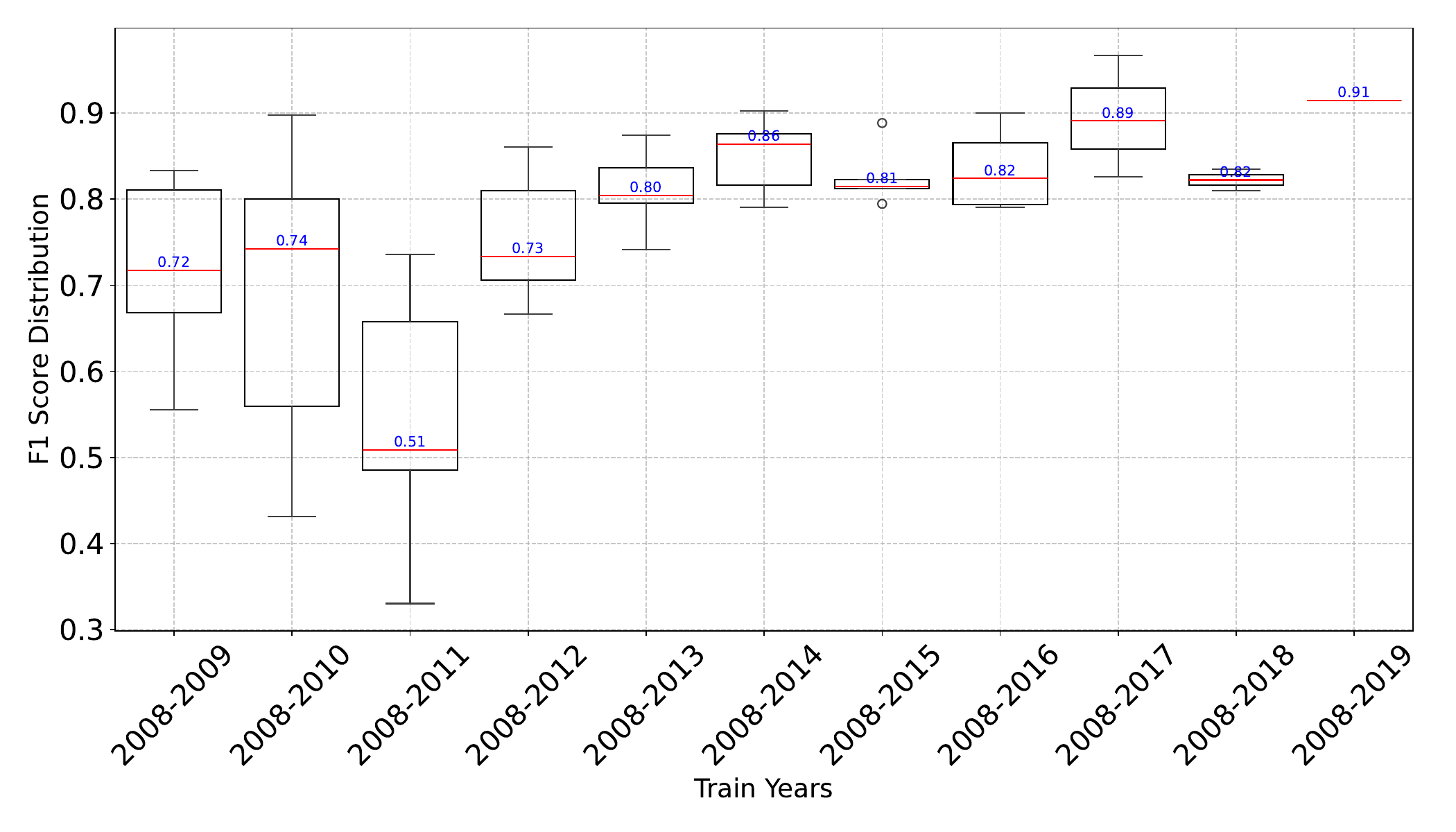}\vspace{-2mm}
        \caption{\normalfont Pre-Emu-F1.}
        \label{fig:BP_F1_Pre_ATAT_RF_V2_Emu_Static_Incremental}
    \end{subfigure}
    ~
    \begin{subfigure}[t]{0.235\textwidth}
        \centering
        \includegraphics[width=0.99\textwidth]{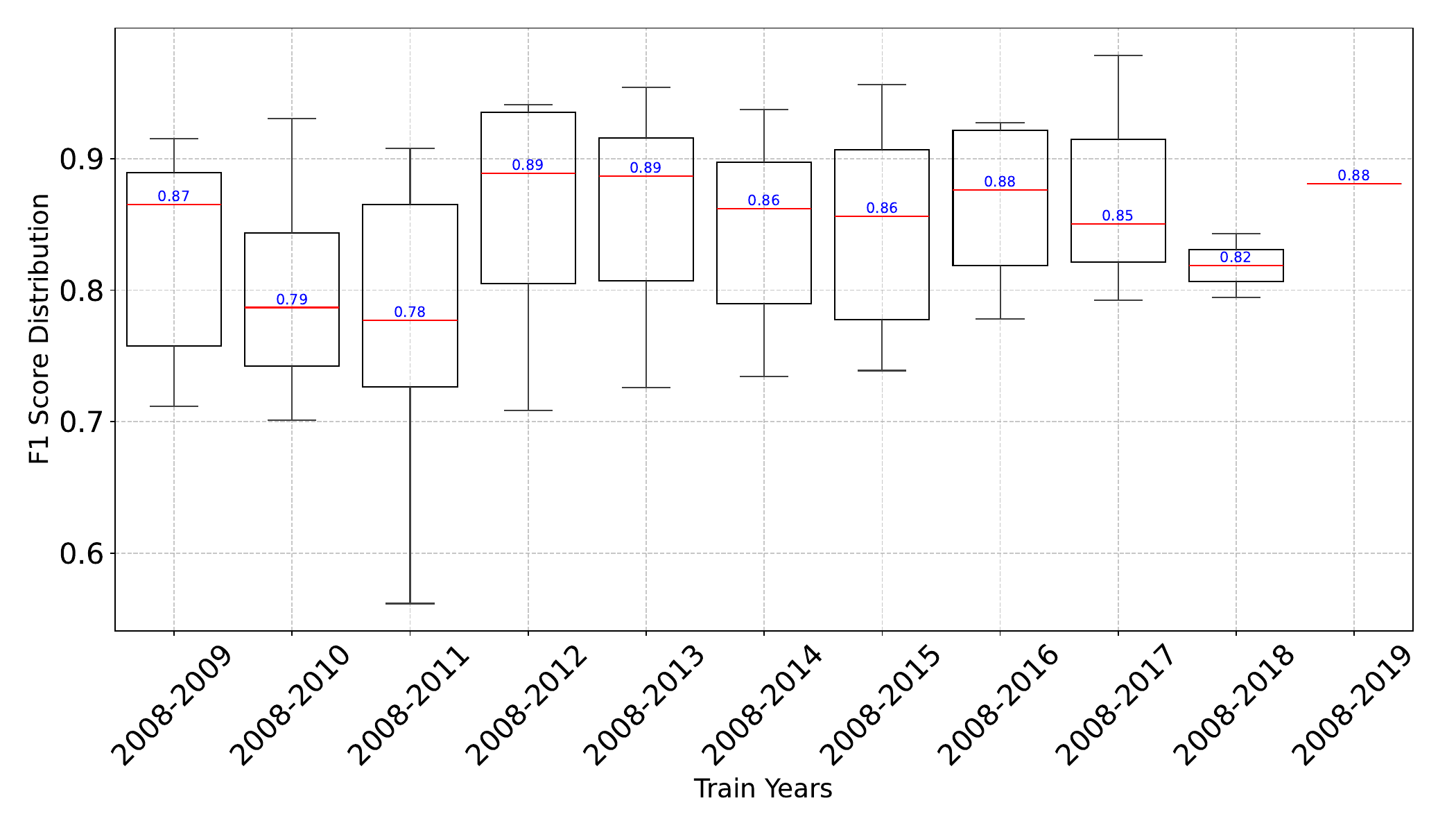}\vspace{-2mm}
        \caption{\normalfont Post-Emu-F1.}
        \label{fig:BP_F1_Post_ATAT_RF_V2_Emu_Static_Incremental}
    \end{subfigure}

\vspace{-2mm}
    \caption{\normalfont The performance of RF algorithm pre- and post-balancing on real and emulator data with static features.}\vspace{-3mm}
    \label{fig:BoxPlot_RF_Static_Incremental_Emu_Real_F1}
\end{figure}

\begin{figure}[H]
    \centering
    \begin{subfigure}[t]{0.235\textwidth} 
        \centering
        \includegraphics[width=0.99\textwidth]{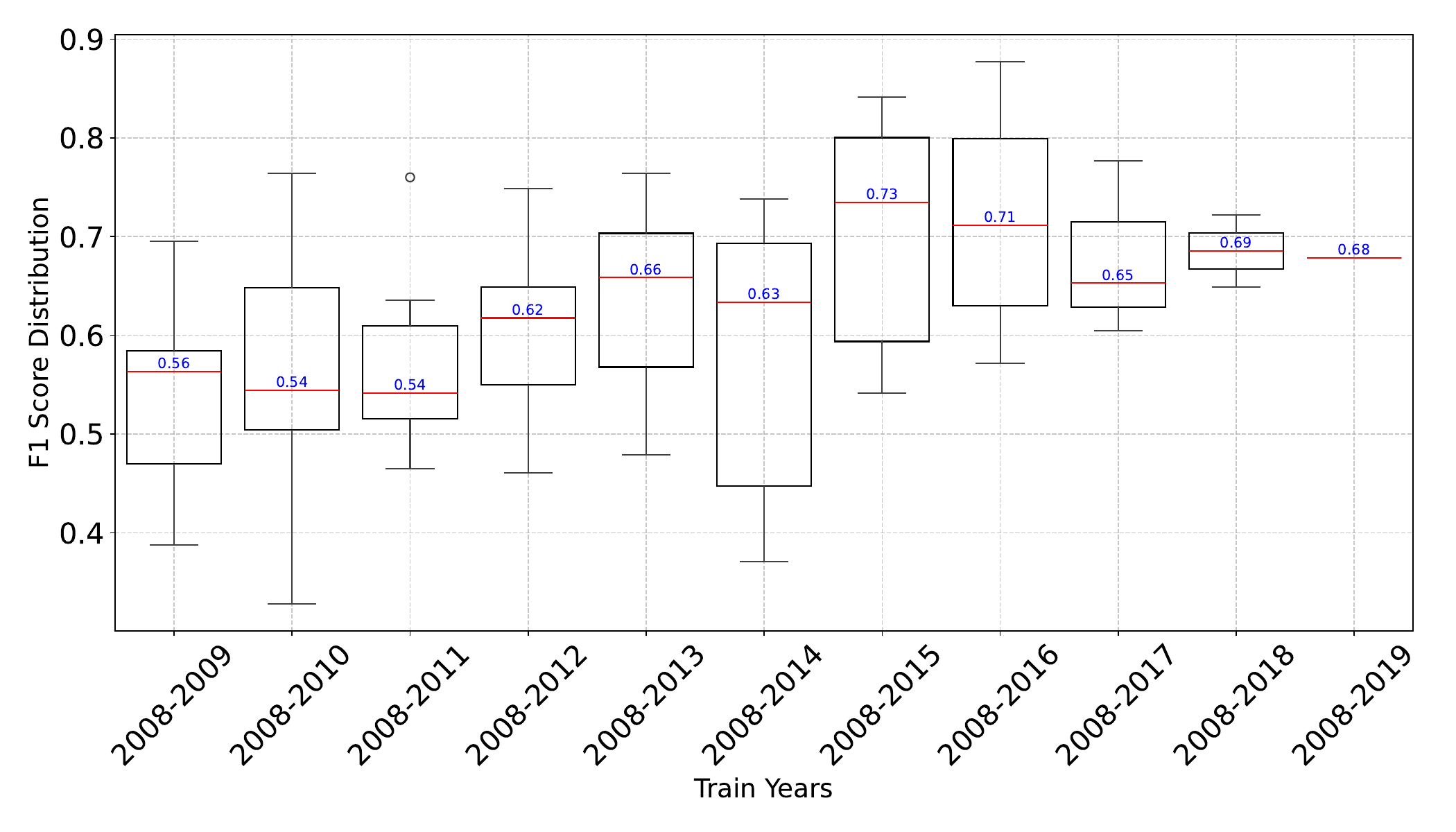}\vspace{-2mm}
        \caption{\normalfont Pre-Real-F1.}
        \label{fig:BP_F1_Pre_ATAT_GRU_V2_Real_Dynamic_Incremental}
    \end{subfigure}
    ~
    \begin{subfigure}[t]{0.235\textwidth}  
        \centering
        \includegraphics[width=0.99\textwidth]{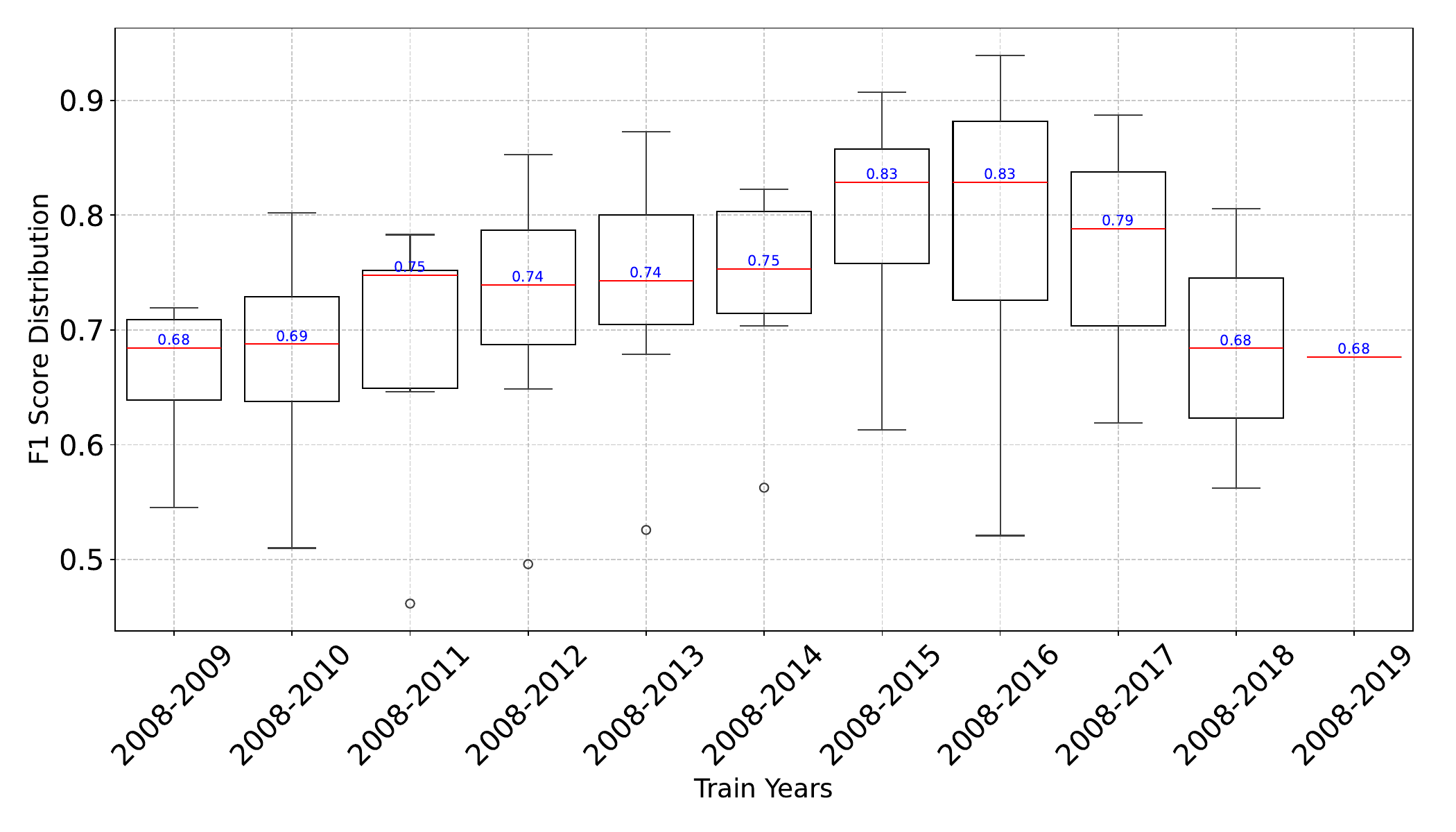}\vspace{-2mm}
        \caption{\normalfont Post-Real-F1.}
        \label{fig:BP_F1_Post_ATAT_GRU_V2_Real_Dynamic_Incremental}
    \end{subfigure}
    ~
    \begin{subfigure}[t]{0.235\textwidth}
        \centering
        \includegraphics[width=0.99\textwidth]{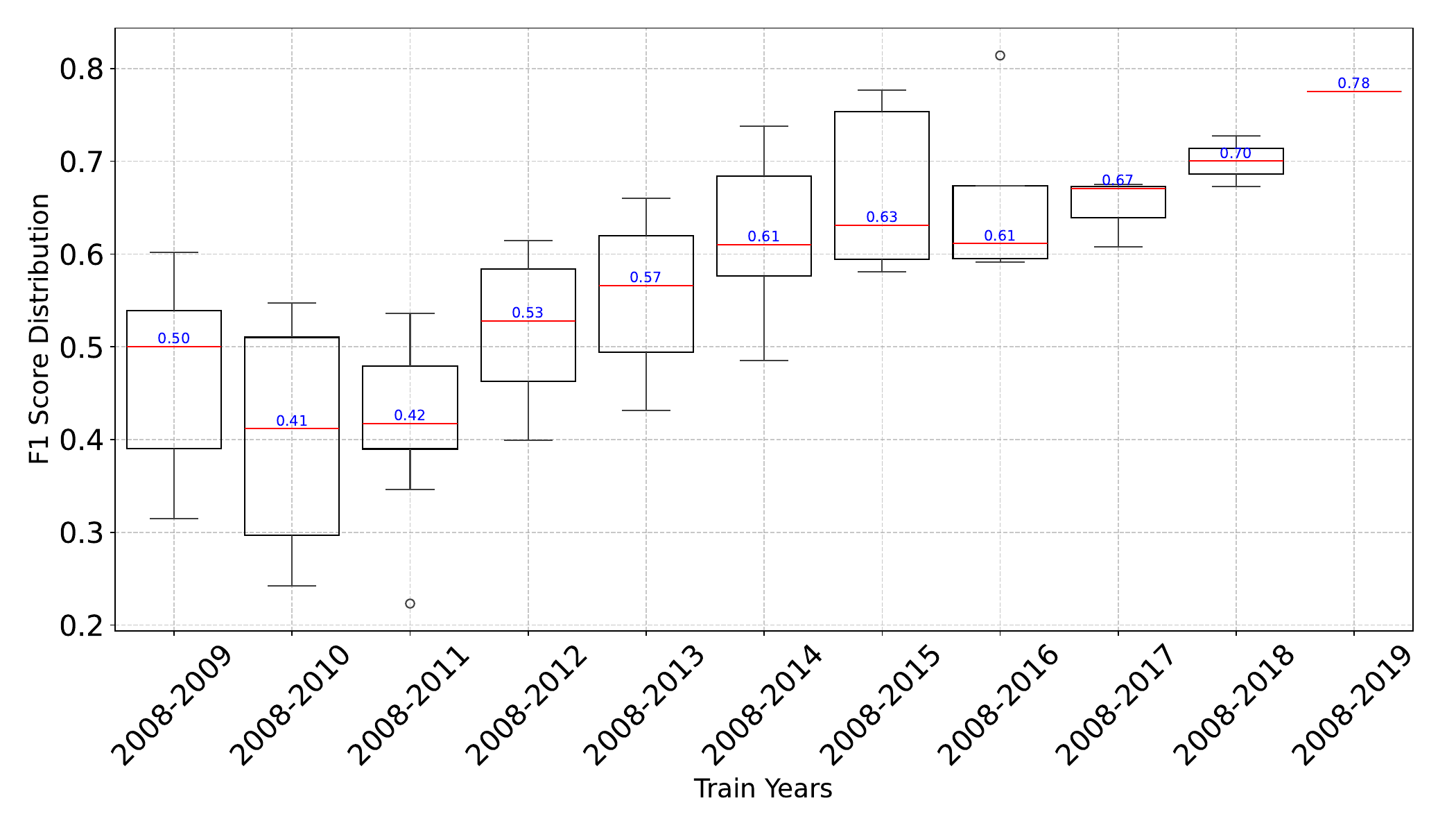}\vspace{-2mm}
        \caption{\normalfont Pre-Emu-F1.}
        \label{fig:BP_F1_Pre_ATAT_GRU_V2_Emu_Dynamic_Incremental}
    \end{subfigure}
    ~
    \begin{subfigure}[t]{0.235\textwidth}
        \centering
        \includegraphics[width=0.99\textwidth]{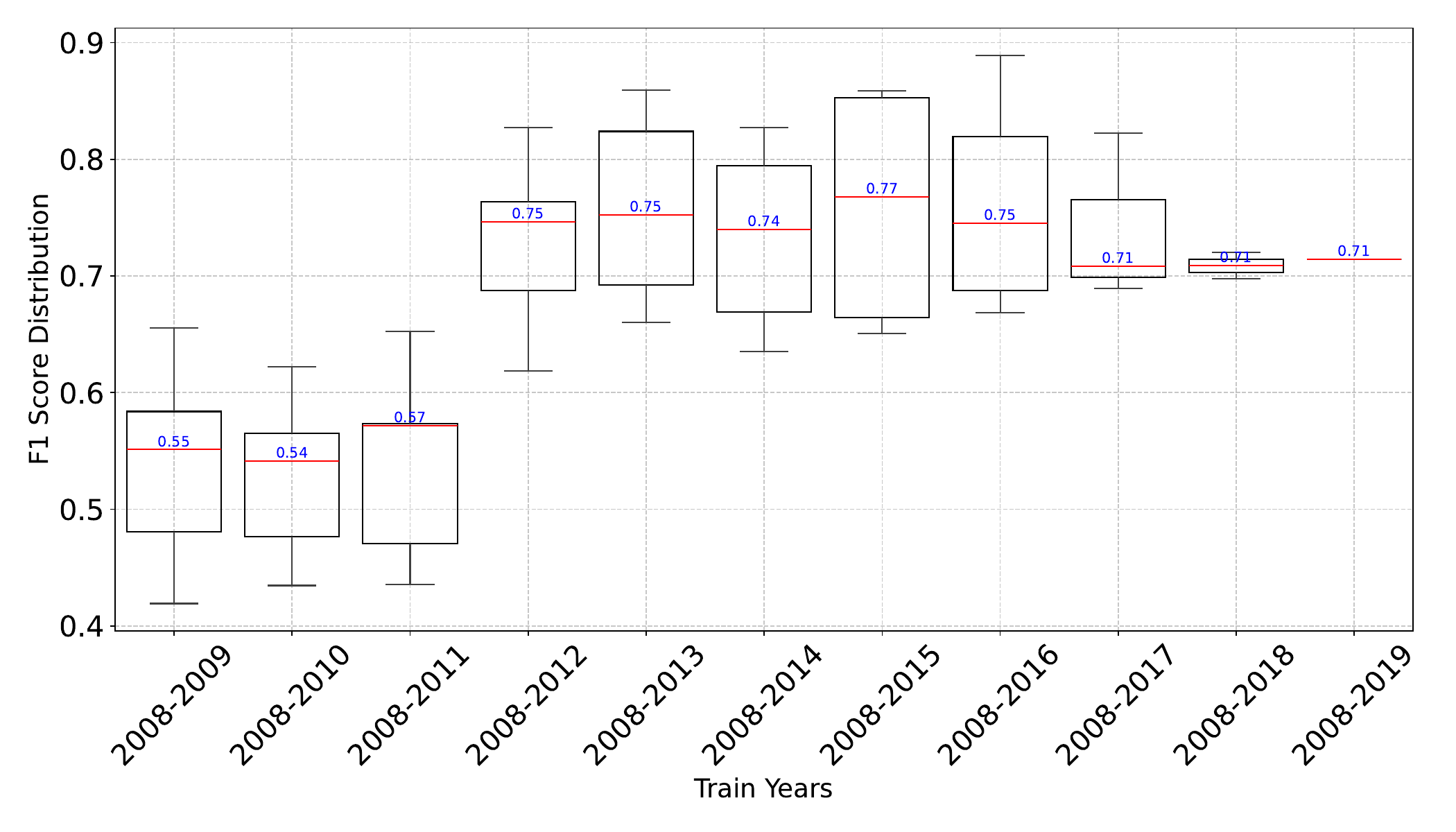}\vspace{-2mm}
        \caption{\normalfont Post-Emu-F1.}
        \label{fig:BP_F1_Post_ATAT_GRU_V2_Emu_Dynamic_Incremental}
    \end{subfigure}

    \vspace{-2mm}

    \caption{\normalfont The performance of GRU algorithm pre- and post-balancing on real and emulator data with dynamic features.}
    \label{fig:BoxPlot_RF_Incremental_Emu_Real_F1}\vspace{-3mm}
\end{figure}

\begin{figure}[H]
    \centering
    \begin{subfigure}[t]{0.235\textwidth} 
        \centering
        \includegraphics[width=0.99\textwidth]{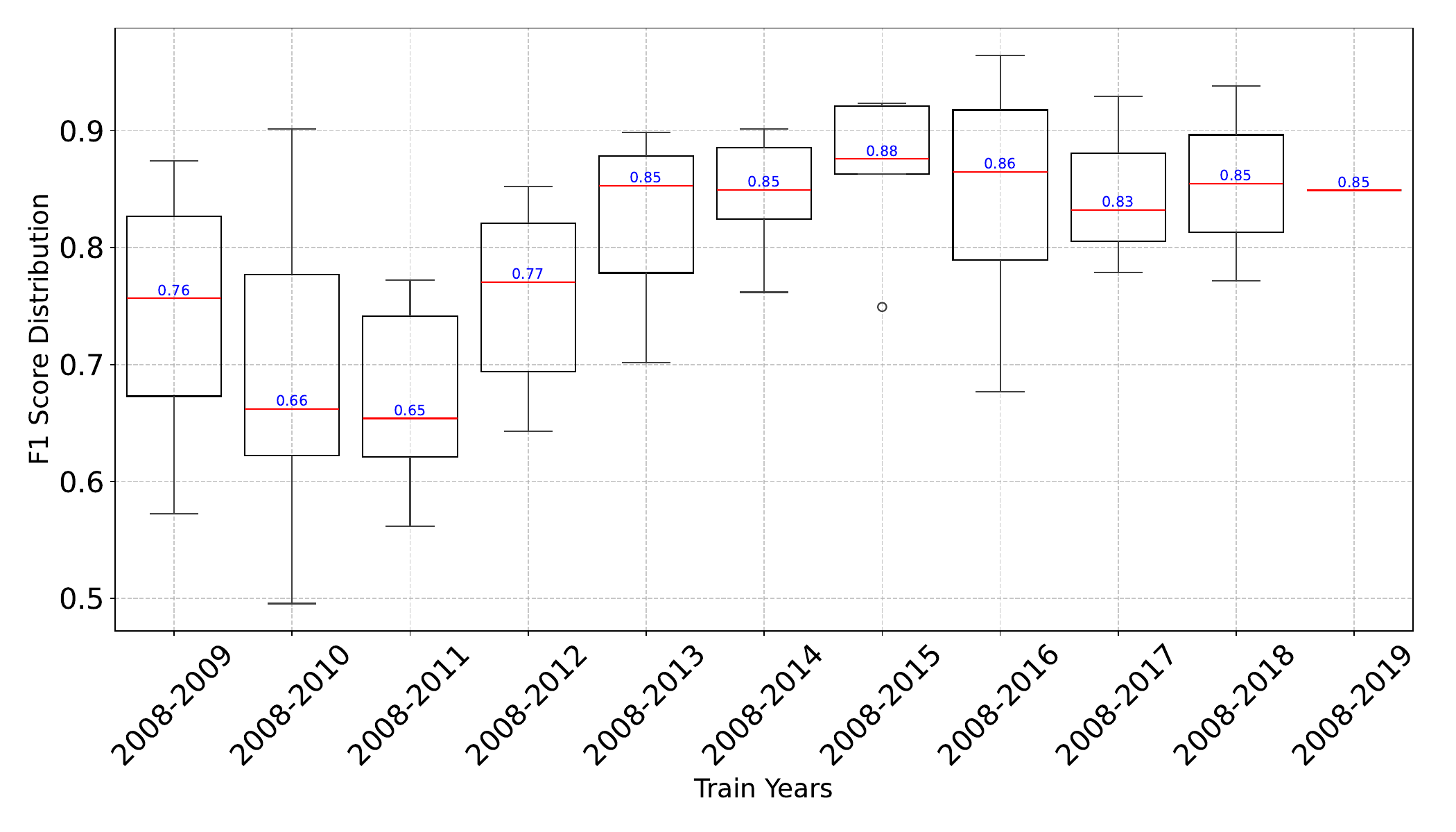}\vspace{-2mm}
        \caption{\normalfont Pre-Real-F1.}
        \label{fig:BP_F1_Pre_ATAT_RNN_V2_Real_Hybrid_Incremental}
    \end{subfigure}
    ~
    \begin{subfigure}[t]{0.235\textwidth}  
        \centering
        \includegraphics[width=0.99\textwidth]{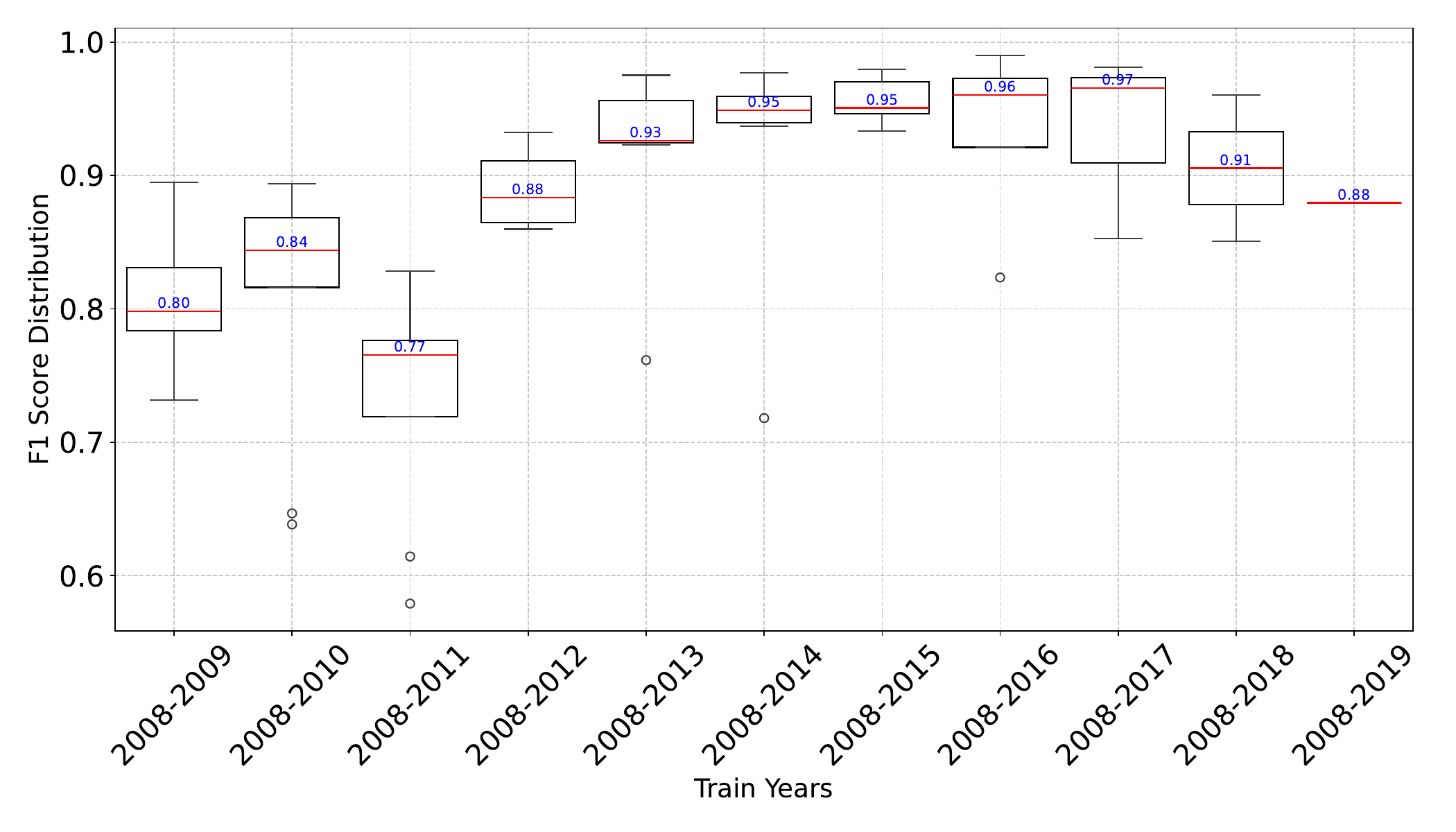}\vspace{-2mm}
        \caption{\normalfont Post-Real-F1.}
        \label{fig:BP_F1_Post_ATAT_RNN_V2_Real_Hybrid_Incremental}
    \end{subfigure}
    ~
    \begin{subfigure}[t]{0.235\textwidth}
        \centering
        \includegraphics[width=0.99\textwidth]{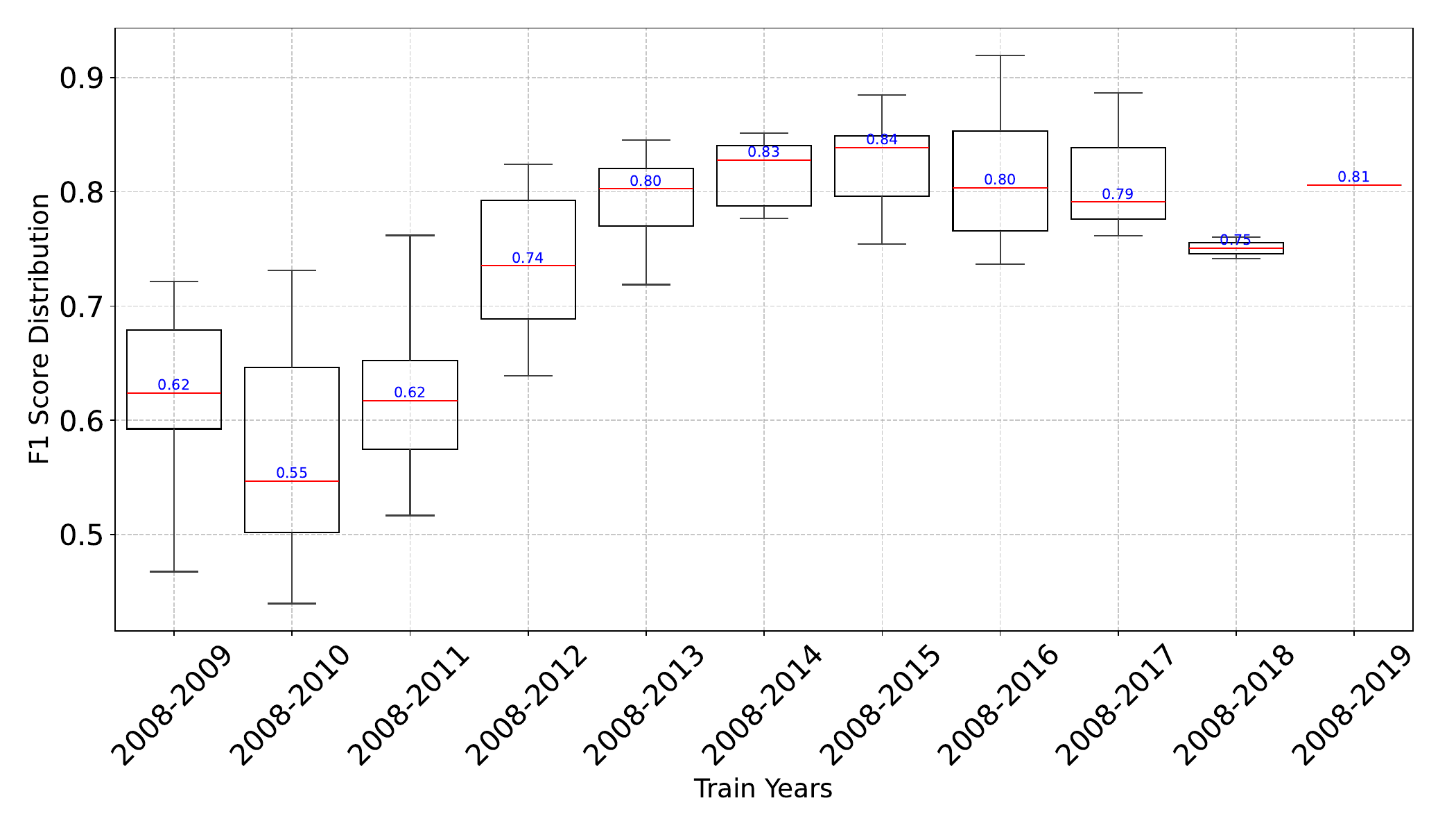}\vspace{-2mm}
        \caption{\normalfont Pre-Emu-F1.}
        \label{fig:BP_F1_Pre_ATAT_RNN_V2_Emu_Hybrid_Incremental}
    \end{subfigure}
    ~
    \begin{subfigure}[t]{0.235\textwidth}
        \centering
        \includegraphics[width=0.99\textwidth]{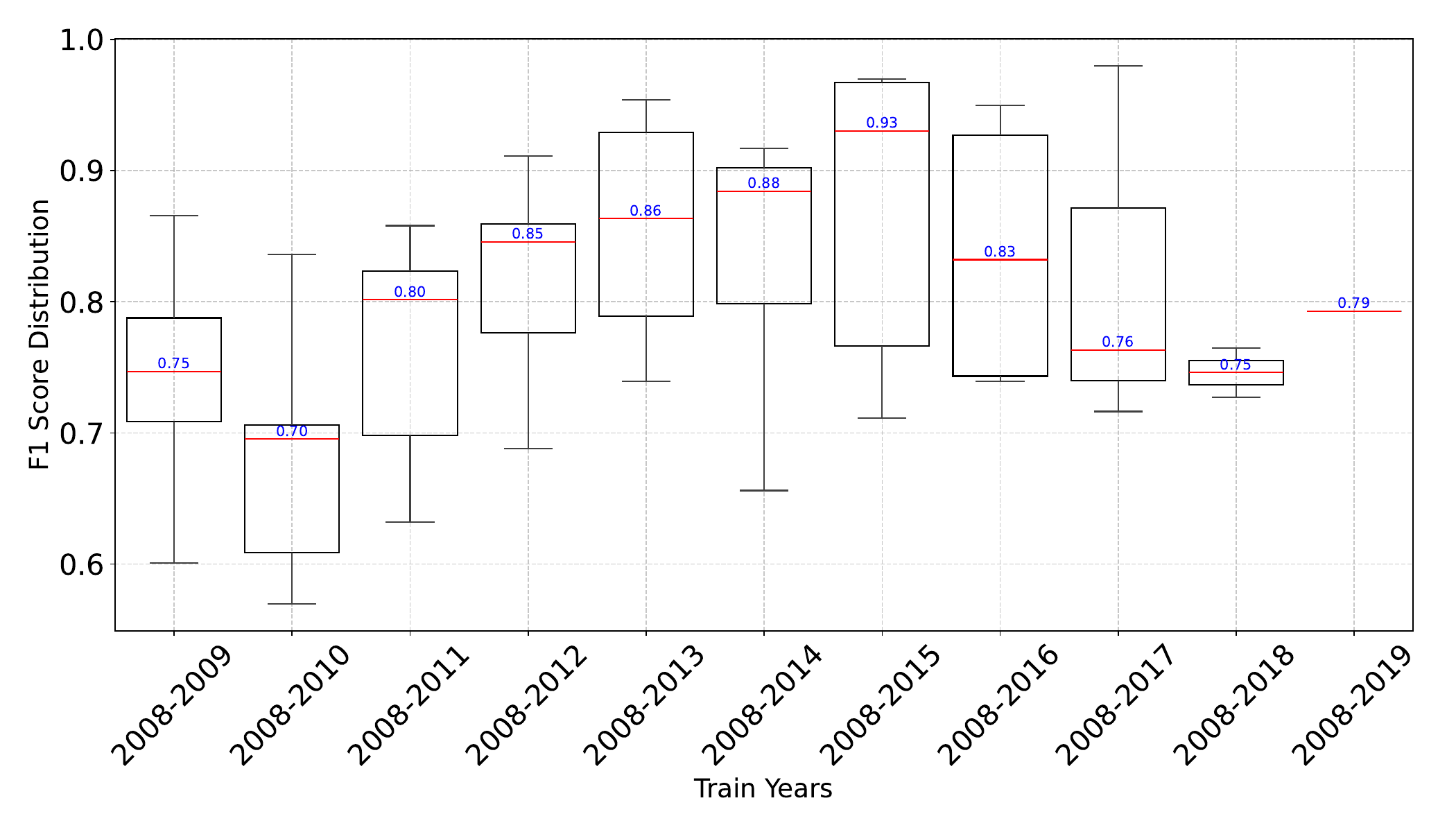}\vspace{-2mm}
        \caption{\normalfont Post-Emu-F1.}
        \label{fig:BP_F1_Post_ATAT_RNN_V2_Emu_Hybrid_Incremental}
    \end{subfigure}

    \vspace{-2mm}
    \caption{RNN performance under incremental training with real and emulator data and hybrid features.}
    \label{fig:BoxPlot_RNN_Incremental_Emu_Real_F1} 
\end{figure}

Figures~\ref{fig:BoxPlot_RF_Static_Incremental_Emu_Real_F1}--\ref{fig:BoxPlot_RNN_Incremental_Emu_Real_F1} present the F1 score distributions for the RF, GRU, and RNN models under an incremental training strategy across training years, evaluating both real and emulator data before and after class balancing. In Figure~\ref{fig:BoxPlot_RF_Static_Incremental_Emu_Real_F1}, the RF model trained with static features shows high and stable performance across time. In Fig.~\ref{fig:BoxPlot_RF_Static_Incremental_Emu_Real_F1}(a), \textit{Pre-Real-F1}, the model starts with moderate performance (F1$\approx$0.7) and gradually improves to near-perfect scores (F1$\approx$0.99) by the later years. Fig.~\ref{fig:BoxPlot_RF_Static_Incremental_Emu_Real_F1}(b), \textit{Post-Real-F1}, confirms this trend with reduced variance and consistently strong scores across all years. The emulator-based settings in Fig.~\ref{fig:BoxPlot_RF_Static_Incremental_Emu_Real_F1}(c)--(d) show similar gains, although the variance is notably higher before balancing. After balancing, F1 scores stabilize, indicating improved emulator generalization.

In contrast, Figure~\ref{fig:BoxPlot_RF_Incremental_Emu_Real_F1} shows the GRU model trained with dynamic features. Fig.~\ref{fig:BoxPlot_RF_Incremental_Emu_Real_F1}(a), \textit{Pre-Real-F1}, reveals an upward trend in performance as more training years are accumulated, though variance remains high. Fig.~\ref{fig:BoxPlot_RF_Incremental_Emu_Real_F1}(b), \textit{Post-Real-F1}, shows improved medians and compressed box ranges, reflecting enhanced learning consistency due to balancing. The emulator-based settings (Fig.~\ref{fig:BoxPlot_RF_Incremental_Emu_Real_F1}(c)--(d)) again highlight the benefits of balancing, where pre-balancing F1 scores fluctuate more wildly, while post-balancing yields smoother, upward trajectories with less dispersion.

Figure~\ref{fig:BoxPlot_RNN_Incremental_Emu_Real_F1} illustrates similar incremental training results for the RNN model using hybrid features. Fig.~\ref{fig:BoxPlot_RNN_Incremental_Emu_Real_F1}(a), \textit{Pre-Real-F1}, starts from modest performance (F1$\approx$0.5) and climbs to around 0.9 as training years accumulate, with noticeable year-to-year variance. Fig.~\ref{fig:BoxPlot_RNN_Incremental_Emu_Real_F1}(b), \textit{Post-Real-F1}, improves both central tendency and spread, showing that balancing increases robustness. Fig.~\ref{fig:BoxPlot_RNN_Incremental_Emu_Real_F1}(c)--(d) demonstrate similar emulator dynamics: pre-balancing performance is noisy and less predictable, while post-balancing greatly enhances consistency and average F1 scores.

Comparatively, the RF model benefits most prominently from the incremental strategy, achieving near-perfect scores and low variance with static features. Both GRU and RNN models show progressive gains over time but with more fluctuation, especially when using emulator data. Class balancing consistently improves all models across data types by reducing variance and boosting median performance. However, the absolute F1 scores and convergence speed differ across models and feature types: RF with static features reaches saturation quickly, while GRU and RNN require more training years to stabilize, particularly under emulator data where performance gains are more gradual and sensitive to imbalance. RNN exhibits more stable and higher performance than GRU, especially after balancing, indicating better generalization with hybrid features.

\end{document}